\documentclass{article}

\usepackage[margin=.75in]{geometry}

\usepackage{authblk}

\usepackage[english=american]{csquotes} 

\usepackage{amsmath,amssymb,amsfonts,amsthm,mathtools} 

\usepackage{graphicx}
\usepackage[labelfont=bf]{caption} 
\usepackage{subcaption}
\usepackage{float} 
\usepackage{chngcntr}
\counterwithin{figure}{section}
\counterwithin{table}{section}

\usepackage[dvipsnames,table]{xcolor}

\usepackage[numbers,sort]{natbib}

\usepackage{array, multirow, longtable, booktabs} 
\newcommand{\grayhline}{\\[-10pt]\arrayrulecolor{gray!60}\specialrule{0.1pt}{0pt}{0pt}\arrayrulecolor{black}\\[-10pt]} 

\usepackage{textcomp, gensymb} 
\usepackage{enumitem} 
\usepackage[normalem]{ulem} 

\usepackage{tikz}
\usetikzlibrary{shadows,calc,arrows.meta,positioning,backgrounds}
\usepackage{adjustbox}
\newcommand{\quartettree}[1]{%
  \adjustbox{valign=m, raise=0.5ex, scale=0.6}{%
    \begin{tikzpicture}[
      grow=up,
      level distance=2mm,
      sibling distance=6mm,
      every node/.style={fill, circle, inner sep=0pt, minimum size=1.2mm},
      edge from parent/.style={draw, -}
    ]
      \ifcase#1
        \node {}
          child [xshift=-1.5mm] {node {}
            child [xshift=-1.5mm] {node {}
              child [xshift=-1.5mm] {node {}}
              child [xshift=1.5mm] {node {}}
            }
            child [xshift=-0.5mm, yshift=2mm] {node {}}
          }
          child [xshift=-2mm, yshift=4mm] {node {}};
      \or
        \node {}
          child {node [xshift=-4mm] {}
            child [xshift=-1.7mm] {node {}}
            child [xshift=1.7mm] {node {}}
          }
          child [xshift=-1.6mm, yshift=2mm] {node {}}
          child [xshift=1.5mm, yshift=2mm] {node {}};
      \or
        \node {}
          child [xshift=-0.3mm, yshift=-0.4mm] {node {}
            child [xshift=-2.8mm, yshift=0.4mm] {node {}}
            child [xshift=0mm, yshift=0.4mm] {node {}}
            child [xshift=2.8mm, yshift=0.5mm] {node {}}
          }
          child [xshift=-1.5mm, yshift=2mm] {node {}};
      \or
        \node {}
          child {node {}
            child [xshift=-1.3mm, yshift=0.5mm] {node {}}
            child [xshift=1.3mm, yshift=0.5mm] {node {}}
          }
          child {node {}
            child [xshift=-1.3mm, yshift=0.5mm] {node {}}
            child [xshift=1.3mm, yshift=0.5mm] {node {}}
          };
      \or
        \node {}
          child [xshift=-3.5mm, yshift=2mm] {node {}}
          child [xshift=-1mm, yshift=2mm] {node {}}
          child [xshift=1mm, yshift=2mm]  {node {}}
          child [xshift=3.5mm, yshift=2mm]  {node {}};
      \fi
    \end{tikzpicture}
  }%
}

\theoremstyle{definition}

\usepackage[listings,breakable]{tcolorbox}
\definecolor{verylightgray}{rgb}{0.95,0.95,0.95}
\usepackage{xurl} 
\usepackage[hidelinks]{hyperref}

\title{Assessing 3D Tree Model Quality and Species Classification Using Imbalance Indices}

\author[1]{Sophie J. Kersting\thanks{\url{sophie.kersting@uni-greifswald.de}}}
\author[1]{Mareike Fischer\thanks{Corresponding author: \url{mareike.fischer@uni-greifswald.de, email@mareikefischer.de}}}

\affil[1]{Institute of Mathematics and Computer Science, University of Greifswald, Greifswald, Germany}

\date{}

\begin{document}
\maketitle

\begin{abstract}
We investigate the use of additional 3D and phylogenetic non-3D tree balance indices for analyzing and monitoring forests using an exemplary \enquote{virtual forest} dataset from the Wytham Woods, Oxford, UK. This study assesses 3D model quality, species classification performance, and the relevance of these indices.

Our study shows that indices stemming from the study of ancestry trees of species can be successfully applied to 3D models of organic trees and, accompanied with recently introduced 3D imbalance indices, offer a complementary perspective on 3D tree models and improve the detection of deviations. Their computational efficiency combined with the simple and reproducible workflow presented in this manuscript form a computationally feasible quality control step in the 3D model construction.

Species classification models reached an estimated accuracy of up to 81.8\% and allowed to make confident species predictions for a large portion of the unlabeled trees in the dataset. While conventional tree metrics can already provide strong predictive performance, the addition of filtered 3D and non-3D statistics improved results consistently, particularly for minority species classes.

Alongside this manuscript, we provide updated functionality in the \textsf{R} package \textsf{treeDbalance} to include the necessary functionalities and release the derived index datasets and species predictions.
\end{abstract}


\section{Introduction} \label{sec:introduction}
Plant architecture, especially the branching structure of trees, contains valuable information about the plants' development, health, and environmental interactions \citep{jackson_finite_2019, schelhaas_introducing_2007, rid_apple_2016, baltenberger_reactions_1987, kunz_neighbour_2019, lau_quantifying_2018}. In the context of climate change, high-quality estimates of stored CO$_2$, the monitoring of tree health, and the study of adaptive responses to heat and drought are becoming increasingly important, where the 3D architecture of trees provide key structural indicators for these analyses. To address one aspect of 3D plant architecture, a set of 3D imbalance indices was recently introduced that can measure asymmetry in so-called rooted 3D tree models by considering subtree centroids and growth directions \citep{kersting_measuring3D_2024}. These 3D indices were inspired by physical pendulum models and can quantify both \enquote{external} imbalance describing how much the plant has grown unevenly with respect to the horizontal plane, i.e., how much the whole plant leans to one side, as well as \enquote{internal} imbalance characterizing how irregular, crooked, and twisted the plant's various parts have grown. Furthermore, in the literature there exists a multitude of topological (im)balance indices from phylogenetics (see \cite{fischer_tree_2023} for an overview of at least 30 (families of) indices), which disregard spatial orientation and focus on the branching patterns alone. 

In this study, we apply the diverse range of 3D and non-3D tree balance indices (all have linear computation time) to a large dataset of 3D tree models from a \enquote{virtual forest} dataset from the Wytham Woods, Oxford, UK \cite{calders_laser_2022}. Each tree is represented by ten alternative 3D tree model reconstructions (QSMs), allowing us to assess the consistency of QSM reconstruction and the indices' sensitivity to QSM reconstruction variation. We pursue three key questions: 

\begin{enumerate}
\item Can deviating index values across reconstructions indicate flaws and uncertainty in the QSM reconstruction and guide the selection of the most representative QSM per tree?
\item Can balance indices help infer the species identity of trees labeled \enquote{Unknown}, and how big is the impact of the addition of 3D and non-3D indices?
\item Which indices contribute most strongly to these tasks and what insights can be drawn from their relations and comparative performance?
\end{enumerate}

To this end, we updated our openly-available \textsf{R} package \textsf{treeDbalance} to include functionality for transforming QSMs into rooted 3D trees, allowing the application of 3D imbalance indices, as well as extracting the corresponding topological trees, allowing the application of standard (im)balance indices. We combine this with statistical and machine learning techniques to assess 3D tree model robustness, perform species classification, and evaluate index importance.

Analyzing the consistency of the ten QSMs per tree provided the necessary information to sort out \enquote{faulty} QSMs and ultimately select a \enquote{best} QSM for each tree. This uncovered common construction errors that can be confirmed with the naked eye, but also error patterns captured by some new indices that cannot be identified, e.g., by the computationally involved comparison with the corresponding point cloud.

Training random forest and gradient boosting models allowed to make confident species predictions for a large portion of trees with unknown species label. While the addition of the 3D and non-3D statistics improved this performance consistently when filtered for redundancy, commonly used tree statistics (DBH, height, volume, crown area, etc.) can already form the basis for sufficiently well performing models. The resulting datasets of index values and other statistics of the QSMs, as well as the species predictions have been made available at \url{https://github.com/SophieKersting/SupplementaryMaterial/blob/main/WW\_QSMquality\_SpeciesClass}.

The manuscript is structured as follows: Section~\ref{sec:materialsmethods} introduces the Wytham Woods dataset and describes the QSM-to-rooted-3D-tree-model conversion pipeline, followed by explanations regarding the indices/statistics and methods employed. In Section~\ref{sec:results}, we present our findings on QSM quality, species classification, and importance of the individual statistics. Section~\ref{sec:discussion} discusses the implications of these results, and concludes with ideas for future research. Appendix~\ref{sec:software} provides an overview of the new additions to the software package \textsf{treeDbalance} while Appendix~\ref{sec:tablesAndfigures} contains additional tables and figures.

\section{Materials and Methods} \label{sec:materialsmethods}

This section provides all necessary information on the data used for this study, including explanations of the (3D) tree model formats and statistics that can be derived from those as well as a description of the workflow that produced the final dataset, but also a list of methods used for the various research questions.

Since the main focus of the present  manuscript lies in the evaluation of the dataset rather than in the mathematical details of the individual tree model formats and features/statistics, we keep their descriptions and lists as simple as possible, using only the most essential mathematical notations. The next paragraphs summarize the formats as well as the whole data collection process, entailed by the statistics and methods.

\subsection{The Wytham Woods dataset} \label{sec:dataset}
\begin{figure}[ht]
\centering
\includegraphics[width=\textwidth]{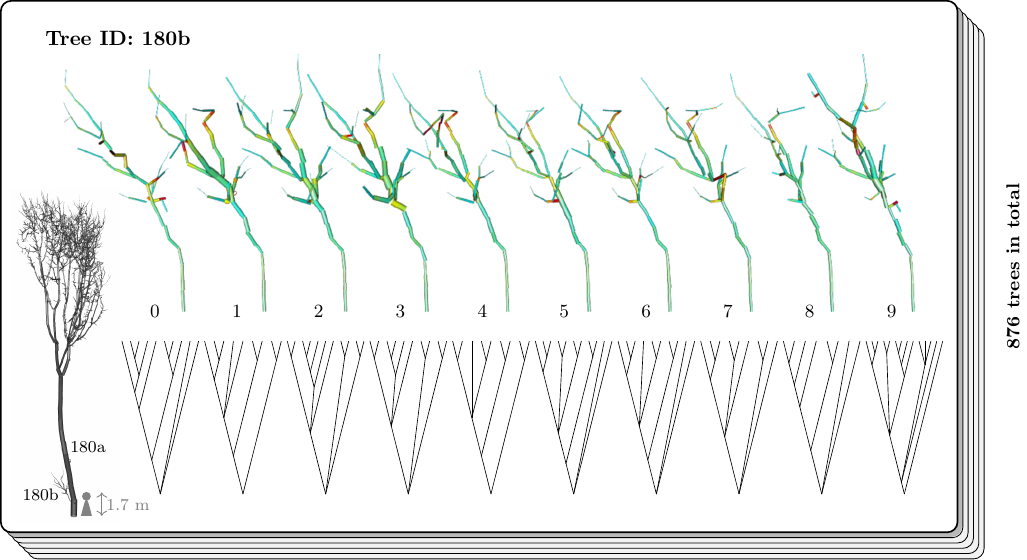}
\caption{A visualization of the dataset and the workflow: Each tree has ten different 3D tree models (in the format of a QSM or a rooted 3D tree model) as well as the ten corresponding extracted non-3D topologies, here exemplarily depicted for the sycamore (ACERPS) with ID 180b. The rooted 3D trees are colored according to their internal $\mathcal{A}$ imbalance (see Section~\ref{sec:statistics3DT}), where lighter blue colors show low and darker red tones higher imbalance. This helps to highlight some common flaws/uncertainties of QSMs: cylinders forming a bend/angle in the stem (e.g. 180b$^7$), protruding and non-connecting cylinders (e.g. 180b$^3$ and 180b$^4$) or the absence of certain tree parts (e.g. 180b$^0$). \\
This example was chosen since it is still small and clear (height of $\approx 3.5$ m and $9$ to $14$ tips/leaves in the non-3D topologies) in contrast to the average tree height ($\approx 14$ m) and average number of leaves (791) in the dataset. 
Tree 180b has a sister stem 180a (height $\approx 25.3$ m, with around 1526 leaves). Both are depicted on the left side with a human figure for scale. This also explains why 180b is leaning to the side / has a high external imbalance. \\
In the QSM quality assessment process, 180b$^2$ was chosen as the best and 180b$^9$ as the worst QSM (QSM iterations 2, 6, 5, and 4$\in B_{\text{180b}}$ were candidates for the best QSM, while 1, 7, 8, 3, 0, and 9 were filtered out in the negative selection since they each were outlier-QSMs for 1 to 11 statistics).}
\label{fig:workflow}
\end{figure}

We used a \enquote{virtual forest} dataset which was made available alongside a study by \citet{calders_laser_2022} revealing that current standard estimates potentially underestimate the biomass and bound CO$_2$ in temperate forests. It contains 3D models of the trees inside a 1.4-ha study area located within Wytham Woods, Oxford, UK, derived from 3D point clouds with the software TreeQSM \cite{raumonen_fast_2013, calders_nondestructive_2015}. These point clouds were created from terrestrial laser scanning (TLS) -- a technique which captures the 3D shapes of trees by emitting millions of laser pulses and measuring the time differences between the emission and the detection of the reflected pulses \cite{calders_terrestrial_2020}. 
The dataset covers 876 trees of 6 different species and provides ten 3D models (QSMs) per tree specimen (see Figure~\ref{fig:workflow}). This enables us to assess both intra-tree variability and inter-species differences. We used information about the given QSM models while also transforming them into two different formats which allow us to compute further statistics that provide information about the 3D and non-3D imbalance of the tree models.

\subsection{The three different tree model formats} \label{sec:treeformats}
This section provides an overview of the three tree formats that were investigated. More information on how these formats are related and can be transformed into one another can be found in Appendix \ref{sec:software} (Figures \ref{fig:ex_transformation} and \ref{fig:ex_extraction} provide a quick summary).

\paragraph{Quantitative structure models (QSMs):} 
The 3D tree models of \cite{calders_laser_2022} are given as quantitative structure models (QSMs), a widely-used 3D model format that uses cylinders to represent tree branches and also stores the information on how these are connected \citep{hackenberg_simpleforest_2021, hackenberg_simpletree_2015, raumonen_fast_2013, calders_laser_2022}. These QSMs were derived from terrestrial laser scanning data  and provide precise digital representations of the individual trees. The Wytham Woods dataset contains 10 QSMs each for 876 tree specimens, and thus, 8,760 QSMs in total. The already published datasets on these 3D tree models range from species, location, trunk diameter at breast height, height of the tree, estimation of the crown area, total volume, to estimations of the volume of tree parts of certain diameters.

\paragraph{(Graph-theoretical) rooted 3D trees} 
In this present study, these QSMs were transformed into so-called rooted 3D trees, i.e.,  rooted graph-theoretical trees with vertices that have 3D coordinates and edges with volume/thickness \cite{kersting_measuring3D_2024}. This format holds similar information as a QSM while focusing more on the graph-theoretical branching structure. Thus, converting a QSM into a rooted 3D tree is straightforward without loss of branching information (see Figure~\ref{fig:ex_transformation}). Given the rooted 3D trees, the external and internal imbalance of the trees can be measured. As there are only formal differences between QSMs and rooted 3D trees, they cannot be visually distinguished. Thus, the 3D tree models depicted in Figure~\ref{fig:workflow} in the top row can be interpreted both as depictions of QSMs or rooted 3D trees.

In order to understand the details of the 3D imbalance statistics in Section~\ref{sec:statistics3DT} and non-3D statistics in Section~\ref{sec:statisticsTOP}, we next provide the precise mathematical definitions of rooted 3D trees and the most crucial concepts,  beginning with some general notes: 

A (graph-theoretical) \emph{rooted tree} or \emph{topology}\footnote{In the main part of this manuscript, apart from the mathematical definitions, we will primarily use the term \emph{topology} to reduce confusion with the actual trees.} is a connected directed graph $T$ with finite \emph{vertex/node} set $V(T)$ and \emph{edge} set $E(T)\subseteq V(T)^2$, containing precisely one vertex, called the \emph{root} and denoted $\rho$, such that all edges are directed away from the root. The degree of a vertex is the number of incident edges, consisting of the incoming edges, yielding the \emph{in-degree}, and the edges leaving the vertex, yielding the \emph{out-degree}. The root is the only vertex with in-degree 0. Moreover, there are no vertices with in-degree $>1$. The \emph{leaf}/tip set, which is denoted by $V_L(T)$, is the set of all nodes in $T$ that have out-degree 0, and $n$ is used to refer to its cardinality, i.e., $n=\vert V_L(T)\vert $.
The set of \emph{inner vertices}, denoted by $\mathring{V}(T)$, is the set of all vertices with out-degree $\geq 1$, i.e., $\mathring{V}(T)=V(T)\setminus V_L(T)$. Note that $\vert V(T)\vert= 1$ is the only case where the root is the only vertex in the tree and thus also a leaf. For $\vert V(T)\vert \geq 2$, the root is an inner vertex. 

The \emph{neighbors/adjacent} vertices of a node $v$ are all vertices that connected with $v$ by its incident edges. The root-induced partial order on the vertices (stemming from all edges being  directed away from the root) allows us to refer to a neighbor of a vertex $v$, as its \emph{parent} $p(v)$ if it is closer to the root than $v$ or as its \emph{child} otherwise. Whenever there exists a directed path from a vertex $u$ to a vertex $v$, we say that $u$ is an
\emph{ancestor} of $v$ and $v$ is a \emph{descendant} of $u$ (a node is its own ancestor/descendant). $children(v)$ and $descs(v)$ of a vertex $v$ refer to the set of its children and the set of all its descendant vertices, respectively. $n_v$ denotes the number of descending leaves of vertex $v$. The \emph{depth} $\delta_T(v)$ of a vertex $v$ is its distance to the root measured as the number of edges on the unique path from the root to $v$. The width $w_T(d)$ of tree $T$ at a given depth $d$ is the number of vertices at depth $d$, and the height of a tree is the maximal depth of its leaves: $h(T)=\max\limits_{l\in V_L(T)} \delta_T(l)$. By $T_v$ we denote the \emph{pending subtree} of a rooted tree $T$ rooted in $v$, i.e., $T_v$ consists of all of $v$'s descendants (includes $v$ itself) and all edges connecting them. The \emph{last/lowest common ancestor} $LCA(u,v)$ of two vertices $u$ and $v$ is the unique vertex in the intersection of both vertices' respective set of ancestors -- their common ancestors -- with greatest depth.
Two leaves, say $x,y \in V_L(T)$ are said to form a \emph{cherry} if they have the same parent. Similarly, the unique tree with $n$ leaves is also called a cherry tree.

Now, a \emph{rooted 3D tree} $\mathsf{T}=(T,w)$ is a pair consisting of a rooted tree $T=(V,E)$, the topology, in which $V(T)$ is a subset of $\mathbb{R}^3$, i.e., each vertex $v$ is a distinct point in the three dimensional space, combined with a weight function $w$ that assigns a weight $w(e)$ to each edge $e \in E(T)$. Depending on the application, the weight of an edge $e$ is typically its volume or its physical heaviness, i.e., volume multiplied by density. Similar to non-3D trees, a pending subtree $\mathsf{T}_v=(T_v, w\vert _{E(T_v)})$ consists of the topology $T_v$ and the weight function restricted to the edges in $T_v$, such that $\mathsf{T}_v$ is again a 3D tree.

\paragraph{Non-3D topology}
From rooted 3D trees we can extract their non-3D topologies, which only keeps their branching structure but drops all information on the 3D positions and weights of the edges (see Figure~\ref{fig:ex_extraction}). This allows us to focus on this structural aspect alone and to apply topological imbalance indices, which were initially invented and used in fields like phylogenetics and population genetics. Figure~\ref{fig:workflow} in the bottom row exemplarily shows extracted non-3D topologies.

In the extraction process, the 3D coordinate information is dropped, any \enquote{stem} consisting of a path from the root to the first branching vertex is removed, and other non-branching inner vertices are suppressed (see Appendix \ref{sec:software}). Thus, an extracted non-3D topology $T^*$ from a rooted 3D tree $\mathsf{T}=(T,w)$ is a rooted tree whose vertices are not points in 3D space anymore and which does not contain vertices of both in- and out-degree 1, i.e., all inner nodes have at least 2 children each.

\subsection{Statistics} \label{sec:statistics}
This section gives an overview of all statistics this study is based on -- statistics of the trees, their individual QSMs as well as their corresponding rooted 3D trees and non-3D topologies: Those showcased in the lists are used to address at least one of the research questions. All others, i.e., statistics that were omitted or replaced as explained in Section~\ref{sec:methods}, are only mentioned in the text. Each feature is given an abbreviated name (in bold letters) to be used in the figures throughout this manuscript. Table~\ref{tab:statisticsColumns} provides a comprehensive summary of the features, including their names in the (original) csv-files and their usage in this manuscript (removed, replaced, or used to address one of the research questions).

\subsubsection{General features} \label{sec:statisticsGEN}
First, we go over some general features of the trees that are not directly connected to the 3D tree model (given in the \texttt{analysis\_and\_figures} directory of the supplementary material of \cite{calders_laser_2022}, and there mainly in the \texttt{tls\_summary.csv} file): 

Every tree (and its corresponding point cloud) has a unique \textbf{Tree ID}, and the 10 different QSM versions corresponding to the tree are indicated by a number ranging from 0-9, their \textbf{QSM iteration} number. Throughout this manuscript we refer to a specific QSM of a tree with ID$^\text{iteration}$, e.g., 60$^4$ for the QSM number 4 of the tree with ID 60. Several trees in the study area of the Wytham Woods consisted of multiple stems and the stems were treated as individual trees if the split into single stems occurred below 1.3 m \cite{calders_laser_2022}. These individual stem-trees have the same ID number followed by a letter, e.g., 180a and 180b (see Figure~\ref{fig:workflow}). Thus, we introduced the statistic \textbf{StemCount} -- the number of such \enquote{sister}-stems in the tree. Furthermore, we know for each tree specimen its stem's precise location (\textbf{Locx} and \textbf{Locy}, i.e., x- and y-coordinate) in the rectangular study area in the Wytham Woods \cite{calders_laser_2022}.

Of the 876 trees only 835 were actually regarded as \enquote{inside} of this study area, i.e., at least half of their stem was inside the boundaries. For these trees -- the census trees -- there is more information available (called census data in \cite{calders_laser_2022}, see \texttt{trees\_summary.csv}): A manual measurement of the diameter at breast height (\textbf{DBHc}) which will be used to assess the validity of the QSMs (available for only 695 trees), as well as the \textbf{Species} of each tree and whether each tree was alive or \textbf{Dead} at the time of investigation (815 alive, 20 dead).

\begin{figure}[ht]
\centering
\includegraphics[width=\textwidth]{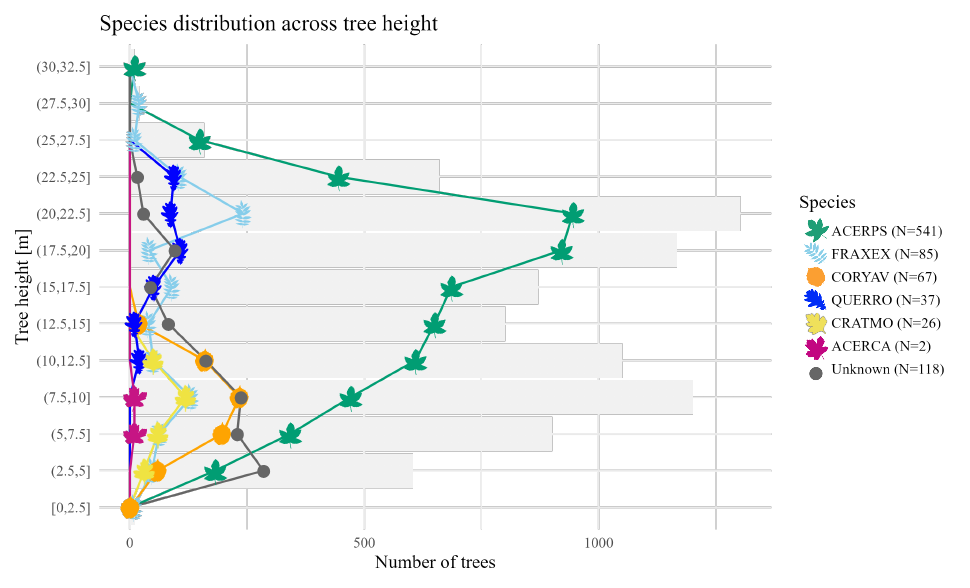}
\caption{Species distribution across tree height. The barplot in the background shows the total number of trees per height class and the points the respective numbers for the individual species. The points are shaped as the leaves of the corresponding tree species and are only depicted for values $>0$. The legend also provides the respective total number of trees per species (N). The colors for the species are the same as in \cite[e.g., Fig.~4]{calders_laser_2022} to allow easy transfer between visualized results between both studies.}
\label{fig:speciesDistrib}
\end{figure}

The tree species of the 835 census trees -- including their number of trees $N$ in the dataset as well as their Latin name and abbreviations in capital letters -- range from field maple (N=2, ACER CAmpestre), sycamore (N=541, ACER PSeudoplatanus), common hazel (N=67, CORYlus AVellana), common hawthorn (N=26, CRATaegus MOnogyna), European/common ash (N=85, FRAXinus EXcelsior), to pedunculate/English oak (N=37, QUERcus RObur)\footnote{At least 23 tree species have been recorded in the larger 18-ha long-term forest inventory plot of the Wytham Woods run by Oxford University \cite{butt_initial_2009}, but only these six species have been identified in the scanned 1.4-ha area within the larger plot.}. 78 trees (71 living + 7 dead specimens) of the 835 census trees could not be assigned a species \cite[p.~4]{calders_laser_2022}. Combined with the 41 remaining non-census trees for which no species information is available, this yields 119 trees with species label \enquote{unknown} (see Figure~\ref{fig:speciesDistrib}). Apart from these two features, Dead and Species, all other information is available for all trees. Thus, there is no further difference between the 835 census trees and the 41 other trees in our study.

\subsubsection{QSM statistics} \label{sec:statisticsTLS}
Regarding features of the actual 3D models of the trees, many were already provided in the \texttt{tls\_summary.csv} of \cite{calders_laser_2022}.
The downside of these features is that they have been averaged over all ten QSMs per tree specimen (or have only been estimated from the tree's TLS point cloud) and as such do not display any variation between different QSMs of the same tree and, thus, could not be used regarding the first key question of this manuscript: evaluating the quality of the QSMs with the goal of finding a \enquote{best} QSM for each tree. Thus, we recalculated all statistics for which it was possible per QSM from the rooted 3D tree models (see second list below). The following list contains all existing features used in this present study that show no differences between QSMs of the same tree. Each feature is given an abbreviated name to be used in the figures throughout this manuscript.

\begin{description}
\item [DBHsd] Standard deviation of trunk diameter at breast height (that is at 1.3 m) in m over all 10 QSM. 
\item [CrownArea]: Estimated area of the crown in m$^2$.
\item [Volsd]: Standard deviation of total volume in m$^3$ over all 10 QSM iterations.
\item [Vol0-25sd, etc.]: Standard deviation of total volume of QSM parts/cylinders with a diameter between 0 and 25mm in m$^3$ over all 10 QSM iterations. Similar statistics exist for the diameter ranges 25-50 mm, 50-75 mm, 75-100 mm, 100-200 mm.
\end{description}

The census DBH measurement (\texttt{DBH\_census\_m}) and three further estimations of the DBH from the TLS point clouds (\texttt{DBH\_TLS\_m} and \texttt{DBH\_pts\_m}) and the QSM averages (\texttt{DBH\_QSM\_avg\_m}), as well as a tree height estimation from the TLS point clouds (\texttt{Hgt\_pts\_m}), a volume estimation averaged over all QSMs (\texttt{Vol\_QSM\_avg\_m3}), and volume estimations of QSM parts with a certain diameter averaged over all QSMs (\texttt{Vol\_QSM\_D0\_25\_avg\_m3} etc.) were replaced by the QSM-specific recomputed statistics as listed below.

The following list contains the features that were recalculated (in cases where only the average over all QSMs is known) or additionally computed per QSM. From this point onward in Section~\ref{sec:statistics}, all features are \emph{QSM-specific}.

\begin{description}
    \item [DBH] Diameter of the main stem of the QSM at breast height in m.
    \item [Height] Height of the tree (model) in m (difference between maximal and minimal $z$-coordinates in the QSM).
    \item [Volume] Total volume of the tree (model) in m$^2$ (volume of all cylinders in the QSM).
    \item [Vol0-25, etc.]: Total volume of QSM cylinders with a diameter between 0 and 25 mm in m$^3$. Similar statistics exist for the diameter ranges 25-50 mm, 50-75 mm, 75-100 mm, 100-200 mm, and $>200$ mm.
    \item [Length] Summed up length of all branches in the tree (model) in m (length of all QSM cylinders and connecting branches).
    \item [ZeroCyl] The number of cylinders in the 3D tree model with zero weight/volume. A cylinder has zero weight or volume if it has no length or no radius.
\end{description}

\subsubsection{3D imbalance statistics} \label{sec:statistics3DT}
Several 3D imbalance statistics measuring the internal and external balance were computed from the rooted 3D trees. The final selection of 3D measures used in this study is:
\begin{description}
    \item [Ext(A)(A/alpha/M/mu)] External imbalance, i.e., how far the tree leans to the side, as measured by different approaches: The root's \emph{centroid angle} $\mathcal{A}(\rho)$ calculates the angle $\mathcal{A}(\rho) \coloneqq \angle(\mathcal{C}(\mathsf{T})-\rho, (0,0,1)^t)$ between the line from the base of the stem (the \enquote{root} of the rooted 3D tree) to the centroid of the tree and the vertical axis. With the \emph{minimal centroid angle} approach we have $\alpha(\rho) \coloneqq \mathcal{A}_{\mathsf{T},e_v}(v)$ if $0\leq \mathcal{A}(\rho)\leq \frac{\pi}{2} = 90\degree$ and $\pi-\mathcal{A}(\rho)$ if $\frac{\pi}{2} < \mathcal{A}(\rho)\leq \pi = 180\degree$. The \emph{relative centroid distance} yields $\mu(\rho) \coloneqq \frac{d(\mathcal{C}(\mathsf{T}), g_{vert})}{d(\mathcal{C}(\mathsf{T}),\rho)}$, where $g_{vert}$ is the vertical line going through the root $\rho$, and the \emph{expanded relative centroid distance} yields $\mathcal{M}(\rho) \coloneqq \mu(\rho)$ if $0\leq \mathcal{A}(\rho)\leq \frac{\pi}{2} = 90\degree$ and $2-\mu(\rho)$ if $\frac{\pi}{2} < \mathcal{A}(\rho)\leq \pi = 180\degree$.

    \item [Int-w(A)(A/alpha/M/mu)] Internal imbalance as measured by a weight-weighted integral-based index using different imbalance measuring approaches:\\
    The centroid angle yields the index $\widetilde{\mathcal{A}}^{w}$, which calculates the average 3D imbalance over the volume (\enquote{weights} of the edges) of the tree, where the 3D imbalance at a certain point $v$ of a branch is measured similarly to the external imbalance: It is the angle between the branch leading up to this point and the line to the centroid $\mathcal{C}(\mathsf{T}_v)$ of the part of the tree that starts from there.\\
    The formal definition is as follows: Let $\mathsf{T}=((V,E),w)$ be a rooted 3D tree, then $\widetilde{\mathcal{A}}^{w}(\mathsf{T}) \coloneqq 0$ if  $\vert V\vert  = 1$ and otherwise
    \[\widetilde{\mathcal{A}}^{w}(\mathsf{T})  \coloneqq \frac{1}{\sum\limits_{v\in V\setminus\{\rho\}}{w(e_v)}} \sum_{v\in V\setminus\{\rho\}}{\left(w(e_v) \cdot \int_{0}^{1} \mathcal{A}_{\mathsf{T},e_v}(v+(p(v)-v)\cdot x) \,\mathrm{d}x \right)},\]
    where $e_v=(p(v),v)$ denotes the incoming edge of $v$ and $\mathcal{A}_{\mathsf{T},e_v}: V(\mathsf{T})\setminus\{\rho\} \to [0,\pi]$ is the \emph{centroid angle} of a node $v \neq \rho$ (or edge subdivision) defined as $\mathcal{A}_{\mathsf{T},e_v}(v) \coloneqq 0$ if $\mathcal{C}(\mathsf{T}_v)=v$ and otherwise as the angle
    $\mathcal{A}_{\mathsf{T},e_v}(v) \coloneqq \angle(\mathcal{C}(\mathsf{T}_v)-v, v-p(v))$.\\
    Indices $\widetilde{\alpha}^{w}$, $\widetilde{\mathcal{M}}^{w}$, and $\widetilde{\mu}^{w}$ for the other approaches are formed analogously by exchanging $\mathcal{A}_{\mathsf{T},e_v}$ in the formula above: If $\mathcal{C}(\mathsf{T}_v)\neq v$, the \emph{minimal centroid angle} is $\alpha_{\mathsf{T},e_v}(v) \coloneqq \mathcal{A}_{\mathsf{T},e_v}(v)$ if $0\leq \mathcal{A}_{\mathsf{T},e_v}(v)\leq \frac{\pi}{2}$ and $\pi-\mathcal{A}_{\mathsf{T},e_v}(v)$ if $\frac{\pi}{2} < \mathcal{A}_{\mathsf{T},e_v}(v)\leq \pi$, the relative centroid distance is $\mu_{\mathsf{T},e_v}(v) \coloneqq \frac{d(\mathcal{C}(\mathsf{T}_v), g_{v,p(v)})}{d(\mathcal{C}(\mathsf{T}_v),v)}$, with $d$ being the Euclidean distance and $g_{v,p(v)}$ the line $v+\lambda (p(v)-v)$ with $\lambda \in \mathbb{R}$ going through $v$ and $p(v)$, and the expanded relative centroid distance is $\mathcal{M}_{\mathsf{T},e_v}(v) \coloneqq \mu_{\mathsf{T},e_v}(v)$ if $0\leq \mathcal{A}_{\mathsf{T},e_v}(v)\leq \frac{\pi}{2}$ and $2-\mu_{\mathsf{T},e_v}(v)$ if $\frac{\pi}{2} < \mathcal{A}_{\mathsf{T},e_v}(v)\leq \pi$. All three measures are $0$ if $\mathcal{C}(\mathsf{T}_v)=v$.
    
    \item [Int-l(m)(A/alpha/M/mu)] Internal imbalance as measured by a length-weighted integral-based index using different imbalance measuring approaches. These indices are structurally similar to Int-w(A) but differ in the weighting method. Exemplarily, for the centroid angle based index $\widetilde{\mathcal{A}}^{\ell}$ we have $\widetilde{\mathcal{A}}^{\ell}(\mathsf{T}) \coloneqq 0$ if  $\vert V\vert  = 1$ and otherwise
    \[\widetilde{\mathcal{A}}^{\ell}(\mathsf{T})  \coloneqq \frac{1}{\sum\limits_{v\in V\setminus\{\rho\}}{\ell(e_v)}} \sum_{v\in V\setminus\{\rho\}}{\left(\ell(e_v) \cdot \int_{0}^{1} \mathcal{A}_{\mathsf{T},e_v}(v+(p(v)-v)\cdot x) \,\mathrm{d}x \right)}.\]
    Indices $\widetilde{\alpha}^{\ell}$, $\widetilde{\mathcal{M}}^{\ell}$, and $\widetilde{\mu}^{\ell}$ for the other three approaches are formed analogously by exchanging $\mu_{\mathsf{T},e_v}$ in the formula above.
\end{description}

\subsubsection{Topological/non-3D statistics} \label{sec:statisticsTOP}
This section provides information on all statistics derived from the extracted non-3D topologies. Although the features are known as \enquote{tree shape statistics} or \enquote{tree balance indices}, we use the terms \enquote{non-3D} or \enquote{topological statistics} to avoid confusion with the actual biological trees. 

As basic topological features, the number of leaves/tips $n=|V_L(T)|$ (LeafN) and the number of inner vertices $|\mathring{V}(T)|$ (InnerN) in the topology $T$ were collected. Since both highly correlated with other statistics, their ratio was introduced as a measure of the resolution of the tree, i.e., how little multifurcating nodes there are. $n-1$ is the maximal number of inner vertices (reached in a fully binary tree).
\begin{description}
    \item [TopRes] The topological resolution based on the ratio $\frac{|\mathring{V}(T)|}{(n-1)}$ of the number of inner vertices and the maximal possible number of inner vertices in the topology $T$.
\end{description}

Next, we were also interested in the topological imbalance. Although there is a wide range of tree shape statistics, especially (im)balance indices, for rooted trees in phylogenetics \cite{fischer_tree_2023}, only a subset is applicable to the extracted non-3D topologies of the Wytham Woods trees (see \cite[Table~4.1 \& 4.2]{fischer_tree_2023} for details). Several topological statistics can only be applied to strictly binary trees, i.e., trees in which edges/branches can only split up into two branches and not more at the same time. However, a quick exploration of the 3D tree models revealed that 98.5\% (8,632 of the 8,760 QSMs) contain so-called multifurcating vertices indicating points where an edge branches into more than two edges. Therefore, we only applied statistics which are also suitable for non-binary trees. These are:

\begin{description}
    \item [ALD] The average leaf depth~\cite{sackin_good_1972, shao_tree_1990}: $\overline{N}(T) \coloneqq \frac{1}{n}\cdot\sum\limits_{l \in V_L(T)} \delta_T(l)$.
    
    \item [AVD] The average vertex depth~\cite{herrada_scaling_2011, ford_probabilities_2005, blum_which_2006, hernandez_simple_2010}:
    $AVD(T) \coloneqq \frac{1}{|V(T)|}\cdot\sum\limits_{v \in V(T)} \delta_T(v)$.
    \item [B1] The $B_1$ index~\cite{shao_tree_1990} is the sum of the reciprocals of the heights of the subtrees of $T$ rooted at inner vertices of $T$ (except for $\rho$): $B_1(T) \coloneqq \sum\limits_{v \in \mathring{V}(T) \setminus \{\rho\}} \frac{1}{h(T_v)}$
    \item [B2] $B_2$ index~\cite{agapow_power_2002,hayati_new_2019,kirkpatrick_searching_1993,shao_tree_1990}: Shannon-Wiener information function (measures the equitability of the probabilities $p_l(T)$ of reaching the leaves $l \in V_L$ in a tree $T$ when starting at the root and assuming equiprobable branching at each inner vertex):
    $B_2(T) \coloneqq - \sum\limits_{l \in V_L(T)} p_l \cdot \log_2(p_l)$.
    \item [Cherry] The cherry index~\cite{mckenzie_distributions_2000} is defined as the number $c(T)$ of cherries in the tree: $ChI(T) \coloneqq c(T)$.
    \item [CLe] A Colless-like index $\mathfrak{C}_{D,f}$ with $f=\exp$ and $D=$ mean deviation from the median: A representative of the family of Colless-like indices~\cite{mir_sound_2018}, which calculate the sum of $(D,f)$-balance values of the inner vertices of $T$. This balance value of an inner vertex $v$ is defined as $bal_{D,f}(v)  \coloneqq D(\Delta_f(T_{v_1}), \ldots, \Delta_f(T_{v_k}))$ over the so-called $f$-sizes of its children $v_1,...,v_k$, where $\Delta_f(T_{v_i}) \coloneqq\sum\limits_{v \in descs(v_i)} f(|children(v)|)$. Then,
    $\mathfrak{C}_{D,f}(T) \coloneqq \sum\limits_{v \in \mathring{V}(T)} bal_{D,f}(v)$..
    \item [CLln] Another Colless-like index~\cite{mir_sound_2018} with $f=\ln$ and $D$ being the mean deviation from the median.
    \item [mD] The maximum depth~\cite{colijn_phylogenetic_2014} or height of a tree: $mD(T) \coloneqq \max\limits_{l\in V_L(T)} \delta_T(l) = \max\limits_{v\in V(T)} \delta_T(v) = h(T)$.
    \item [mW] The maximum width~\cite{colijn_phylogenetic_2014}: $mW(T) \coloneqq \max\limits_{i=0,\ldots,h(T)} w_T(i)$.
    \item [mDW] The Modified maximum difference in widths $mDW$~\cite{fischer_tree_2023,colijn_phylogenetic_2014} is the maximum difference in widths of two consecutive depths:
    $mDW(T) \coloneqq \max\limits_{i=0,\ldots,h(T)-1} {w_T(i+1)-w_T(i)}$.
    \item [mWomD] The maximum width over maximum depth~\cite{colijn_phylogenetic_2014} is defined as $W/D(T) \coloneqq \frac{mW(T)}{mD(T)}$.
    \item [mI'] Mean $I'$ index $\overline{I'}$~\cite{fusco_new_1995,purvis_evaluating_2002}: A representative of the family of $I$-based indices which calculates the mean of the $I_v'$ values over all inner vertices $v$ with $n_v \geq 4$ and exactly two children (let $\mathring{V}_{\geq4,2}(T)$ denote this set of vertices). The balance value $I_v$ measures how uneven the descending leaves are split with regard to the most uneven possible split and is defined as $I_v = \frac{n_{v_1}-\lceil\frac{n_v}{2}\rceil}{(n_v-1)-\lceil\frac{n_v}{2}\rceil}$, where $v_1$ is the child of $v$ with the larger  number of descendant leaves $n_{v_1}$. The $I'$-correction, $I_v'=  I_v$ if $n_v$ is odd and $I_v'=\frac{n_v-1}{n_v} \cdot I_v$ else, ensures independence from the \enquote{tree size} $n$ (under certain basic assumptions).  Then, 
    $\overline{I'}(T) \coloneqq \frac{1}{\vert \mathring{V}_{\geq4,2}(T) \vert} \cdot  \sum\limits_{v \in \mathring{V}_{\geq4,2}(T)} I'_v$.
    \item [mIw] Another $I$-based index, namely the mean $I^w$ index~\cite{fusco_new_1995,purvis_evaluating_2002}: $\overline{I^w}(T) \coloneqq \frac{1}{\vert \mathring{V}_{\geq4,2}(T) \vert} \cdot  \sum\limits_{v \in \mathring{V}_{\geq4,2}(T)} I^w_v$. This uses a different correction method for the $I_v$-values namely forming a weighted mean: The $I^w_v$ value of a vertex $v\in\mathring{V}_{bin,\geq 4}$ is defined as
    $I^w_v \coloneqq \frac{w(I_v) \cdot I_v}{\text{mean}_{v \in \mathring{V}_{bin,\geq 4}} w(I_v)}$
     \quad with weights \quad $w(I_v) \coloneqq \begin{cases} 
    1 & \text{if } n_v \text{ is odd} \\ 
    \frac{n_v-1}{n_v} & \text{if }n_v \text{ is even and } I_v>0 \\
    \frac{2\cdot(n_v-1)}{n_v} & \text{if } n_v \text{ is even and } I_v=0. \end{cases}$
    \item [rQi]  Rooted quartet index $rQI$~\cite{coronado_balance_2019} with $q_i=i$: A balance index based on the symmetry of subtrees $T|_{Q}$ induced by quartets $Q$ (sets of four leaves). The possible tree shapes with four leaves have a weight $q_i$ that scales with the number of automorphisms. Let $QC_T(T^*)$ be the quartet count, the number of quartet-induced subtrees in $T$ of shape $T^*$, then,
    $rQI(T) \coloneqq q_0 \cdot QC_T(\,\quartettree{0}) + q_1 \cdot QC_T(\,\quartettree{1}) + q_2 \cdot QC_T(\,\quartettree{2}) + q_3 \cdot QC_T(\,\quartettree{3}) + q_4 \cdot QC_T(\,\quartettree{4})$.
    \item [Sackin] The Sackin index~\cite{sackin_good_1972,shao_tree_1990} (also known as the total external path length) is defined as $S(T) \coloneqq \sum\limits_{l \in V_L(T)} \delta_T(l)$.
    \item [s-shape] The $\widehat{s}$-shape statistic~\cite{blum_which_2006} is the sum of $\log(n_v-1)$ over all inner vertices of $T$: $\widehat{s}(T) \coloneqq \sum\limits_{v \in \mathring{V}(T)} \log(n_v-1)$.
    \item [TC] Total cophenetic index $\Phi$~\cite{mir_new_2013}: The sum of the cophenetic values $\varphi_T(x,y)$ of all distinct pairs of leaves of $T$, where $\varphi_T(x,y) = \delta_T(LCA_T(x,y))$ is the depth of the last common ancestor $LCA_T(x,y)$ of $x$ and $y$. Then, \\
    $\Phi(T) \coloneqq \sum\limits_{\substack{\{x,y\} \in V_L(T)^2 \\ x \neq y}} \varphi_T(x,y)$.
    \item [TIPL] The total internal path length~\cite{knuth_volume3_1998}: $TIP(T) \coloneqq  \sum\limits_{v \in \mathring{V}(T)} \delta_T(v)$.
    \item [TPL] The total path length~\cite{dobrow_total_1999, takacs_total_1992, takacs_total_1994}: $TPL(T) \coloneqq  \sum\limits_{v \in V(T)} \delta_T(v) = S(T)+TIP(T)$.
    \item [VLD] Variance of leaf depths $\sigma_N^2$~\cite{coronado_sackins_2020, sackin_good_1972,shao_tree_1990}: $\sigma_N^2(T) \coloneqq \frac{1}n\cdot\sum\limits_{l \in V_L(T)} \left( \delta_T(l) - \overline{N}(T) \right)^2$.
\end{description}

Detailed definitions and more information on the properties of these indices can be found in \citet{fischer_tree_2023}.

\subsection{Methods and workflow} \label{sec:methods} 

This section provides detailed information on the workflow and the statistical and machine learning methods, mainly outlier detection for analyzing the quality and robustness of QSMs as well as random forests and gradient boosting models for the species classification.

\subsubsection{Data collection} \label{sec:meth_data}

In a parallelized process using the \textsf{R} package \textsf{treeDbalance}, for each tree, each of the 10 QSM files was loaded and transformed into the rooted 3D model format (see Figure~\ref{fig:workflow} top row). This allowed for the computation of the 3D imbalance statistics (Section~\ref{sec:statistics3DT}) as well as several QSM statistics (Section~\ref{sec:statisticsTLS}). Then, the non-3D topology of each rooted 3D model was extracted (see Figure~\ref{fig:workflow} bottom row). Based on this, all non-3D imbalance indices were computed (Section~\ref{sec:statisticsTOP}). All this was combined with the existing data by \citet{calders_laser_2022} to form an initial complete dataset (see \texttt{df\_ww\_all.csv}).

\subsubsection{Data cleaning/preparation (Part I)}  \label{sec:meth_prepI}

In this phase, the dataset was prepared for all subsequent steps yielding \texttt{df\_ww\_prep.csv}. All changes from this procedure are marked in gray in Table~\ref{tab:statisticsColumns}.

Several non-QSM-specific variables of the original datasets of \cite{calders_laser_2022} were replaced by recomputed QSM-specific variables. This affected 4 DBH measurements, as well as height, volume, and tree part volume estimations as explained in Section~\ref{sec:statisticsTLS}.
Furthermore, some sparse or non-informative variables were combined into a new variable (Vol200+ as the sum of the volumes over the diameter ranges 200-500, 500-1000, and $>$1000 mm, where the values for the two thickest categories were both $>0$ for only 9\% of trees) or omitted if this was impossible. The latter was the case for the standard deviations of the volumes of thicker tree parts with diameters 500-1000 and $>$1000 mm, which were both also only non-zero for $<9$\% of trees.

When comparing the pairwise relationships of the variables, most had no clear or a rather linear relation. However, two statistics, the rooted quartet index with $q_i=i$ (rQi) and the Colless-like index with $f=\exp$ (CLe), heavily deviated from this, showing a clear quartic, in the case of rQi, and (at least) quadratic, in the case of CLe, relation with other variables. Thus, these two variables were replaced by \textbf{4thrt-rQi} and \textbf{sqrt-CLe}, respectively. The quartic relationship in rQi is not surprising, since the number of quartets is quartic related to, for example, the number of leaves in a tree ($\binom{n}{4} = \frac{n!}{4!(n-4)!} = \frac{n(n-1)(n-2)(n-3)}{4!} \in O(n^4)$). 

\subsubsection{QSM robustness, quality, and selection}  \label{sec:meth_QSMquali}

In this step, the collected data is used to determine the quality of the individual QSMs with the aim of having a single \enquote{best} QSM per tree. For this, we consider each tree with its 10 QSMs separately. The method is explained in detail below, but the general idea is to let every statistic decide per tree which QSMs are outliers, and then to select the potentially most accurate QSM from all QSMs with minimal outlier count. The process can be traced in Figure~\ref{fig:QSM_quality_results} in the results section.

For this task, only the 46 \enquote{QSM-specific} statistics which have individual values per QSM could be used. Let this set of statistics be denoted by $S$.
This limited the statistics to the re- or newly computed DBH, Height, Length, ZeroCyl, Volume, and the Vol0-25 etc. measurements, as well as the 3D and non-3D statistics (see Table~\ref{tab:statisticsColumns}). 

\paragraph{Method: Outlier detection and selection of the \enquote{best} QSM}
The goal in this part of the study was to establish sensible criteria based on the available 46 statistics for the quality of a QSM in comparison to its 9 other variants -- allowing us to determine the best QSM for a tree. Two main ideas build the basis: 

First, we assume that the QSMs capture the real 3D structure to some degree and that \emph{consensus} between QSMs conveys some degree of truth, and that deviations from this consensus point towards flawed QSMs. Second, we assume that the most \emph{common errors} in QSMs are the absence of tree parts rather than an insertion of new non-existing parts, as well as improperly fitted and connected cylinders (cylinder paths making sharp bends or not being smooth) rather than real zigzag branches in the underlying tree. This second notion arose from the manual investigation of a large number of smaller QSMs, in which these typical errors became apparent (see Figure~\ref{fig:workflow} or also Figure~\ref{fig:CylinderOffsets} for examples). Missing tree parts tend to create higher internal 3D imbalance in the tree since the \enquote{counterweight} for other tree parts is missing (see 180b$^1$, 180b$^4$, or 180b$^7$ in Figure~\ref{fig:workflow}). 

All in all, the QSM quality assessment process is, thus, partly driven by a \enquote{majority vote} of the QSMs but also by favoring QSMs that are fuller -- therefore potentially more complete than others -- and internally more balanced -- therefore potentially with higher cylinder path quality and without loss of tree parts.\\

Following the first \emph{consensus} notion, we implemented a \textbf{negative selection}. For each tree, its 10 QSMs were examined to see which of the statistics classified them as outliers. Let $id$ be a tree, then QSM $j$ is considered an outlier according to a statistic $s$ (a $s$-outlier-QSM), if its value $s(id^j)$ is outside of $[q_{0.25}^{id,s}-1.5\cdot IQR^{id,s}, q_{0.75}^{id,s}+1.5\cdot IQR]$ (boxplot-outlier criterion), where $q_{0.25}^{id,s}$ and $q_{0.75}^{id,s}$ are the quartiles and $IQR^{id,s}=q_{0.75}^{id,s}-q_{0.25}^{id,s}$ the interquartile range of the values $s(id^0),\dots,s(id^9)$. Let the outlier assessement be $out_s(id^j)=1$ if $s$ regards QSM $id^j$ as an outlier, and $=0$ otherwise.

In case of missing values, i.e., if $s(id^j)$ could not be computed (only mI' and TC were affected by this in cases where the tree was too \enquote{small} with regards to number of leaves and inner vertices), as long as the statistic could be computed for at least two of the ten QSMs per tree, all QSMs with missing values were flagged as outliers since this pointed towards erroneously missing tree parts (in accordance with the second \emph{common errors} idea). If all ten QSMs' values of a tree were missing, this information was considered consensus and none of the QSMs was considered an outlier, and if only one QSM value was available, this QSM was considered the outlier.

Based on this, the \emph{outlier count} $o(id^j)=\displaystyle\sum_{s \in S}{out_s(id_j)}$, defined as the number of statistics which flagged this specific QSM $id^j$ as outlier, was chosen as the metric for the negative selection of the QSMs. 

In the subsequent \textbf{positive selection} only QSM versions of a tree $id$ with $o$-ratings $\displaystyle = \min_{j=0,\ldots,9}\{o(id^j)\}$ are considered. Let $B_{id}\subseteq \{0,\ldots,9\}$ be the QSM iteration numbers of these minimal outlier count QSMs for a tree $id$, all other QSMs, the negSel-QSMs, are sorted out. Hence, let the negative selection of a QSM be expressed with $negS_M(id^j)=0$ if $j \in B_{id}$ for the method $M$, and $=1$ otherwise.

Then, the QSM quality is determined based on three statistics: on the volume of smaller tree parts (Vol0-25, Vol25-50) and the internal 3D imbalance Int-w(A), which was chosen over the other internal 3D imablance measures according to the decision tree in \cite[Figure~11]{kersting_measuring3D_2024}. They formed the set $S_B = \{\text{Vol0-25}, \text{Vol25-50}, \text{-Int-w(A)}\}$ with the internal 3D imbalance statistic having a negative sign as it had to be minimized and not maximized like the other two volume statistics. As a simple \emph{quality} criterion which uses the quantitative information within these three statistics, i.e., how much more volume one QSM has than another etc., we have chosen the following formula (given a QSM $id^j$ with $j \in B_{id}$)
\[quality(id^j) \coloneqq \sum_{s\in S_B}{\left(\frac{s(id^j)-\min_{b\in B_{id}}{s(id^b)}}{\max_{b\in B_{id}}{s(id^b)}-\min_{b\in B_{id}}{s(id^b)}}\right)},\]
which maps the values of each statistic to the interval $[0,1]$ and then takes the sum over the three values per QSM. In the case of $\max=\min$ for a set of QSMs for a statistic, the corresponding summand was set to 0.

The \enquote{best} QSM for a given tree is then the QSM which has the maximal $quality$-value among all QSMs $id^b$ with $b \in B_{id}$ (there were no ties). 

The resulting assignment of Tree ID to the best QSM (with the additional outlier information) can be found in \texttt{df\_QSMiteration\_quality.csv}. For the next steps, a dataset that only contains the best QSM per tree (\texttt{df\_ww\_1QSM\ perTree.csv}) was extracted from \texttt{df\_ww\_prep.csv}.

\paragraph{Method evaluation}

Simple indicators of the quality/structure of the method explained above (and variations of it) are the negSel-QSM counts per tree, i.e., the number of QMS of a tree that were removed in the negative selection, as well as the number of rejecting statistics per QSM. To put it precisely, the \emph{negSel-QSM count} of a tree $id$ is defined as $negSQC(id)=\sum_{j \in \{0,\cdots,9\}}{negS_M(id^j)}=10-|B_{id}|$, and the \emph{number of rejecting statistics} of a QSM $id^j$ is $RS(id^j)=\sum_{s \in S}{out_s(id^j)}$. The respective results for the method variations are shown in Tables~\ref{tab:negSelQSM_count} and \ref{tab:QSM_rejectingStats_count}.

To validate and assess the quality of the negative selection of the method, two other available information sources were used: the original tree point clouds and the census DBH measurements done by hand.

\begin{figure}[ht]
\centering
\begin{subfigure}[t]{0.4\textwidth}
\centering
\includegraphics[width=1\textwidth]{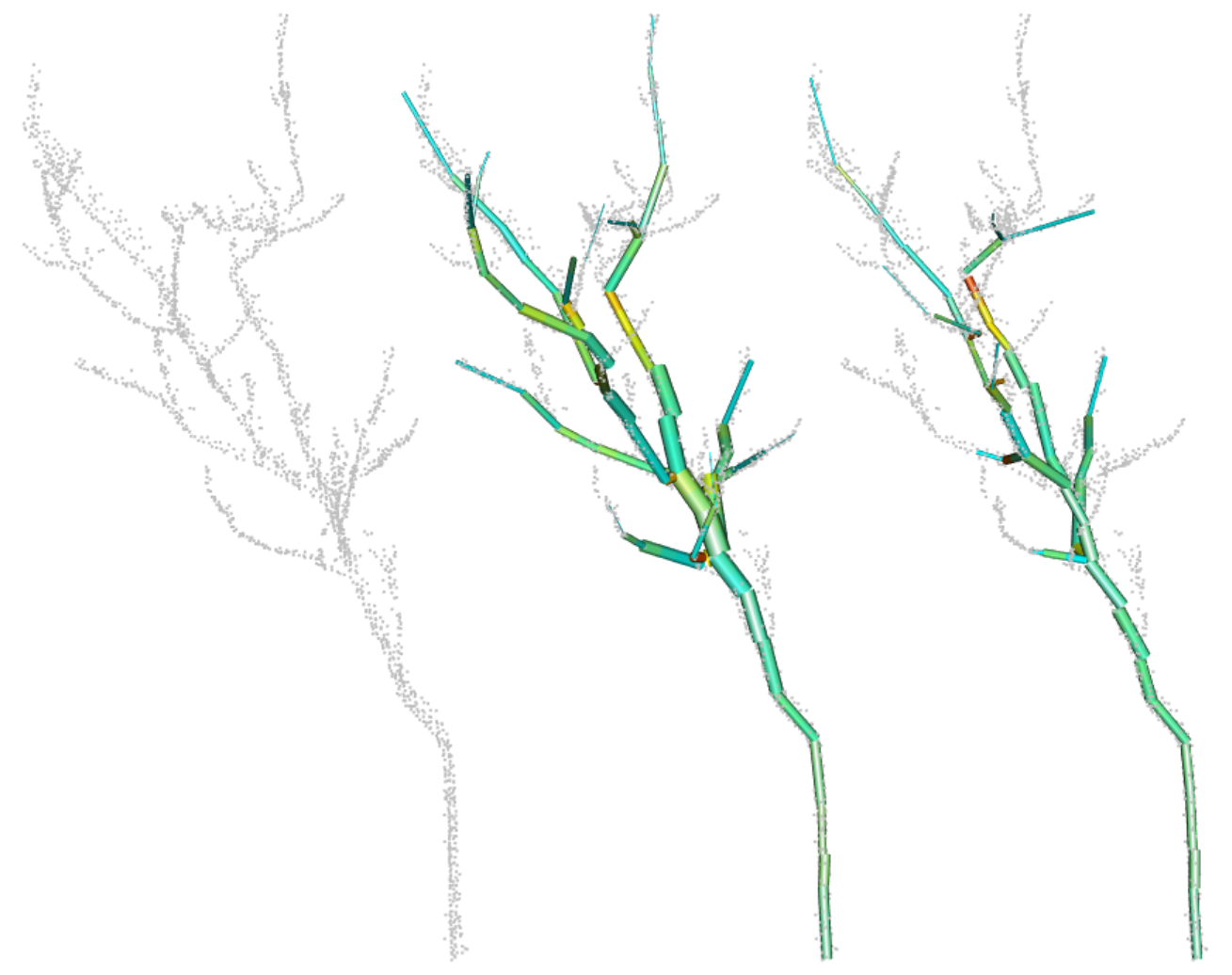}
\caption{}
\end{subfigure}\qquad\qquad
\begin{subfigure}[t]{0.4\textwidth}
\centering
\includegraphics[width=1\textwidth]{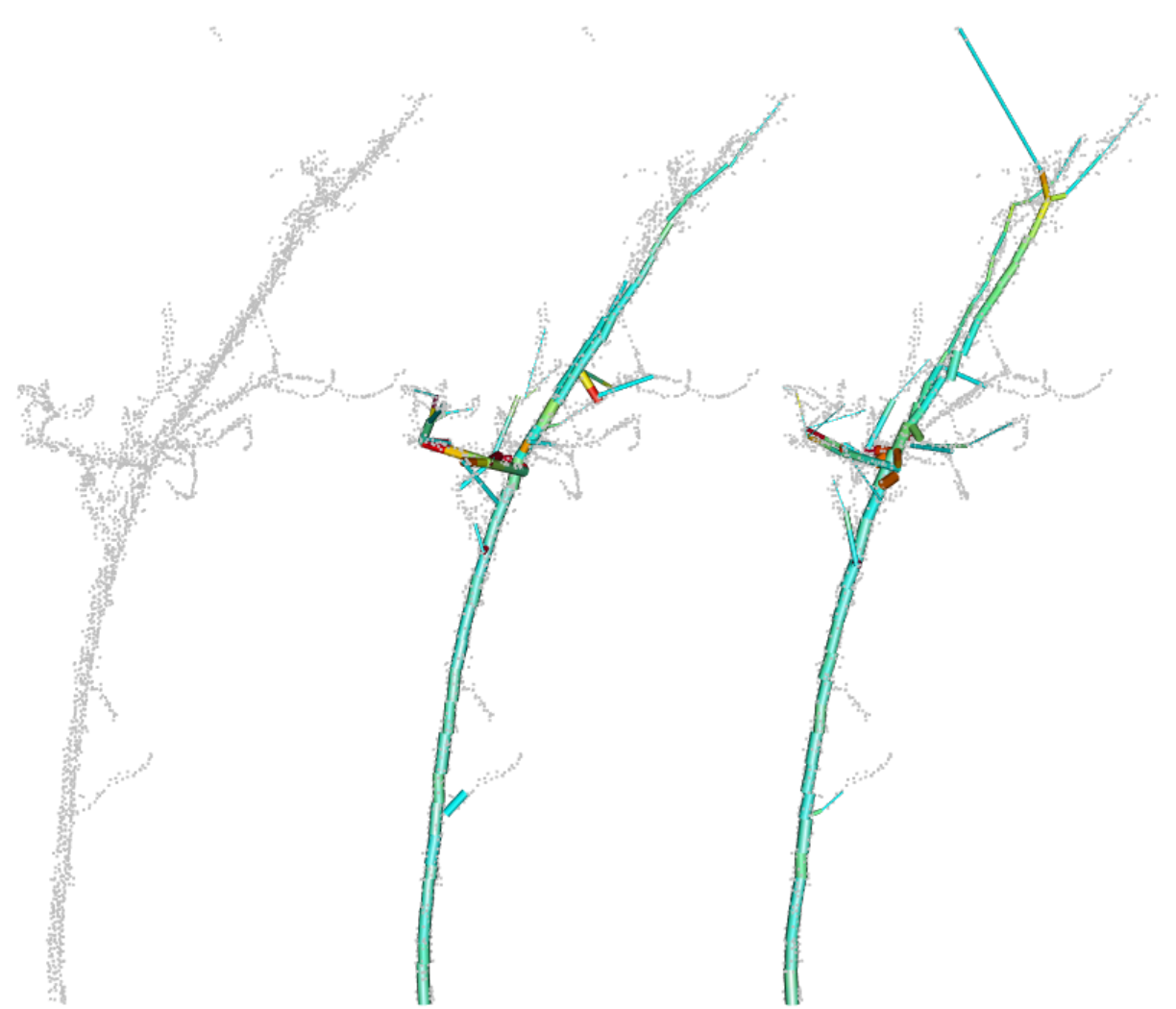}
\caption{}
\end{subfigure} \vspace{-0.1cm}
\caption{Two exemplary point clouds, on the left by themselves, then in the middle with their respective \enquote{best} QSM as decided by our method, and on the right with a flawed QSM. (a) Point cloud of tree 180b consisting of 3,803 points, with QSMs 180b$^2$ (best) and 180b$^8$ with 63 and 42 cylinders, respectively. (b) Point cloud of tree 145c consisting of 3,039 points, with QSMs 145c$^8$ (best) and 145c$^5$ with 66 and 73 cylinders, respectively. The QSMs are colored according to their internal $\mathcal{A}$ imbalance (see Section~\ref{sec:statistics3DT}).}
\label{fig:pointClouds}
\end{figure}

The \textbf{original point clouds} were consulted to have reliable information on the absence of tree parts in the QSMs. For the 200 trees $sT$ with the smallest QSM files (see Figure~\ref{fig:pointClouds} for examples), we computed the minimal distance of each point in the 3D point cloud to its nearest QSM cylinder, where the distance is zero if the point lies within the cylinder and, otherwise, it is the distance to the closest point on one of the outer faces of the cylinder. This procedure is computationally very time expensive as -- in the worst case -- all pairwise distances of all cylinders and points have to be checked (computations for one point can be stopped if one distance to a cylinder is 0). Thus, only a subset of the files was used. However, with 2,000 QSMs this is still large enough to draw conclusions. To provide some perspective on the scales of data, the smallest QSM file contains 13 (original QSM-) cylinders and its corresponding point cloud 1255 points while the largest assessed QSM files have 500 cylinders and around 28,000 corresponding points. The overall largest files in the dataset contain around 70,000 cylinders and 5.75 million points, which would have resulted in up to $400\cdot 10^{9}=$ 400 billion distance computations.

Studying individual trees with QSMs that are known to lack tree parts (see, e.g., 180b and 145c in Figure~\ref{fig:pointClouds}) revealed that several statistics of these minimal point-cylinder distances are meaningful indicators of missing tree parts in a QSM $id^j$. For example, the 90 or 95\%-quantiles or the fraction of points with minimal point-cylinder distances $>0.1$m$ = 10$cm ($R_{md>0.1}(id_j)$). The decision fell on the latter to evaluate different method versions as its meaning of distances $>10$cm can be directly understood. The fraction is tied to the size of the tree model: the bigger the tree, the larger the missing tree part has to be to have the same impact on the value. However, we only do comparisons within the $R_{md>0.1}(id_j)$ values based on the same point cloud and thereby identify QSMs with high values that indicate incompleteness.

The aim was to build a set of true-positive QSMs (flawed QSMs that should be removed in the negative selection) and measure which fraction of these are identified during the negative selection. In detail, such true-positive QSMs were set as the QSMs with a higher value $R_{md>0.1}(id_j)>q_{0.75}^{id,R_{md>0.1}}+0.5\cdot IQR^{id,R_{md>0.1}}$ (stricter upper bound of the boxplot-outlier-criterion) than the other QSM iterations of the same tree $id$. As above, let the outlier assessement be $out_{R_{md>0.1}}(id^j)=1$ if $R_{md>0.1}$ regards QSM $id^j$ as an outlier, and $=0$ otherwise. To measure the \textit{point cloud sensitivity} of a method $M$, we computed the fraction of correctly sorted out PC-outlier-QSMs
\[\displaystyle sens_{\text{PC}}(M) = \frac{\sum_{id^j \in sQ}{out_{R_{md>0.1}}(id^j)\cdot negS_M(id^j)}}{\sum_{id^j \in sQ}{out_{R_{md>0.1}}(id^j)}},\] 
where $sQ$ denotes the QSMs of the 200 trees $sT$ with the smallest files in the dataset. This value lies between 1 (all PC-outlier-QSMs detected) and 0 (none detected). Within this subset of 2,000 QSMs / 200 trees there were $\sum_{id^j \in sQ}{out_{R_{md>0.1}}(id^j)}=129$ PC-outlier-QSMs distributed across 159 trees (41 had no PC-outlier-QSMs).

The \textbf{census DBH measurements} were used to determine if the average QSMs estimated the correct DBH and with that if the outliers in our method were also outliers from the correct DBH. Please note that this information was not available for all trees, just for 695 of the 835 trees in the census area.

In order to build a set of true-positive QSMs for the census DBH data, the outlier-criterion was adapted and relaxed: $out_{\text{DBHc}}(id^j)=1$ if $\text{DBH}(id_j) \not\in [\text{DBHc}(id)-2\cdot IQR^{id,\text{DBH}},\text{DBHc}(id)+2\cdot IQR^{id,\text{DBH}}]$ (DBHc regards QSM $id^j$ as an outlier), and $=0$ otherwise. To measure the \textit{census DBH sensitivity} of a method $M$, we computed the fraction of correctly sorted out census DBH-outlier-QSMs
\[\displaystyle sens_{\text{DBHc}}(M) = \frac{\sum_{id^j \in cQ}{out_{\text{DBHc}}(id^j)\cdot negS_M(id^j)}}{\sum_{id^j \in cQ}{out_{\text{DBHc}}(id^j)}},\] 
where $cQ$ denotes the set of all QSMs of the 695 census trees $cT$ with available DBHc data. All in all, among those 6,950 QSMs there were $\sum_{id^j \in cQ}{out_{\text{DBHc}}(id^j)}=3,677$ DBHc-outlier-QSMs spread across 487 trees (208 had no DBHc-outlier-QSMs). Again, this value lies between 1 (all census DBH-outlier-QSMs detected) and 0 (none detected).

Unfortunately, the available data does not give the option to test the precision or specificity of a method, i.e., how many true best QSMs were identified, and we have to settle with measuring whether bad QSMs are correctly sorted out. However, we can approach a kind of Type II (or $\beta$-) error by investigating how many best QSMs were in fact PC- or DBHc-outlier-QSMs ($\#\beta_{\text{PC}}(M)$ and $\#\beta_{\text{DBHc}}(M)$), which provides an understanding if these two aspects, which can only be obtained through additional effort (computationally expensive or manual measurements), can be correctly accounted for with the given statistics. To put these two counts into perspective, we computed the expected number of best QSMs that would have been PC- or DBHc-outliers under two simple reference methods: The first, Uniform $\{0,\ldots,9\}$ or $U_{10}$ in short, selects each best QSM uniformly at random from the 10 available ones, i.e., the expected number is the sum of probabilities to select a PC-outlier-QSM $\displaystyle E[\#\beta_{\text{PC}}(U_{10})] = \sum_{id \in sT}\frac{\sum_{j = 0}^{9}{out_{R_{md>0.1}}(id^j)}}{10}$; analogously for DBHc $\displaystyle E[\#\beta_{\text{DBHc}}(U_{10})] = \sum_{id \in cT}\frac{\sum_{j = 0}^{9}{out_{DBHc}(id^j)}}{10}$. The second, Uniform $B$ or $U_B$ in short, assumes that the negative selection (of the main method explained above) was correct and only selects the best QSM uniformly at random from the remaining QSMs $id^j$ with $j\in B_{id}$ per tree $id$. The corresponding expected values are $\displaystyle E[\#\beta_{\text{PC}}(U_B)] = \sum_{id \in sT}\frac{\sum_{j \in B_{id}}{out_{R_{md>0.1}}(id^j)}}{|B_{id}|}$) and $\displaystyle E[\#\beta_{\text{DBHc}}(U_B)] = \sum_{id \in cT}\frac{\sum_{j \in B_{id}}{out_{DBHc}(id^j)}}{|B_{id}|}$. If these counts $\#\beta_{\text{PC}}$ and $\#\beta_{\text{DBHc}}$ are smaller than the counts of the reference models, the method $M$ is less, and if larger, more likely to choose a \enquote{faulty} PC- or DBHc-outlier-QSM as best QSM. 

\paragraph{Statistics and their importance} 

Since the negative selection for this dataset is mostly based on a true-false-assessment (whether a statistic was flagged as an outsider by at least one statistic) rather than letting the outlier count have a gradual influence, similarities of the statistics did not affect this selection criterion. If this had not been the case, similarity of the statistics would have had to be addressed as the outlier count would be affected by having subsets of statistics coming to similar conclusions. The aim would be to have the outlier decisions come from different information sources and not from statistics that generally come to the same conclusion regarding the deviation of a QSM. 

Even so, we -- with regard to the third research question concerning the impact and usefulness of the individual statistics -- are interested in similarities and replaceability of the statistics for detecting deviating QSMs. Thus, we introduced the pairwise a \emph{replaceability score} or \emph{outlier agreement ratio} which could also be used to filter the statistics for datasets where a quantitative influence of the outlier count is aimed at. 

An agreement ratio $a(i,j)$ of statistics $i$ and $j$ is the number of times $j$ agrees with $i$ that a QSM is an outlier divided by the total number of times $i$ flags a QSM as outlier. The closer $a(i,j)$ is to 1, the better $i$ can be replaced by $j$ as $j$ detects $i$'s outliers just as well. These scores were computed for all pairs of these 46 QSM-specific statistics as shown in Figure~\ref{fig:agreementRatios}. Row $i$ shows how well $i$ can be replaced by the other statistics. The minimal replaceability score of 0.018 has mIw with Vol200+ -- in other words, Vol200+ catches only around 2\% of mIw's detected outliers. 

To evaluate the effect of filtering and with that the impact of the statistics, we created exemplary filtered sets of statistics for different thresholds. The thresholds $0.99$, $0.85$, and $0.7$ were considered, i.e., the filtering removes all statistics for which at least one other statistic also detected $>99/85/70\%$ of its outlier-QSMs. By iteratively removing the statistic with the highest replaceability score $> 0.99/0.85/0.7$ (and highest $\sum_{j}a(i,j)$ in case of ties), the initial 46 statistics were filtered down to subsets (listed below). Furthermore, different initial subsets of statistics were used, e.g., only QSM statistics or only non-3D statistics, to investigate how much usable information these types of statistics can bring to the table.

The importance of the individual statistics was measured by the \emph{outlier production} $OP(s) = \sum_{id^j \in Q}out_s(id^j)$, defined as the number of QSMs a statistic $s$ flags as outlier with $Q$ being the set of all 8,760 QSMs, and the \emph{impact factor} $IF(s)=\sum_{id^j \in Q}\frac{out_s(id^j)}{\sum_{s'\in S} out_{s'}(id^j)}$, which takes into account both the outlier production as well as the \enquote{uniqueness} of the respective statistics as it rewards higher values if the statistic is one of few that detects a QSM as an outlier. Please note that the impact factor is dependent on the set of observed statistics $S$ which varies in the different method variations.

Similar to the overall sensitivity scores of the method, we can measure each statistic's sensitivity to detect PC- and DBHc-outlier-QSMs, i.e., \\$\displaystyle sens_{\text{PC}}(s) = \frac{\sum_{id^j \in sQ}{out_{R_{md>0.1}}(id^j)\cdot out_s(id^j)}}{\sum_{id^j \in sQ}{out_{R_{md>0.1}}(id^j)}}$ and $\displaystyle sens_{\text{DBHc}}(s) = \frac{\sum_{id^j \in cQ}{out_{\text{DBHc}}(id^j)\cdot out_s(id^j)}}{\sum_{id^j \in cQ}{out_{\text{DBHc}}(id^j)}}$. Last but not least, the impact factors of the statistics restricted to only PC- or census DBH-outlier-QSMs focuses on how uniquely useful a statistic is in detecting precisely these outlier-QSMs. In mathematical terms these impact factors are defined as $IF_{PC}(s)=\sum_{id^j \in sQ}out_{R_{md>0.1}}(id^j)\cdot \frac{out_s(id^j)}{\sum_{s'\in S} out_{s'}(id^j)}$ and $IF_{DBHc}(s)=\sum_{id^j \in cQ}out_{\text{DBHc}}(id^j)\cdot\frac{out_s(id^j)}{\sum_{s'\in S} out_{s'}(id^j)}$.

Overall we tested the following sets of statistics, i.e., instead of $S$ we consider subsets of $S$, with varying filtering options yielding different variations of the method (see also Table~\ref{tab:method_comparison}). Amongst other subsets, we also built some optimized subsets of statistics that allow comparisons with other subsets of the same sizes (after filtering).

\begin{description}
 \item[all-nofil] Based on all statistics with no filtering.\\
 All 46 statistics in $S$: 
    Height, Volume, Length, ZeroCyl, DBH, Vol0-25, Vol25-50, Vol50-75, Vol75-100, Vol100-200, Vol200+, Int-w(A), Int-w(a), Int-w(M), Int-w(m), Int-l(A), Int-l(a), Int-l(M), Int-l(m), Ext(A), Ext(a), Ext(M), Ext(m), LeafN, InnerN, B1, B2, Cherry, CLln, mD, mW, mDW, mI', mIw, Sackin, s-shape, TPL, TC, VLD, ALD, AVD, TIPL, mWomD, 4thrt-rQi, sqrt-CLe, and TopRes.
 \item[all-fil99] Based on all statistics with filtering out all statistics with a replaceability score $> 0.99$.\\
    43 statistics (Same initial 46 statistics except Ext(A), Ext(a), Ext(M)).
 \item[all-fil85] Based on all statistics with filtering out all statistics with a replaceability score $> 0.85$.\\
    41 statistics (Same initial 46 statistics except Ext(A), Ext(a), Ext(M), AVD, and Int-w(m)).
 \item[all-fil70] Based on all statistics with filtering out all statistics with a replaceability score $> 0.7$. \\
    35 statistics (Same initial 46 statistics except Ext(A), Ext(a), Ext(M), AVD, Int-w(m), Int-w(M), Int-w(A), TPL, Int-l(M), Int-l(m), and CLln).
 \item[onlyQSM = onlyQSM-fil70] Based only on QSM statistics with replaceability filtering threshold $0.7$. \\
    11 statistics: Height, Volume, Length, ZeroCyl, DBH, Vol0-25, Vol25-50, Vol50-75, Vol75-100, Vol100-200, and Vol200+.
 \item[only3D-fil70] Based only on 3D imbalance statistics with replaceability filtering threshold $0.7$. \\
    4 statistics: Int-w(a), Int-l(A), Int-l(a), and Ext(m).
 \item[onlyNon3D-fil70] Based only on topological/non-3D statistics with replaceability filtering threshold $0.7$. \\
    20 statistics: LeafN, InnerN, B1, B2, Cherry, mD, mW, mDW, mI', mIw, Sackin, s-shape, TC, VLD, ALD, TIPL, mWomD, 4thrt-rQi, sqrt-CLe, and TopRes.
 \item[all-NOTQSM-fil70] Based on all but QSM statistics with replaceability filtering threshold $0.7$. \\
    24 statistics: Int-w(a), Int-l(A), Int-l(a), Ext(m), LeafN, InnerN, B1, B2, Cherry, mD, mW, mDW, mI', mIw, Sackin, s-shape, TC, VLD, ALD, TIPL, mWomD, 4thrt-rQi, sqrt-CLe, and TopRes.
 \item[all-NOT3D-fil70] Based on all but 3D imbalance statistics with replaceability filtering threshold $0.7$. \\
    31 statistics: Height, Volume, Length, ZeroCyl, DBH, Vol0-25, Vol25-50, Vol50-75, Vol75-100, Vol100-200, Vol200+, LeafN, InnerN, B1, B2, Cherry, mD, mW, mDW, mI', mIw, Sackin, s-shape, TC, VLD, ALD, TIPL, mWomD, 4thrt-rQi, sqrt-CLe, and TopRes.
 \item[all-NOTnon3D-fil70] Based on all but topological/non-3D statistics with replaceability filtering threshold $0.7$. \\
    15 statistics: Height, Volume, Length, ZeroCyl, DBH, Vol0-25, Vol25-50, Vol50-75, Vol75-100, Vol100-200, Vol200+, Int-w(a), Int-l(A), Int-l(a), Ext(m).
 \item[opt4] Based on four variables obtained by a greedy search for the statistics with the highest impact: This search process starts with all statistics $s \in S$ and computes their impact factors. Then, iteratively the statistic with the highest impact factor is chosen as an optimal statistic and the impact factors of all other statistics are recomputed without all QSMs the optimal statistics have already flagged as outliers. This yielded  DBH, Height, sqrt-CLe, and B2.
 \item[opt11] Based on ten variables  with highest impact, obtained as in opt4. These are DBH, Height, sqrt-CLe, B2, mDW, Vol0-25, Vol50-75, ZeroCyl, mI', Vol100-200, and Vol25-50. 
 \item[opt15] Based on 15 variables with highest impact, obtained as in opt4. These are DBH, Height, sqrt-CLe, B2, mDW, Vol0-25, Vol50-75, ZeroCyl, mI', Vol100-200, Vol25-50, TopRes, B1, mW, and Vol75-100. 
\end{description}

\subsubsection{Data cleaning/preparation (Part II)}  \label{sec:meth_prepII}

In this step, the dataset \texttt{df\_ww\_1QSMperTree.csv} was prepared for the species classification under the name \enquote{best} QSM data (yielding \texttt{df\_ww\_1QSMperTree\_prep.csv}). The sparse variable ZeroCyl as well as any standard deviation statistic (DBHsd, Volsd, etc.) were removed at this point as they are only a sign of QSM quality without giving information about the underlying 3D structure.

Some variables had missing values. These were informative for the outlier detection, but would interfere with the species classification as these statistics would have been omitted entirely. Only eight QSMs were affected (1950$^2$, 2024a$^0$, 2053b$^6$, 2316$^6$, 532b$^6$, 8033$^4$, 8171$^0$, 8357$^5$), and for these only the statistics mI' and mIw as well as mWomD. In all cases the reason was that the extracted topology was too small. mI' and mIw need at least $n=4$ leaves in the topology to be computable, but all eight extracted topologies have 1-3 leaves. Since the values of mI' and mIw can range between 0 (balanced) and 1 (imbalanced), the values were set to 0.5 since there is only one topology for each of these tree sizes. mWomD $\frac{mW}{mD}$ cannot be computed if the maximum depth $mD$ is zero, which is the case if the non-3D topology consists of only one (root) vertex, i.e., the 3D models are a single stem without branching. This applies to the trees with IDs 1950$^2$, 2316$^6$, 532b$^6$, 8171$^0$, and 8357$^5$. We set the missing mWomD values to 2, the value of the smallest tree which has a mWomD value, the cherry tree with $n=2$ leaves, since this value is also consistent with the ranges of mWomD for higher $n$.\footnote{The maximal mWomD values for $n=2,3,4,5,...$ are $2,3,4,5,...$ (reached by the \enquote{star} topology consisting of the root and otherwise only leaves) and the minimal values are $\frac{2}{n-1}$ \cite[Cor.~23.4]{fischer_tree_2023}, i.e., for $n=2$ the only possible value is $2$, for $n=3$ values lie between $1$ and $3$, for $n=4$ between $2/3$ and $4$, for $n=5$ between $2/4=1/2$ and $5$, and so on.}

\subsubsection{Species classification}  \label{sec:meth_SpeciesClass}

The goal was to evaluate the predictive power of both the original statistics by themselves and the union of the original and newly introduced statistics for an application example, namely tree species classification. The prediction models were trained and evaluated on the subset of trees for which the species was known, and the resulting models were subsequently applied to trees with unknown species labels.

Due to small sample sizes in some species classes, 2 ACERCA, 26 CRATMO, and 37 QUERRO trees (each with fewer than 40 observations) were merged into a single class, Other, to decrease the high class imbalance. The final labeled data comprised 758 trees with the following species distribution: ACERPS: 541, CORYAV: 67, FRAXEX: 85, Other: 65. The unlabeled part of the dataset consists of 118 trees with species class Unknown.

Two predictor sets were evaluated: i) all 49 available numeric statistics $P_{all}$ and ii) the 12 original numeric statistics $P_{ori}$  (consisting of Locx, Locy, DBH, CrownArea, Height, and the various volume statistics, see Table~\ref{tab:statisticsColumns} for details). Although random forests are generally robust to multicollinearity and tend to down-weight redundant predictors, variable importance estimates may be diluted when predictors are strongly correlated. In order to make more meaningful statements about the importance of individual predictors, an additional iii) filtered predictor set $P_{fill}$ comprising 17 variables was constructed by removing highly correlated features based on the Pearson correlation coefficient, using a threshold of $|r| > 0.9$ (Figures~\ref{fig:correlations_all} and \ref{fig:correlations_allfiltered}). Pearson correlation was chosen over rank-based alternatives (e.g., Spearman) because the cases of clear non-linear relations (rQi and CLe) were already addressed in Section~\ref{sec:meth_prepI}.

\paragraph{General procedure}

Species classification was performed using two methods to assess robustness with respect to classifier choice: Random forests (RF) \cite[Sec.~15]{hastie_elements_2009}, an ensemble learning method that combines predictions from a large number of decision trees trained on bootstrapped samples and random subsets of variables, as well as gradient boosting models (GB) \cite[Sec.~10.9]{hastie_elements_2009}, a sequential ensemble approach in which trees are combined iteratively to correct the errors of previous ones.

To assess the predictive performance of the RF and GB models based on the three predictor sets (while avoiding overfitting), we used nested cross-validation \cite[Sec.~5.1]{james_introduction_2013}. This was done with the \textsf{R} packages \textsf{randomForest} \cite{liaw_classification_2002}, \textsf{gbm} \cite{ridgeway_gbm_2026}, and \textsf{caret} \cite{kuhn_building_2008}. This procedure consists of two loops:

An outer loop (10 folds) used for performance evaluation: The labeled dataset is split into 10 subsets. In each iteration, one fold is held out as a test set and the remaining folds are combined, then used for parameter tuning in the inner loop and as a training set for a final model with optimized parameters which is evaluated on the test set. 

An inner loop (5 folds) used for hyperparameter tuning: The training portion from the outer loop is split into 5 folds. In each iteration, again, one fold is held out for evaluation and the remaining folds serve as training data for each hyperparameter configuration. The inner loop returns the best parameter setting over all iterations according to a performance metric, in this case log-loss.

The log-loss, in short $\displaystyle-\frac{1}{t}\sum_{i=1}^{t}{log(p_{true}(i))}$, where $t$ is the number of observations/trees and $p_{true}(i)$ is the probability that tree $i$ is assigned its true class, was chosen as an optimization metric. It is more suitable for this case as, for example, the accuracy of the predictions, since it is sensitive to minority classes and punishes over-confident mistakes. For instance, a model that assigns ACERPS to all trees with certainty 0.99 would have a good accuracy of around 71\% but extremely high log-loss as $p_{true}(i)$ is very small for the remaining 29\% of trees.

Overall, this procedure ensures that the final evaluation on the outer test folds reflects out-of-sample performance, avoiding over-optimistic estimates that occur if tuning and evaluation are done on the same data. 

We report the following performance metrics derived from the pooled out-of-sample predictions across the test set of the outer folds of the nested cross validation (each tree of the dataset appears exactly once): Accuracy $A$, the percentage of correctly predicted tree species; Cohen’s $\kappa$ defined as $\displaystyle \kappa = \frac{A-A_E}{1-A_E}$, where $A_E$ is the expected accuracy by random guessing given the observed class frequencies, to quantify classification agreement beyond chance; log-loss (see above); class-wise sensitivities; and macro-averaged sensitivity, i.e., the mean across all class-wise sensitivities.

Subsequently, a final model was trained on the entire predictor and dataset (using the best hyperparameter configuration from the nested cross-validation) for the prediction of the unlabeled trees' species classes.

\paragraph{Details on species classification with random forests}

Exploring a broad range of numbers of trees \texttt{ntree} in the random forest, the number was fixed at 1,000 as the performance plateaus for $\geq$800 trees. The only hyperparameter for RF is \texttt{mtry}, which determines the size of the subsets of randomly chosen predictors each split in each decision tree is allowed to consider. A small value ($\ll|P_{\ldots}|$) decreases correlations between the individual decision trees, leading to reduced variance and more stable ensemble classification performance. The configurations for tuning were $\{2,3,\ldots,8\}$, which includes the default value $\approx\sqrt{|P_{\ldots}|}$, which is between $\sqrt{|P_{ori}|}=\sqrt{12}\approx3.5$ and $\sqrt{|P_{all}|}=\sqrt{49}=7$ for the three predictor sets.

For the final RF model based on $|P_{fil}|$, the best performing version, the hyperparameter \texttt{mtry} was set to 6 -- one of the most frequently selected  \texttt{mtry} values (and relatively close to the mean value $6.8$) across the outer folds for this predictor set. The species classification results with all class probabilities can be found in \texttt{species\_class\_RF.csv}.

Variable importance was assessed using the mean decrease in accuracy, which quantifies the reduction in model accuracy when the values of a given predictor are randomly permuted while all other predictors are held constant.

\paragraph{Details on species classification with gradient boosting models (GB)}

For GB models there are several hyperparameters \cite[Sec.~8.2.3]{james_introduction_2013}, which makes tuning more involved. \texttt{n.trees} controls the total number of sequential decision trees that are added and the \texttt{interaction.depth} specifies the number of splits\footnote{Each split (binary vertex) in a decision tree represents a decision rule of the form $p \leq c$, where a single predictor $p$ and a threshold value $c$ partition the feature space into two disjoint regions (left/right child nodes). The parameter name here is misleading as \texttt{interaction.depth} does not refer to the decision tree's depth but to its number of inner (binary) vertices.} in the individual decision trees, while the learning rate (\texttt{shrinkage}) manages the contribution of each individual tree to the ensemble and \texttt{n.minobsinnode} limits the minimum number of observations per terminal node.
The following tuning configurations were used: $\{800,\ 1,000,\ 1,200\}$ for \texttt{n.trees} to be sufficient for common (default) \texttt{shrinkage} values $\{0.1,\  0.01\}$, $\{1,\  2,\  3\}$ for \texttt{interaction.depth} to capture only main effects and small interactions (reduces prediction variance), and $\{3,\  5,\  7\}$ for \texttt{n.minobsinnode} to prevent overly specific and unsupported splits which reduces overfitting and sensitivity to noise.

The most frequently chosen hyperparameters in the outer folds which were used in the final GB model based on $|P_{fil}|$, the best performing version, were \texttt{n.trees} = 800, \texttt{interaction.depth}= 3, \texttt{shrinkage} = 0.01, as well as \texttt{n.minobsinnode}=7. \texttt{species\_class\_GB.csv} contain the species predictions with all class probabilities.

For the GB models, variable importance was assessed using the built-in metric from the \textsf{caret} package, which quantifies the contribution of each predictor to reducing the model’s loss function (multinomial deviance, i.e., the negative log-likelihood of the true class labels under the predicted probabilities) across all trees in the ensemble.

\paragraph{Data visualization}
A principal component analysis (PCA) \cite[Sec.~10.2]{james_introduction_2013}, a method that allows summarizing a large set of statistics (here $P_{all}$) with a smaller set of representative variables (linear combinations of the statistics) that collectively explain the most variance in the data, was performed on the\enquote{best} QSM data using the \textsf{R} package \textsf{stats} \cite{RCoreTeam2025}. Prior to the PCA, all variables were centered and scaled to unit variance. 

Furthermore, uniform manifold approximation and projection (UMAP) \cite{mcinnes_umap_2018} was used as a nonlinear embedding \cite[Sec.~14.9]{hastie_elements_2009} for the visualization of the (local) neighborhood structure of the scaled data in two dimensions. UMAP embeddings were computed using the \textsf{R} package \textsf{umap} (version 0.2.10.0) \cite{konopka_umap_2023} with Euclidean distance, \texttt{n\_neighbors = 20} and \texttt{min\_dist = 0.1}. A fixed random seed was used to ensure reproducibility. 

To assess the contribution of the new 3D and non-3D statistics to the original set of QSM statistics and general features, UMAP embeddings were computed for all three sets of statistics i)-iii). In all cases, identical UMAP parameters and random seeds were used, such that differences between embeddings reflect changes in the input feature space rather than tuning of the embedding algorithm.

For both, PCA and UMAP, samples were colored according to their species class for visualization purposes only.

\section{Results} \label{sec:results}
This section discusses the results of this study with regards to the three main research questions regarding 1) the QSM quality, 2) the species classification as an application example, and 3) the importance and relations of the statistics.

\subsection{QSM quality assessment} \label{sec:results_quality}

The two interesting aspects here are, firstly, what we uncover about the 3D tree models themselves: What are common QSM construction errors, how can they be detected, and how high is the variation between different QSM versions of one tree? Secondly, the method of quality assessment based on the (outliers of the) statistics is evaluated regarding how well it can detect flaws in QSMs and select a \enquote{best} QSM per tree. While discussing the details of these two main aspects, the validity of the two foundational assumptions, \textit{consensus} and the \textit{common errors}, will be addressed as well. Overall, this aims at narrowing down how an efficient quality control within the QSM construction process could look like to immediately flag or discard highly flawed QSMs.

\paragraph{Quality of the QSMs}

\begin{figure}[htbp!]
\centering
\begin{subfigure}[t]{\textwidth}
\centering
\includegraphics[width=\textwidth]{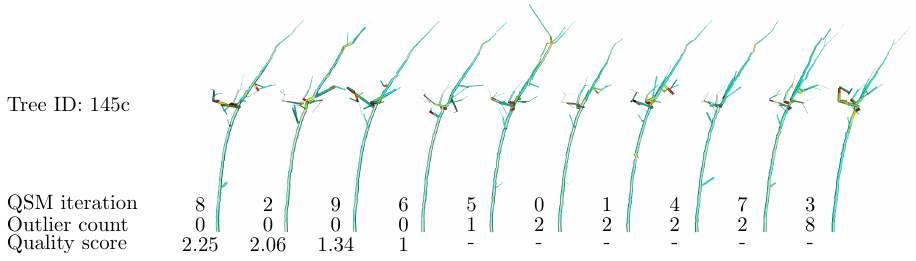}
\caption{}
\end{subfigure}

\vspace{0.2cm}
\begin{subfigure}[t]{\textwidth}
\centering
\includegraphics[width=\textwidth]{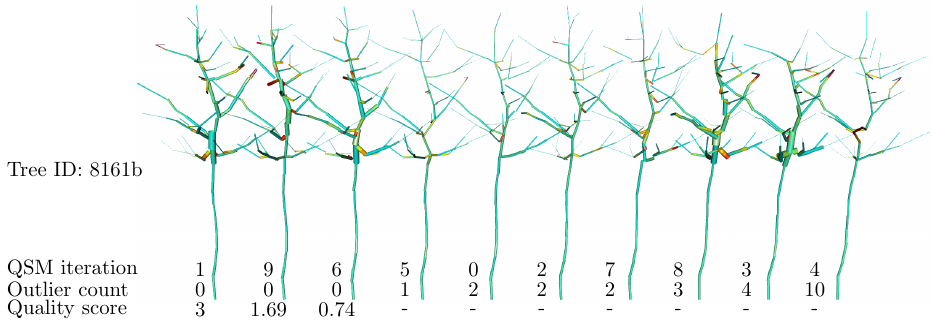}
\caption{}
\end{subfigure}

\vspace{0.2cm}
\begin{subfigure}[t]{\textwidth}
\centering
\includegraphics[width=\textwidth]{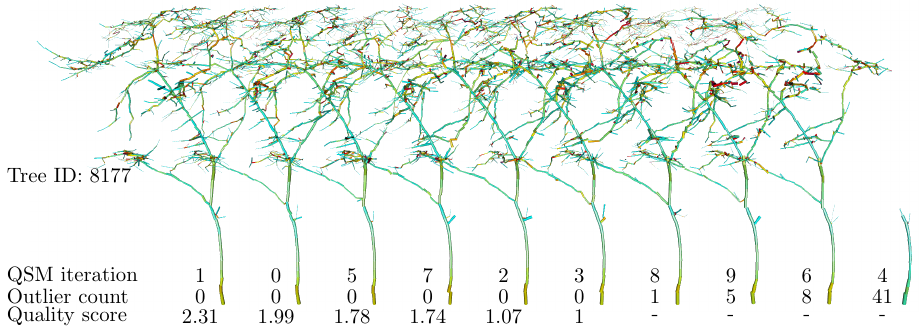}
\caption{}
\end{subfigure}

\caption{Exemplary results of the QSM quality assessment for (a) the sycamore (ACERPS) with ID 145c (145c$^8$ has height 3.42 m, DBH 0.05 m and 16 leaves), (b) the sycamore with ID 8161b (8161b$^1$ has height 5.26 m, DBH 0.04 m and 23 leaves), and (c) the common hazel (CORYAV) with ID 8177 (8177$^1$ has height 5.9 m, DBH 0.06 m and 292 leaves). The QSMs are sorted by outlier count and then quality score. The QSMs are colored according to their internal $\mathcal{A}$ imbalance (see Section~\ref{sec:statistics3DT}).}
\label{fig:QSM_quality_results}
\end{figure}

The most important result is that -- in the overwhelming majority of cases -- the QSMs are relatively well fitting models of the 3D structure and, while there are notable quality differences, most of the time several of the 10 QSMs per tree are good and non-deviating representations of the 3D architecture (see Figures~\ref{fig:QSM_quality_results} or also \ref{fig:workflow}). On average around 6.6 but always at least 1 of the 10 QSMs were filtered out in the negative selection (see Table~\ref{tab:negSelQSM_count}). This left 3.5 candidates for the \enquote{best} QSMs on average, with numbers ranging between 1 to 6 in most cases. 
The finally chosen \enquote{best} QSMs per tree were mostly considered non-deviating by all statistics. 843 of the 876 \enquote{best} QSMs have an outlier count of 0, and the remaining 33 only one of 1. 

Another main discovery is that the QSM flaws detected within this investigation do not hinder the quantitative structure models' -- as the name suggests -- primary usage as a foundation for above-ground volume and biomass estimations. On the contrary, the findings suggest that the results by \citet{calders_laser_2022} were rather conservative and point to a slight underestimation of the total volume and with that the biomass.

After summarizing these most relevant results and discoveries, the remainder of this section showcases and discusses the more detailed observations. 

Figure~\ref{fig:QSM_quality_results} shows three exemplary trees with the outlier counts and quality scores of their QSMs. Some reasons why certain QSMs were filtered out in the negative selection can be observed there: for example increased internal 3D imbalance in 8177$^8$,  8177$^9$, and  8177$^6$, or missing tree tips in 8161b$^8$, 8161b$^3$, and 8161b$^4$. The results for the sycamore with ID 180b depicted in Figure~\ref{fig:workflow} are as follows: QSM iteration 2 is the best QSM. Iterations 2, 4, 5, and 6 were left after the negative selection which sorted out 180b$^1$ with one,  180b$^7$ with two, 180b$^8$ with three, 180b$^3$ with four, 180b$^0$ with five, and 180b$^9$ with eleven statistics regarding them as outlier-QSMs.

\begin{table}[htbp!]
    \centering
    \caption{The number of trees with the respective negSel-QSM counts. The negSel-QSM count of a tree $id$ is the number of QSMs  $negSQC(id)=\sum_{j \in \{0,\cdots,9\}}{negS_M(id^j)}$ of the tree that were filtered out in the negative selection as being an outlier for too many statistics and with that were no candidates for the best QSM of $id$. The last column holds the respective mean negSel-QSM counts (rounded to two decimal places). For readability, entries with $0$ are not shown.}
    \label{tab:negSelQSM_count}
    \small
    \begin{tabular}{|@{\hskip 4pt}l@{\hskip 6pt}c@{\hskip 6pt}c@{\hskip 6pt}c@{\hskip 6pt}c@{\hskip 6pt}c@{\hskip 6pt}c@{\hskip 6pt}c@{\hskip 6pt}c@{\hskip 6pt}c@{\hskip 6pt}c@{\hskip 4pt}|@{\hskip 4pt}c@{\hskip 4pt}}
negSel-QSM count & 0 & 1 & 2 & 3 & 4 & 5 & 6 & 7 & 8 & 9 & Mean\\
    \noalign{\vskip 2pt}\hline\noalign{\vskip 2pt}
all-nofil &  & 1 & 6 & 18 & 56 & 126 & 191 & 199 & 185 & 94 & 6.61  \\
all-fil99 &  & 2 & 5 & 18 & 56 & 126 & 191 & 199 & 186 & 93 & 6.60 \\
all-fil85 &  & 2 & 5 & 18 & 56 & 128 & 193 & 198 & 184 & 92 & 6.59 \\
all-fil70 &  & 2 & 5 & 18 & 65 & 145 & 181 & 203 & 173 & 84 & 6.50 \\
    \noalign{\vskip 2pt}\hline\noalign{\vskip 2pt}
onlyQSM-fil70 & 26 & 78 & 167 & 189 & 175 & 121 & 81 & 23 & 9 & 7 & 3.50 \\
only3D-fil70 & 226 & 286 & 214 & 101 & 40 & 8 & 1 &  &  &  & 1.40 \\
onlyNon3D-fil70& 3 & 27 & 69 & 149 & 177 & 190 & 149 & 78 & 29 & 5 & 4.55 \\
    \noalign{\vskip 2pt}\hline\noalign{\vskip 2pt}
all-NOTQSM-fil70 &  & 11 & 38 & 102 & 176 & 172 & 172 & 127 & 58 & 20 & 5.16 \\
all-NOT3D-fil70 &  & 3 & 8 & 35 & 93 & 153 & 202 & 185 & 130 & 67 & 6.18 \\
all-NOTnon3D-fil70 & 11 & 27 & 102 & 175 & 206 & 145 & 130 & 59 & 12 & 9 & 4.20\\
    \noalign{\vskip 2pt}\hline\noalign{\vskip 2pt}
opt4 & 100 & 227 & 243 & 164 & 89 & 42 & 8 & 3 & & & 2.1\\
opt11 & 9 & 40 & 128 & 174 & 184 & 162 & 106 & 45 & 20 & 8 & 4.05\\
opt15 & 3 & 13 & 65 & 117 & 178 & 190 & 159 & 101 & 33 & 17 & 4.83\\
    \end{tabular}
\end{table}

There are cases of total failure, e.g. 8177$^4$ as shown in Figure~\ref{fig:QSM_quality_results} (c), where next to none of the 3D structure is captured by the QSM cylinders, but these are not typical and especially for larger tree models there are no examples of comparable severity as 8177$^4$. Fortunately, many statistics (41 different statistics in case of 8177$^4$) can identify such extreme cases as deviating, e.g., Volume which is already commonly computed. Consulting Table~\ref{tab:QSM_rejectingStats_count}, roughly one third of the QSMs were not considered an outlier-QSM by any statistics, more than one third was flagged by only 1 or 2 statistics (these could still be considered candidates for the \enquote{best} QSM in a less strict approach), and around 15\% of the QSMs were flagged by at least 5 statistics.

It would be advisable to employ a small set of variables at the minimum, so that at least such QSMs, whose assessment as an outlier is supported by several or all of them, can be removed or replaced. However, the smaller the set of statistics, the stricter and more generalized the filtering turns out. While it could be discussed to keep QSMs as candidates if only a few statistics identified them as outlier-QSMs as mentioned above, the usage of the opt11 statistics, for instance, flags only around 11\% of the QSMs by more than 2 statistics and 30\% by 1 statistic (see Table~\ref{tab:QSM_rejectingStats_count}). This does not leave a lot of room for deciding on a threshold of rejecting statistics for the negative selection of the most extreme cases, since a threshold of 2 would mean that at least 5 of the 15\% the main method identified as well-supported outlier-QSMs are not removed, whereas a threshold of 1 would remove even mild cases of deviation (which would be valid if intended). Therefore, only using the already commonly computed statistics like Height and Volume would, at the one hand, already be extremely helpful to identify a large part of the most heavily deviating QSMs, but, on the other hand, imply a strict removal of any flagged QSMs and not allow for a more relaxed approach to the filtering if that is desired. The addition of 3D and non-3D statistics could give this relaxation option. Furthermore, it would grant the ability to identify various other deviation patterns because most statistics -- apart from a few subsets like Int-l, Int-w, and Ext -- have extremely low pairwise replaceability scores, meaning that they capture unique outlier information (see Figure~\ref{fig:agreementRatios}).

\begin{table}[htbp!]
    \centering
    \caption{The number of QSMs with the respective $RS$ counts. The number of rejecting statistics $RS(id^j)=\sum_{s \in S}{out_s(id^j)}$ of a QSM $id^j$ counts how many statistics regard this QSM as an outlier. For readability, entries with $0$ are not shown.}
    \label{tab:QSM_rejectingStats_count}
    \small    
    \begin{tabular}{|@{\hskip 4pt}l@{\hskip 6pt}c@{\hskip 6pt}c@{\hskip 6pt}c@{\hskip 6pt}c@{\hskip 6pt}c@{\hskip 6pt}c@{\hskip 6pt}c@{\hskip 6pt}c@{\hskip 6pt}c@{\hskip 6pt}c@{\hskip 6pt}c@{\hskip 6pt}c@{\hskip 6pt}c@{\hskip 6pt}c@{\hskip 6pt}c@{\hskip 6pt}c@{\hskip 6pt}c@{\hskip 6pt}c@{\hskip 6pt}c@{\hskip 4pt}|}
Num. of rejecting statistics & 0 & 1 & 2 & 3 & 4 & 5 & 6 & 7 & 8 & 9 & 10 & 11 & 12 & 13 & 14 & 15 & 16 & 17 & $\geq 18$ \\
    \noalign{\vskip 2pt}\hline\noalign{\vskip 2pt}
all-nofil & 2857 & 2146 & 1224 & 664 & 527 & 363 & 241 & 182 & 134 & 99 & 83 & 60 & 50 & 21 & 20 & 15 & 18 & 14 & 42 \\
all-fil99 & 2857 & 2236 & 1308 & 728 & 480 & 309 & 206 & 161 & 121 & 89 & 69 & 46 & 39 & 22 & 19 & 9 & 15 & 16 & 30 \\
all-fil85 & 2867 & 2284 & 1308 & 754 & 485 & 315 & 188 & 165 & 114 & 69 & 62 & 37 & 28 & 16 & 17 & 16 & 8 & 6 & 21 \\
all-fil70 & 2944 & 2378 & 1439 & 807 & 458 & 277 & 161 & 96 & 65 & 37 & 34 & 24 & 14 & 8 & 7 & 4 & 2 & 1 & 4 \\
    \noalign{\vskip 2pt}\hline\noalign{\vskip 2pt}
onlyQSM-fil70 & 5623 & 2185 & 634 & 222 & 59 & 19 & 11 & 6 & 1 & & & & & & & & & & \\
only3D-fil70 & 7537 & 852 & 280 & 81 & 10 & & & & & & & & & & & & & &\\
onlyNon3D-fil70 & 4765 & 2089 & 902 & 418 & 248 & 139 & 78 & 48 & 41 & 15 & 9 & 2 & 2 & 2 & 1 & & & & 1\\
    \noalign{\vskip 2pt}\hline\noalign{\vskip 2pt}
all-NOTQSM-fil70 & 4228 & 2194 & 1071 & 540 & 293 & 166 & 102 & 66 & 38 & 23 & 15 & 12 & 5 & 3 & 2 & & 1 & & 1 \\
all-NOT3D-fil70 & 3249 & 2443 & 1372 & 717 & 395 & 218 & 125 & 99 & 41 & 37 & 24 & 21 & 6 & 3 & 3 & 2 & 1 & 1 & 3\\
all-NOTnon3D-fil70 & 5021 & 2262 & 879 & 342 & 150 & 61 & 24 & 9 & 7 & 3 & 2 & & & & & & & &\\
    \noalign{\vskip 2pt}\hline\noalign{\vskip 2pt}
opt4 & 6920 & 1662 & 169 & 9 & & & & & & & & & & & & & & & \\
opt11 & 5139 & 2691 & 731 & 160 & 29 & 5 & 3 & 1 & 1 & & & & & & & & & &\\
opt15 & 4441 & 2904 & 1023 & 290 & 76 & 15 & 7 & 1 & 1 & 1 & & 1 & & & & & & &\\
    \end{tabular}
\end{table}

Apart from these rare notable outliers, there are also several more subtle flaws in some QSMs which could impact their usage for some applications that focus more on the details rather than the overall volume of the 3D tree models. 
These are, for example, smaller missing tree parts like the top right branch of tree 180b, which is not represented by cylinders in QSM iterations four to eight (see Figure~\ref{fig:workflow}). A comparison with the corresponding point cloud depicted in Figure~\ref{fig:pointClouds} (a) shows that, indeed, these parts are genuinely absent rather than being imagined in the 3D model construction of the other QSMs. Because the remaining QSM iterations agree on the branch's existence, this provides some validity to the \emph{consensus} claim. 
QSM \emph{consensus} can also clear up the more seldom opposite cases of falsely added branches: The top-most twig of 145c$^5$ only appears in this specific QSM shown in Figure~\ref{fig:QSM_quality_results} (a) and no other. Consulting the point cloud (see Figure~\ref{fig:pointClouds} (b)) reveals that this branch, in fact, is not well supported and thereby with great certainty faulty. This deviation was successfully identified by the statistic Height, which flagged 145c$^5$ as an outlier-QSM. While potentially problematic for smaller trees, where these parts can make up a significant portion of the whole structure, larger trees will not be impacted notably by the absence of smaller branches.

Another observation regarding detailed QSM quality, were the frequent cylinder offsets, i.e., the end points of QSM cylinders not matching or following the same line (see Figure~\ref{fig:CylinderOffsets}). This prompted us to use a different parameter for the transformation into the rooted 3D tree format, which allowed the Int-w indices to ignore the imbalance resulting from the mulitude of small highly imbalance cylinder connections. Again, this flaw became apparent but does not affect most usages. For some special cases, it might be of interest to experiment with running a \enquote{smoothing} algorithm along the stems of the 3D tree models to both reduce cylinder offsets but also catch erroneously protruding cylinder peaks as shown in Figure~\ref{fig:CylinderOffsets} (a).

One of the most practically irrelevant observations is the occurrence of \enquote{non-existing} cylinders, cylinders which have either no length or nor radius and with that no volume, as tracked by the statistic ZeroCyl. These do not affect the 3D tree models usage but are unnecessary information within the model.

All in all, the method showed reliable results in deciding on plausible \enquote{best} QSMs per tree. While the volume of smaller tree parts was a factor in the positive selection to choose QSMs with fewer losses of tree parts, the total volume was not. When comparing the total volume of the \enquote{best} QSMs, 1104.71m$^3$, with the total of average volumes over each set of ten QSMs, 1092.44m$^3$, we see a 1.12\% increase. When restricted to only census trees the total volume increased by 1.06\% from 1054.46$^3$ to 1065.65$^3$ through the selection process of \enquote{best} QSMs. This change is plausible as we have seen examples of considerable losses ($8177^4$ in Figure~\ref{fig:QSM_quality_results} (c)). Simultaneously the difference is not high enough to attribute a significant bias (favoring overly voluminous 3D tree models) to the method itself. 

As mentioned at the beginning of this section, the Wytham Woods QSM dataset was used to estimate the above-ground biomass with the tree volumes averaged over each set of ten QSMs \cite{calders_laser_2022}. The fact that the \enquote{best} QSMs have a similar or slightly higher total volume might indicate that the volume that had been used as a basis was slightly underestimated and supports or probably even strengthens \citet{calders_laser_2022}'s result that the widely used allometric model underestimates the above-ground biomass of a typical UK temperate forest.

For most other statistics, the total averaged values and the total best values only differed by -2.7\% to 2.4\%, i.e., the selection process did not impact these statistics on a dataset level. For four statistics the best-QSM-selection resulted in more significant decreases: ZeroCyl $\approx-16\%$, sqrt-CLe $\approx-11\%$, TC $\approx-7\%$, and VLD $\approx-6\%$. The latter three statistics also have high outlier production (see Table~\ref{tab:outlier_prod} discussed in more detail in Section~\ref{sec:results_importance}), which could mean that, in particular, QSMs with especially large sqrt-CLe and VLD were negatively selected and not chosen as best QSM.

\paragraph{Evaluation of the method}

The entire method was based on two key assumptions, \textit{consensus} and the \textit{common errors}, that are broadly confirmed by the results. The \textit{consensus} worked and prevailed for the manually inspected trees, as already exemplary discussed for 145c$^5$ or 6177$^4$ in Figure~\ref{fig:QSM_quality_results}.

The comparisons with the point clouds (examples in Figure~\ref{fig:pointClouds}) revealed that missing branches are the most prevalent errors as assumed in the \textit{common errors} notion. Most often even the best QSMs do not account for every small twig while worse QSMs might be missing larger parts of the structures. Erroneously added tree parts like with 145$^5$ are extremely rare. Internal imbalance as another common error and also as an indication of the absence of tree parts stemming from a manual inspection of smaller example trees has also been supported on a broader scale. One example being tree 8177 (see Figure~\ref{fig:QSM_quality_results} (c)) with more and more red colored, i.e., highly imbalanced, and less full branches the higher the outlier count of the corresponding QSMs (except 8177$^4$ which is lacking too many branches to even be imbalanced).

Now, how well did the method do regarding the available performance metrics? The main method based on all suitable statistics without filtering was able to detect 86\% of the QSMs the point cloud statistic $R_{md>0.1}$ flagged as outlier and 69.2\% of the consensus DBH-outlier-QSMs (see Table~\ref{tab:method_comparison}).
The quality and degree of deviation of the DBHc- and PC-outlier-QSMs was checked on several small examples, amongst others on the four depicted trees in Figure~\ref{fig:workflow} and \ref{fig:QSM_quality_results}: The three smaller trees 180b, 145c, and 8161b were within the subset of trees for which the QSMs were compared with their point clouds. Thus, we have PC-outlier-QSM information for them: QSM iterations 0 and 8 are PC-outlier-QSMS for 180b, and for 8161b these are 3 and 0 (both successfully filtered out in the negative selection for both trees). 145c has no QSMs flagged as outlier by $R_{md>0.1}$. Please note that $R_{md>0.1}$ can only detect the absence and not the faulty addition of branches as, for example, the one in QSM 145c$^5$ (see Figure~\ref{fig:QSM_quality_results} (a)), since the distances of the points to their nearest cylinders was recorded. Thus, missing cylinders would leave traces in the data where additional cylinders would not. The lack of PC-outlier-QSMs for tree 145c, however, aligns with the additional branch being incorrect, since otherwise all but QSM 5 would have to be outliers. The tree 8177 is the only one of these four which was inside the census area. Its DBHc-outlier-QSMs are 2, 5, 8, 0, 3, and 9 (ordered by decreasing DBH), where, in fact, all its QSMs had a DBH lower than the census DBH value, but these particular QSMs had the lowest values. From these 6 QSMs only 8 and 9 were filtered out in the negative selection but the best QSM $8177^1$ is no DBHc-outlier-QSM. 

The collected data on the Type II error counts $\#\beta_{\text{PC}}$ and $\#\beta_{\text{DBHc}}$ shown in Table~\ref{tab:method_comparison} also allows to evaluated the two steps, the negative selection and the decision on the \enquote{best} QSM according to the quality score, of the process. For PC, we expect around 24 PC-outlier QSMs as \enquote{best} QSM if they were chosen at random from all 10 QSMs and 17 if chosen from the candidate set $B$. This implies that the negative selection already sorts out a fraction of these outlier-QSMs. All in all, our method, regardless of the underlying set of statistics, successfully reduces this number of false-negatives to 3-8 by deciding against PC-outlier QSMs. 

For the DBHc-outlier-QSMs, our method as a whole also performs better than random guessing. However, the negative selection step seems to be slightly more permeable for DBHc-QSMs as the expected $\#\beta_{DBHc}$ is lower for $U_{10}$ than $U_B$. The quality score step is mostly responsible for the decreased false-negative counts. On a dataset level the mean DBH over the 10 QSMs seems to be a good estimator of the real DBH of a tree as the distance of the mean DBH to the census DBH measurement is close to zero on average over all census trees with over $50\%$ of the trees showing a difference of $\leq 1$cm and only a handful of cases ranging from an overestimation of 11cm to an underestimation of 25cm. However, on a tree level, the DBHc values often lie at the periphery of the DBH values estimated from the QSMs. Ordering all ten DBH values and the DBHc value shows that for more than half of the trees the DBHc is either larger (304 cases) or smaller (92 cases) than all 10 DBH values. This is a distortion within the data that our method is able to mitigate at least slightly.

Across all tested sets of statistics, the main chosen method base on all statistics performed best regarding PC- and DBHc-outlier-sensitivity (see Table~\ref{tab:method_comparison}). For some method variations the false-negative counts  $\#\beta_{PC}$ and  $\#\beta_{DBHc}$ were lower at the cost of decreased sensitivity. This suggests that, although they ultimately selected less known outlier-QSMs, they overall did not identify most outlier-QSMs and instead also chose unrecognized outlier-QSMs as \enquote{best} whose deviation could only have been detected by other statistics.

\begin{table}[htbp!]
    \centering
    \caption{Evaluation of the different method versions described in Section~\ref{sec:meth_QSMquali}. The number of statistics is shown before and $\rightarrow$ after filtering. The similarity and sensitivity values are shown as percent (\%), where the similarity of best QSMs is the fraction of best QSMs the respective method decided on, which is also among the best QSMs of the main method used in this study (all-nofil). $\#\beta_{\text{PC}}$ and $\#\beta_{\text{DBHc}}$ are the (rounded expected) counts of respective outlier-QSMs that the method chose as best QSM. Uniform $\{0,\ldots,9\}$ and $B$ denote the two reference methods where the best QSM is chosen uniformely at random from the complete set of 10 QSMs or from the set of the candidates remaining after the negative selection $id^j$ with $j \in B_{id}$, respectively. Note that these values are based on 200 trees for PC (81 trees with at least one PC-outlier QSM) with 240 PC-outlier-QSMs in total and 695 trees for DBHc (487 with at least one DBHc-outlier QSM) with 3,677 DBHc-outlier-QSMs in total. For the expected counts of the Uniform $B$ based on other methods the values did vary a little, ranging from 16.5 to 21.6 and from 370 to 373.2, respectively.}
    \label{tab:method_comparison}
    \small    
    \begin{tabular}{|lcrrrrr|}
Method $M$ & \parbox[t]{2cm}{\centering Number of statistics} & \parbox[t]{2cm}{\raggedleft Similarity of best QSMs}  & $sens_{\text{PC}}$ & $sens_{\text{DBHc}}$ & \parbox[t]{1.5cm}{\raggedleft (expected) $\#\beta_{\text{PC}}$} & \parbox[t]{1.5cm}{\raggedleft (expected) $\#\beta_{\text{DBHc}}$} \\
    \noalign{\vskip 2pt}\hline\noalign{\vskip 2pt}
all-nofil & 46 & reference & 75.4 & 69.2 & 7 & 348 \\
all-fil99 & 46 $\rightarrow$ 43 & 100.0 & 75.4 & 69.2 & 7 & 348 \\
all-fil85 & 46 $\rightarrow$ 41 & 99.5 & 75.4 & 69.1 & 7 & 348 \\
all-fil70 & 46 $\rightarrow$ 35 & 96.0 & 74.2 & 68.3 & 8 & 349 \\
onlyQSM-fil70 & 11 $\rightarrow$ 11 & 47.0 & 50.0 & 39.8 & 6 & 359 \\
only3D-fil70 & 12 $\rightarrow$ 4 & 34.2 & 20.8 & 14.9 & 3 & 368 \\
onlyNon3D-fil70 & 23 $\rightarrow$ 20 & 53.3 & 53.3 & 47.5 & 6 & 361 \\
all-NOTQSM-fil70 & 35 $\rightarrow$ 24 & 63.0 & 60.8 & 53.9 & 6 & 364 \\
all-NOT3D-fil70 & 34 $\rightarrow$ 31 & 85.2 & 71.7 & 65.2 & 9 & 346 \\
all-NOTnon3D-fil70 & 23 $\rightarrow$ 15 & 54.9 & 57.1 & 46.3 & 5 & 357 \\
opt4 & 4 & 34.6 & 29.2 & 24.7 & 6 & 353 \\
opt11 & 11 & 52.2 & 44.2 & 44.4 & 7 & 353 \\
opt15 & 15 & 63.6 & 52.9 & 52.1 & 7 & 352 \\
    \noalign{\vskip 2pt}\hline\noalign{\vskip 2pt} 
Uniform $\{0,\ldots,9\}$ &  &  &  &  & 24 & 368 \\
Uniform $B$ &  &  &  &  & 17 & 373 \\
    \end{tabular}
\end{table}

\subsection{Species Classification} \label{sec:results_classification}

Overall the RF and GB species classification models performed very similar within a narrow range of performance metrics and achieved up to 81.8\% accuracy with a maximal macro-averaged sensitivity of up to 61.1\%. Nevertheless, since the simplest reference model \enquote{all ACERPS} would already reach 71\% accuracy (albeit with high log-loss), even a consistent improvement of few percent points is meaningful.

While the RF models based on the three predictor sets i) $P_{all}$, ii) $P_{ori}$, and iii) $P_{fil}$ performed relatively equally, across all performance metrics model ii) was the worst, followed by model i) and then model iii) the best: accuracy $P_{ori}$: 79\%, $P_{all}$: 80.5\%, $P_{fil}$: 81\%; Cohen's $\kappa$ $P_{ori}$: 0.46, $P_{all}$: 0.5, $P_{fil}$: 0.51; log-loss $P_{ori}$: 0.53, $P_{all}$: 0.52, $P_{fil}$: 0.49; and macro-averaged sensitivity $P_{ori}$: 53.9\%, $P_{all}$: 55.5\%, $P_{fil}$: 56.4\%.

In comparison, the GB models performed slightly better than the RF models, in particular regarding the macro-averaged sensitivity, suggesting that GB models are more sensitive towards minority classes. Here, too, the performance ranking yielded the same order: iii) outperformed i) which in turn outperformed ii) across all performance metrics: accuracy $P_{ori}$: 80.7\%, $P_{all}$: 81.3\%, $P_{fil}$: 81.8\%; Cohen's $\kappa$ $P_{ori}$: 0.51, $P_{all}$: 0.53, $P_{fil}$: 0.55; log-loss $P_{ori}$: 0.52, $P_{all}$: 0.48, $P_{fil}$: 0.47; and macro-averaged sensitivity $P_{ori}$: 57.8\%, $P_{all}$: 57.9\%, $P_{fil}$: 61.1\%.

Consulting the RF models' confusion matrices in Tables~\ref{tab:confusionMat_fil}, \ref{tab:confusionMat_all}, and \ref{tab:confusionMat_ori}  regarding the sensitivities per species class, none of the three models outperforms the others in every regard. While RF models i) and iii) are more sensitive towards the classes ACERPS and Other, RF model ii) is better in detecting FRAXEX. This pattern does not appear for the GB models, where i) is more sensitive regarding the classes FRAXEX and Other, ii) regarding CORYAV, and iii) is better in detecting the minority classes than both of them (see Tables~\ref{tab:confusionMat_boost_fil}, \ref{tab:confusionMat_boost_all}, and \ref{tab:confusionMat_boost_ori}).

\begin{figure}[H]
\centering
\includegraphics[width=\textwidth]{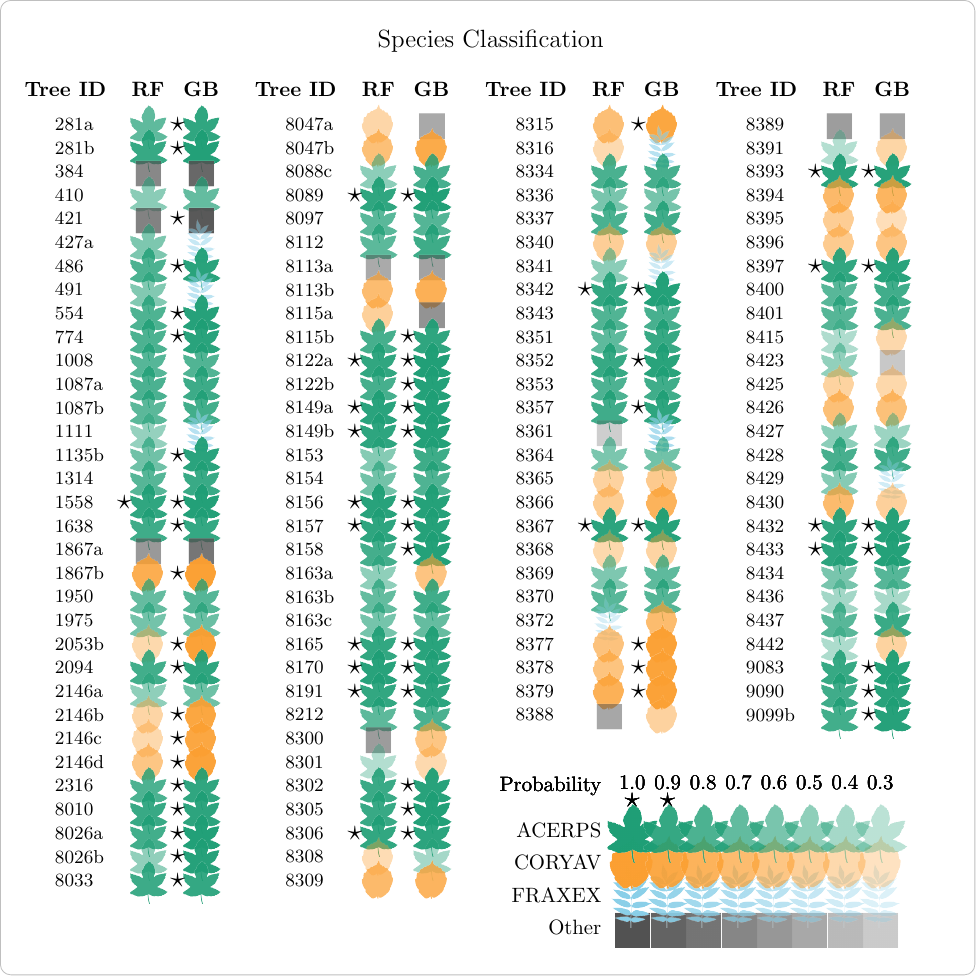}
\caption{Species predictions of the unlabeled trees based on the final RF and GB models iii). Each tree is shown with its corresponding most probable of the four species labels, i.e., the corresponding probability must be $\geq 25\%$ (for this data it is $>0.3$). Transparency provides insight into the probability/certainty of the predictions. Predictions with probability $\geq0.9$ are marked with a $\star$. Note that the class Other comprises of QUERRO, CRATMO, and ACERCA. Class counts for RF/GB: ACERPS 83/73, CORYAV 26/30, FRAXEX 1/7, and Other 8/8.}
\label{fig:species_class}
\end{figure}

All in all, even though the differences between the models are moderate, for both RF and GB, model iii) based on the filtered predictor set performed best and ii) based on the original predictor set worst. 
Because performance was assessed using nested cross-validation, and the same pattern was observed for both modeling approaches, this can be interpreted as a robust signal rather than a resampling artifact. 
On the one hand this suggests that, for this dataset, removing highly correlated predictors (Pearson correlation $\leq0.9$) does not reduce but slightly improve predictive performance. These slight gains indicate that redundant predictors may introduce minor instability or noise in the tree-based ensembles. For the final species prediction of the unlabeled trees we, thus, used the RF and GB models iii) (see Figure~\ref{fig:species_class}).
On the other hand, this shows that the newly introduced 3D and non-3D statistics can enhance performance and prove to be valuable, in particular, when filtered and combined with the stronger modeling approach -- albeit the original statistics by themselves already achieved good overall results.

\begin{table}[htbp!]
    \centering
    \caption{Confusion matrices for the random forest model iii) based on the 17 filtered numeric statistics $P_{fil}$ showing (a) absolute classification counts and (b) row-normalized percentages (per true species). Overall performance: Accuracy: 81\%, log-loss: 0.49, Cohen’s $\kappa$: 0.51, macro-averaged sensitivity: 56.4\%.}
    \label{tab:confusionMat_fil}
    \begin{subtable}[t]{0.495\textwidth}
\centering
\caption{}
\begin{tabular}{@{\hskip 4pt}c@{\hskip 6pt}l@{\hskip 6pt}c@{\hskip 6pt}c@{\hskip 6pt}c@{\hskip 6pt}c@{\hskip 4pt}}
 &  & \multicolumn{4}{c}{Prediction} \\
 &  & ACERPS & CORYAV & FRAXEX & Other \\
    \noalign{\vskip 2pt}\hline\noalign{\vskip 2pt}
\multirow{4}{*}{\rotatebox{90}{Reference}}
& ACERPS & \textbf{528} & 5 & 6 & 2 \\ 
& CORYAV & 17 & \textbf{42} & 1 & 7 \\ 
& FRAXEX & 72 & 5 & \textbf{7} & 1 \\ 
& Other & 15 & 13 & 0 & \textbf{37} \\ 
    \end{tabular}
\end{subtable}
    \begin{subtable}[t]{0.495\textwidth}
\centering
\caption{}
\begin{tabular}{@{\hskip 4pt}c@{\hskip 6pt}l@{\hskip 6pt}c@{\hskip 6pt}c@{\hskip 6pt}c@{\hskip 6pt}c@{\hskip 4pt}}
 &  & \multicolumn{4}{c}{Prediction} \\
 &  & ACERPS & CORYAV & FRAXEX & Other \\
    \noalign{\vskip 2pt}\hline\noalign{\vskip 2pt}
\multirow{4}{*}{\rotatebox{90}{Reference}}
& ACERPS & \textbf{97.6} & 0.9 & 1.1 & 0.4 \\ 
& CORYAV & 25.4 & \textbf{62.7} & 1.5 & 10.4 \\ 
& FRAXEX & 84.7 & 5.9 & \textbf{8.2} & 1.2 \\ 
& Other & 23.1 & 20 & 0 & \textbf{56.9} \\
    \end{tabular}
\end{subtable}
\end{table}

\begin{table}[htbp!]
    \centering
    \caption{Confusion matrices for the gradient boosting model iii) based on the 17 filtered numeric statistics $P_{fil}$ showing (a) absolute classification counts and (b) row-normalized percentages (per true species). Overall performance: Accuracy: 81.8\%, log-loss: 0.47, Cohen’s $\kappa$: 0.55, macro-averaged sensitivity: 61.1\%.}
    \label{tab:confusionMat_boost_fil}
    \begin{subtable}[t]{0.495\textwidth}
\centering
\caption{}
\begin{tabular}{@{\hskip 4pt}c@{\hskip 6pt}l@{\hskip 6pt}c@{\hskip 6pt}c@{\hskip 6pt}c@{\hskip 6pt}c@{\hskip 4pt}}
 &  & \multicolumn{4}{c}{Prediction} \\
 &  & ACERPS & CORYAV & FRAXEX & Other \\
    \noalign{\vskip 2pt}\hline\noalign{\vskip 2pt}
\multirow{4}{*}{\rotatebox{90}{Reference}}
& ACERPS & \textbf{518} & 8 & 13 & 2 \\ 
& CORYAV & 14 & \textbf{48} & 3 & 2 \\ 
& FRAXEX & 59 & 6 & \textbf{17} & 3 \\ 
& Other & 17 & 9 & 2 & \textbf{37} \\
    \end{tabular}
\end{subtable}
    \begin{subtable}[t]{0.495\textwidth}
\centering
\caption{}
\begin{tabular}{@{\hskip 4pt}c@{\hskip 6pt}l@{\hskip 6pt}c@{\hskip 6pt}c@{\hskip 6pt}c@{\hskip 6pt}c@{\hskip 4pt}}
 &  & \multicolumn{4}{c}{Prediction} \\
 &  & ACERPS & CORYAV & FRAXEX & Other \\
    \noalign{\vskip 2pt}\hline\noalign{\vskip 2pt}
\multirow{4}{*}{\rotatebox{90}{Reference}}
& ACERPS & \textbf{95.7} & 1.5 & 2.4 & 0.4 \\ 
& CORYAV & 20.9 & \textbf{71.6} & 4.5 & 3 \\ 
& FRAXEX & 69.4 & 7.1 & \textbf{20} & 3.5 \\ 
& Other & 26.2 & 13.8 & 3.1 & \textbf{56.9} \\
    \end{tabular}
\end{subtable}
\end{table}

As expected from the overlapping of the clusters in the PCA (Figure~\ref{fig:PCA}) and UMAP (Figure~\ref{fig:umap_proj}), which are further discussed in Section~\ref{sec:results_importance}, the species class predictions shown in Figure~\ref{fig:species_class} do not all have high certainty whereby GB makes more confident predictions than RF overall. For the RF model 14\%  and for the GB model 42\% species probabilities reached or exceeded $0.9$ (marked with $\star$). Each of the confident RF predictions was also a confident GB prediction with no contradictions. For RF 36\% of the predictions had a lower certainty of $\leq0.6$ -- 19\%  for GB -- and with an estimated accuracy around 81\% we expect some misclassifications. Some of these are easy to identify since sister stems should typically be of the same tree species. For example, 1867a being predicted as Other and 1867b as CORYAV, three of the four stems of tree 2146 being assigned CORYAV and one ACERPS, or 8115a predicted as CORYAV/Other and 8115b as ACERPS, are most likely wrong in at least one of the sister stem predictions. Since parts of these predictions have high probability, we can speculate that, e.g., 2146a is of species CORYAV. However, there are also several trees whose sister stems have been assigned to the same species, e.g., 1087, 8026, and 8149, where this fact might support the validity of the predictions.

\newpage

\subsection{Statistics and their importance} \label{sec:results_importance}

This section addresses the importance and differences of the individual statistics but also the corresponding subsets, QSM, 3D, and non-3D. By introducing the 3D and non-3D statistics as new tools for this type of data, it is necessary to investigate if their contribution warrants the increased effort of computing them. In fact, they do appear to be moderately to extremely useful depending on the task, as the following paragraphs explain in more detail. The QSM statistics that are usually already collected can also be the basis for well-performing models but have shortcomings in some areas when used by themselves. All 3D and non-3D statistics can be computed in linear time \cite{fischer_tree_2023} and the means to compute them are provided with our software package \textsf{treeDbalance} as explained in Section~\ref{sec:software}.

\paragraph{Outlier-detection}

The filtering of statistics according to the replaceability scores/agreement ratios depicted in Figure~\ref{fig:agreementRatios} can already have an impact even with high thresholds. For example, the number of QSMs per tree that were rejected as not having minimal outlier count changed for all thresholds (see Table~\ref{tab:negSelQSM_count}). As this also applies to the highest threshold of $0.99$ which only filtered out three of the external 3D imbalance statistics, this implies that even those statistics were important for capturing at least one QSM as deviating that would have been missed by others. A filtering threshold of $0.99$ did not change the number of QSMs with no rejecting statistics but already increased the number of outlier-QSMs that were only identified by 1 and not more statistics. However, the decision on the \enquote{best} QSM per tree was relatively robust to the filtering: For threshold $0.99$ the method decided on the same set of best QSMs as when using all statistics, for $0.85$ still $\approx 99.5\%$ of best QSMs matched, and for $0.7$ it was $\approx 96\%$. For this task, especially, if the number of rejecting statistics of a QSM are supposed to have a gradual influence, filtering should be used with caution.

If the set of statistics has to be reduced, three of the four Ext 3D imbalance measurements are the first that can be dispensed with (all-fil99). Next, AVD and representatives of each set of four internal 3D imbalance indices are least important (all-fil80/70). While there are some subsets of statistics that are highly replaceable most are not as they identify unique deviation patterns (Figure~\ref{fig:agreementRatios}).

The various importance scores listed in Table~\ref{tab:outlier_prod} give a more nuanced picture of the individual statistics and their strengths. The impact scores based on the filtered set provide a good general ranking of importance in outlier detection as the normal impact scores are inflated for highly agreeing statistics. While the Ext statistics have small normal impact scores, Ext(m) as their representative is among the most crucial statistics, showing that they should not be removed entirely even if one decides to omit several. The most relevant statistics overall are DBH, sqrt-CLe, Height, B2, mDW, Ext(m), and several Volume measurements. These mostly match the statistics with the highest DBHc-impact score. Interestingly, the statistics Length, LeafN, Height, InnerN, and mI' that have the highest impact in identifying PC-outlier-QSMs are (except for Height) ranked low or medium in the normal impact scores or outlier production and seem to be a subset of statistics especially suited to identify deviation regarding the point cloud fit. Larger portions of the order of PC/DBHc-impact scores and -sensitivities remain similar (except for the impact score inflation of agreeing groups of statistics), but there are a few statistics that jump in ranks which indicates uniqueness: mI' with a medium PC-sensitivity, for example, is among the statistics with highest PC-impact score which means that it identifies more PC-outlier-QSMs that few others would have detected.

The greedy selection for the statistics with highest impact for method versions opt4, opt11, and opt15, worked exactly the opposite as the filtering which iteratively removed the most replaceable statistic. Instead the statistic with the best impact score on the trees that were not yet flagged by the already selected statistics was added, one after the other. Thereby, it became apparent that 3D imbalance statistics were relatively weak contenders and other statistics could take their place. Only the 25th highest impact statistic would have been a 3D imbalance index, namely Int-l(A). Int-w(m) and Ext(M) followed on place 29 and 31, respectively. However, several of the non-3D indices were extremely relevant. The order of opt15 was: DBH, Height, sqrt-CLe, B2, mDW, Vol0-25, Vol50-75, ZeroCyl, mI', Vol100-200, Vol25-50, TopRes, B1, mW, and Vol75-100, which matches the impact score ordering relatively well.

\begin{table}[htbp!]
    \centering
    \caption{Outlier production and (PC- as well as census DBH-)impact scores of the statistics in decreasing order and rounded to two decimal places (see Section~\ref{sec:meth_QSMquali} for the definitions).}
    \label{tab:outlier_prod}
    \footnotesize
\begin{tabular}{*{46}{c@{\hskip 3pt}}} 
\hline
\noalign{\vskip 4pt}
\multicolumn{46}{l}{\textbf{Outlier production}}\\
\addlinespace
\rotatebox{90}{DBH} & \rotatebox{90}{sqrt-CLe} & \rotatebox{90}{VLD} & \rotatebox{90}{TC} & \rotatebox{90}{Int-w(A)} & \rotatebox{90}{Int-w(m)} & \rotatebox{90}{Int-w(M)} & \rotatebox{90}{Height} & \rotatebox{90}{Volume} & \rotatebox{90}{mD} & \rotatebox{90}{Int-w(a)} & \rotatebox{90}{CLln} & \rotatebox{90}{Vol0-25} & \rotatebox{90}{Ext(A)} & \rotatebox{90}{Ext(a)} & \rotatebox{90}{Ext(M)} & \rotatebox{90}{Ext(m)} & \rotatebox{90}{AVD} & \rotatebox{90}{TPL} & \rotatebox{90}{Sackin} & \rotatebox{90}{B2} & \rotatebox{90}{ALD} & \rotatebox{90}{TIPL} & \rotatebox{90}{Vol25-50} & \rotatebox{90}{mWomD} & \rotatebox{90}{4thrt-rQi} & \rotatebox{90}{mW} & \rotatebox{90}{Vol100-200} & \rotatebox{90}{Int-l(A)} & \rotatebox{90}{Int-l(M)} & \rotatebox{90}{Vol50-75} & \rotatebox{90}{mDW} & \rotatebox{90}{Cherry} & \rotatebox{90}{Int-l(m)} & \rotatebox{90}{LeafN} & \rotatebox{90}{mI'} & \rotatebox{90}{Int-l(a)} & \rotatebox{90}{B1} & \rotatebox{90}{Length} & \rotatebox{90}{Vol75-100} & \rotatebox{90}{InnerN} & \rotatebox{90}{s-shape} & \rotatebox{90}{Vol200+} & \rotatebox{90}{TopRes} & \rotatebox{90}{ZeroCyl} & \rotatebox{90}{mIw} \\
 \rotatebox{90}{562} & \rotatebox{90}{556} & \rotatebox{90}{542} & \rotatebox{90}{490} & \rotatebox{90}{485} & \rotatebox{90}{476} & \rotatebox{90}{474} & \rotatebox{90}{473} & \rotatebox{90}{469} & \rotatebox{90}{460} & \rotatebox{90}{459} & \rotatebox{90}{455} & \rotatebox{90}{453} & \rotatebox{90}{450} & \rotatebox{90}{450} & \rotatebox{90}{449} & \rotatebox{90}{449} & \rotatebox{90}{449} & \rotatebox{90}{442} & \rotatebox{90}{441} & \rotatebox{90}{436} & \rotatebox{90}{435} & \rotatebox{90}{431} & \rotatebox{90}{421} & \rotatebox{90}{413} & \rotatebox{90}{413} & \rotatebox{90}{412} & \rotatebox{90}{408} & \rotatebox{90}{407} & \rotatebox{90}{403} & \rotatebox{90}{400} & \rotatebox{90}{399} & \rotatebox{90}{395} & \rotatebox{90}{386} & \rotatebox{90}{382} & \rotatebox{90}{381} & \rotatebox{90}{380} & \rotatebox{90}{365} & \rotatebox{90}{364} & \rotatebox{90}{361} & \rotatebox{90}{358} & \rotatebox{90}{348} & \rotatebox{90}{337} & \rotatebox{90}{331} & \rotatebox{90}{318} & \rotatebox{90}{309} \\
\hline
\noalign{\vskip 4pt}
\multicolumn{46}{l}{\textbf{Impact score (based on all 46 statistics)}}\\
\addlinespace
\rotatebox{90}{DBH} & \rotatebox{90}{Height} & \rotatebox{90}{sqrt-CLe} & \rotatebox{90}{B2} & \rotatebox{90}{mDW} & \rotatebox{90}{Vol0-25} & \rotatebox{90}{Vol50-75} & \rotatebox{90}{Vol25-50} & \rotatebox{90}{ZeroCyl} & \rotatebox{90}{VLD} & \rotatebox{90}{Cherry} & \rotatebox{90}{mW} & \rotatebox{90}{Vol100-200} & \rotatebox{90}{mI'} & \rotatebox{90}{Vol75-100} & \rotatebox{90}{TopRes} & \rotatebox{90}{B1} & \rotatebox{90}{Volume} & \rotatebox{90}{mWomD} & \rotatebox{90}{4thrt-rQi} & \rotatebox{90}{mD} & \rotatebox{90}{Vol200+} & \rotatebox{90}{mIw} & \rotatebox{90}{Length} & \rotatebox{90}{InnerN} & \rotatebox{90}{TC} & \rotatebox{90}{LeafN} & \rotatebox{90}{Int-l(A)} & \rotatebox{90}{s-shape} & \rotatebox{90}{Int-w(A)} & \rotatebox{90}{Int-w(m)} & \rotatebox{90}{Int-l(M)} & \rotatebox{90}{Int-w(M)} & \rotatebox{90}{Int-w(a)} & \rotatebox{90}{Int-l(m)} & \rotatebox{90}{AVD} & \rotatebox{90}{Int-l(a)} & \rotatebox{90}{CLln} & \rotatebox{90}{Ext(M)} & \rotatebox{90}{Ext(m)} & \rotatebox{90}{Ext(A)} & \rotatebox{90}{Ext(a)} & \rotatebox{90}{ALD} & \rotatebox{90}{TIPL} & \rotatebox{90}{Sackin} & \rotatebox{90}{TPL} \\
 \rotatebox{90}{273.97} & \rotatebox{90}{242.24} & \rotatebox{90}{239.96} & \rotatebox{90}{222} & \rotatebox{90}{205.63} & \rotatebox{90}{186.24} & \rotatebox{90}{178.18} & \rotatebox{90}{167.33} & \rotatebox{90}{165.15} & \rotatebox{90}{165.01} & \rotatebox{90}{163.14} & \rotatebox{90}{162.84} & \rotatebox{90}{161.8} & \rotatebox{90}{157.4} & \rotatebox{90}{157.04} & \rotatebox{90}{152.47} & \rotatebox{90}{149.98} & \rotatebox{90}{149.89} & \rotatebox{90}{145.4} & \rotatebox{90}{142.69} & \rotatebox{90}{141.15} & \rotatebox{90}{130.09} & \rotatebox{90}{121.93} & \rotatebox{90}{119.71} & \rotatebox{90}{118.66} & \rotatebox{90}{109.28} & \rotatebox{90}{108.63} & \rotatebox{90}{99.47} & \rotatebox{90}{98.41} & \rotatebox{90}{91.92} & \rotatebox{90}{84.43} & \rotatebox{90}{83.09} & \rotatebox{90}{80.93} & \rotatebox{90}{75.32} & \rotatebox{90}{75.28} & \rotatebox{90}{74.99} & \rotatebox{90}{73.8} & \rotatebox{90}{72.62} & \rotatebox{90}{71.61} & \rotatebox{90}{71.61} & \rotatebox{90}{71.53} & \rotatebox{90}{71.53} & \rotatebox{90}{71.38} & \rotatebox{90}{69.5} & \rotatebox{90}{67.43} & \rotatebox{90}{60.33} \\
\noalign{\vskip 4pt}
\multicolumn{46}{l}{\textbf{Impact score (based on the 35 filtered statistics obtained with filtering threshold 0.7)}}\\
\addlinespace
\rotatebox{90}{DBH} & \rotatebox{90}{sqrt-CLe} & \rotatebox{90}{Height} & \rotatebox{90}{B2} & \rotatebox{90}{mDW} & \rotatebox{90}{Ext(m)} & \rotatebox{90}{Vol0-25} & \rotatebox{90}{Vol50-75} & \rotatebox{90}{VLD} & \rotatebox{90}{Vol25-50} & \rotatebox{90}{Vol100-200} & \rotatebox{90}{Cherry} & \rotatebox{90}{ZeroCyl} & \rotatebox{90}{mW} & \rotatebox{90}{Vol75-100} & \rotatebox{90}{Int-w(a)} & \rotatebox{90}{mI'} & \rotatebox{90}{Volume} & \rotatebox{90}{TopRes} & \rotatebox{90}{B1} & \rotatebox{90}{mWomD} & \rotatebox{90}{mD} & \rotatebox{90}{4thrt-rQi} & \rotatebox{90}{Int-l(A)} & \rotatebox{90}{Vol200+} & \rotatebox{90}{TC} & \rotatebox{90}{Length} & \rotatebox{90}{Int-l(a)} & \rotatebox{90}{mIw} & \rotatebox{90}{InnerN} & \rotatebox{90}{LeafN} & \rotatebox{90}{ALD} & \rotatebox{90}{s-shape} & \rotatebox{90}{TIPL} & \rotatebox{90}{Sackin} \\
 \rotatebox{90}{289.95} & \rotatebox{90}{256.49} & \rotatebox{90}{254.36} & \rotatebox{90}{234.97} & \rotatebox{90}{214.36} & \rotatebox{90}{200.38} & \rotatebox{90}{199.89} & \rotatebox{90}{190.5} & \rotatebox{90}{184.07} & \rotatebox{90}{181.55} & \rotatebox{90}{176.89} & \rotatebox{90}{172.77} & \rotatebox{90}{172.6} & \rotatebox{90}{170.24} & \rotatebox{90}{169.58} & \rotatebox{90}{166.62} & \rotatebox{90}{165.57} & \rotatebox{90}{164.81} & \rotatebox{90}{162.1} & \rotatebox{90}{156.01} & \rotatebox{90}{155.42} & \rotatebox{90}{154.45} & \rotatebox{90}{153.88} & \rotatebox{90}{146.26} & \rotatebox{90}{140.16} & \rotatebox{90}{131.61} & \rotatebox{90}{130.45} & \rotatebox{90}{128.18} & \rotatebox{90}{128.06} & \rotatebox{90}{125.63} & \rotatebox{90}{116.63} & \rotatebox{90}{115} & \rotatebox{90}{113.08} & \rotatebox{90}{97} & \rotatebox{90}{96.48} \\
\hline
\noalign{\vskip 4pt}
\multicolumn{46}{l}{\textbf{PC-impact score (based on all 46 statistics but only on the 129 PC-outlier-QSMs)}}\\
\addlinespace
\rotatebox{90}{Length} & \rotatebox{90}{LeafN} & \rotatebox{90}{Height} & \rotatebox{90}{InnerN} & \rotatebox{90}{mI'} & \rotatebox{90}{B1} & \rotatebox{90}{DBH} & \rotatebox{90}{Vol0-25} & \rotatebox{90}{Ext(A)} & \rotatebox{90}{Ext(a)} & \rotatebox{90}{Ext(M)} & \rotatebox{90}{Ext(m)} & \rotatebox{90}{Volume} & \rotatebox{90}{mIw} & \rotatebox{90}{4thrt-rQi} & \rotatebox{90}{mD} & \rotatebox{90}{TIPL} & \rotatebox{90}{VLD} & \rotatebox{90}{Vol50-75} & \rotatebox{90}{CLln} & \rotatebox{90}{Int-w(M)} & \rotatebox{90}{Int-w(m)} & \rotatebox{90}{Vol100-200} & \rotatebox{90}{Vol25-50} & \rotatebox{90}{sqrt-CLe} & \rotatebox{90}{Int-w(A)} & \rotatebox{90}{TC} & \rotatebox{90}{ALD} & \rotatebox{90}{s-shape} & \rotatebox{90}{B2} & \rotatebox{90}{Int-w(a)} & \rotatebox{90}{AVD} & \rotatebox{90}{mDW} & \rotatebox{90}{TopRes} & \rotatebox{90}{Int-l(a)} & \rotatebox{90}{mWomD} & \rotatebox{90}{Sackin} & \rotatebox{90}{mW} & \rotatebox{90}{TPL} & \rotatebox{90}{Int-l(M)} & \rotatebox{90}{Int-l(m)} & \rotatebox{90}{Int-l(A)} & \rotatebox{90}{Cherry} & \rotatebox{90}{ZeroCyl} & \rotatebox{90}{Vol75-100} & \rotatebox{90}{Vol200+} \\
 \rotatebox{90}{8.46} & \rotatebox{90}{5.87} & \rotatebox{90}{4.55} & \rotatebox{90}{3.43} & \rotatebox{90}{3.34} & \rotatebox{90}{3.33} & \rotatebox{90}{3.15} & \rotatebox{90}{3.1} & \rotatebox{90}{2.96} & \rotatebox{90}{2.96} & \rotatebox{90}{2.96} & \rotatebox{90}{2.96} & \rotatebox{90}{2.96} & \rotatebox{90}{2.62} & \rotatebox{90}{2.41} & \rotatebox{90}{2.33} & \rotatebox{90}{2.19} & \rotatebox{90}{2.1} & \rotatebox{90}{2.08} & \rotatebox{90}{1.91} & \rotatebox{90}{1.86} & \rotatebox{90}{1.82} & \rotatebox{90}{1.76} & \rotatebox{90}{1.69} & \rotatebox{90}{1.68} & \rotatebox{90}{1.6} & \rotatebox{90}{1.6} & \rotatebox{90}{1.6} & \rotatebox{90}{1.55} & \rotatebox{90}{1.51} & \rotatebox{90}{1.48} & \rotatebox{90}{1.43} & \rotatebox{90}{1.33} & \rotatebox{90}{1.32} & \rotatebox{90}{1.13} & \rotatebox{90}{1.11} & \rotatebox{90}{0.96} & \rotatebox{90}{0.94} & \rotatebox{90}{0.74} & \rotatebox{90}{0.63} & \rotatebox{90}{0.63} & \rotatebox{90}{0.46} & \rotatebox{90}{0.4} & \rotatebox{90}{0.09} & \rotatebox{90}{0} & \rotatebox{90}{0} \\
\noalign{\vskip 4pt}
\multicolumn{46}{l}{\textbf{DBHc-impact score (based on all 46 statistics but only the 441 DBHc-outlier-QSMs)}}\\
\addlinespace
\rotatebox{90}{DBH} & \rotatebox{90}{Height} & \rotatebox{90}{mDW} & \rotatebox{90}{VLD} & \rotatebox{90}{B2} & \rotatebox{90}{Vol25-50} & \rotatebox{90}{mW} & \rotatebox{90}{mWomD} & \rotatebox{90}{Length} & \rotatebox{90}{mD} & \rotatebox{90}{Cherry} & \rotatebox{90}{Vol100-200} & \rotatebox{90}{mI'} & \rotatebox{90}{Vol50-75} & \rotatebox{90}{InnerN} & \rotatebox{90}{TopRes} & \rotatebox{90}{Int-w(m)} & \rotatebox{90}{LeafN} & \rotatebox{90}{TIPL} & \rotatebox{90}{mIw} & \rotatebox{90}{B1} & \rotatebox{90}{sqrt-CLe} & \rotatebox{90}{Volume} & \rotatebox{90}{Int-w(M)} & \rotatebox{90}{Vol0-25} & \rotatebox{90}{Sackin} & \rotatebox{90}{Int-l(A)} & \rotatebox{90}{TC} & \rotatebox{90}{4thrt-rQi} & \rotatebox{90}{Vol75-100} & \rotatebox{90}{ALD} & \rotatebox{90}{Ext(A)} & \rotatebox{90}{Ext(a)} & \rotatebox{90}{Ext(M)} & \rotatebox{90}{Ext(m)} & \rotatebox{90}{Int-w(a)} & \rotatebox{90}{AVD} & \rotatebox{90}{Int-l(a)} & \rotatebox{90}{Int-w(A)} & \rotatebox{90}{Int-l(m)} & \rotatebox{90}{Int-l(M)} & \rotatebox{90}{TPL} & \rotatebox{90}{CLln} & \rotatebox{90}{s-shape} & \rotatebox{90}{ZeroCyl} & \rotatebox{90}{Vol200+} \\
 \rotatebox{90}{29.02} & \rotatebox{90}{17.25} & \rotatebox{90}{12.92} & \rotatebox{90}{11.54} & \rotatebox{90}{11.45} & \rotatebox{90}{10.16} & \rotatebox{90}{9.11} & \rotatebox{90}{9.1} & \rotatebox{90}{8.73} & \rotatebox{90}{8.69} & \rotatebox{90}{8.36} & \rotatebox{90}{8.35} & \rotatebox{90}{7.97} & \rotatebox{90}{7.84} & \rotatebox{90}{7.77} & \rotatebox{90}{6.94} & \rotatebox{90}{6.85} & \rotatebox{90}{6.56} & \rotatebox{90}{6.39} & \rotatebox{90}{6.21} & \rotatebox{90}{6.13} & \rotatebox{90}{6.04} & \rotatebox{90}{5.91} & \rotatebox{90}{5.14} & \rotatebox{90}{4.64} & \rotatebox{90}{4.63} & \rotatebox{90}{4.45} & \rotatebox{90}{4.14} & \rotatebox{90}{4.13} & \rotatebox{90}{4.12} & \rotatebox{90}{3.94} & \rotatebox{90}{3.86} & \rotatebox{90}{3.86} & \rotatebox{90}{3.86} & \rotatebox{90}{3.86} & \rotatebox{90}{3.85} & \rotatebox{90}{3.21} & \rotatebox{90}{2.98} & \rotatebox{90}{2.87} & \rotatebox{90}{2.83} & \rotatebox{90}{2.66} & \rotatebox{90}{2.56} & \rotatebox{90}{2.38} & \rotatebox{90}{1.79} & \rotatebox{90}{0.96} & \rotatebox{90}{0} \\
\hline
\noalign{\vskip 4pt}
\multicolumn{46}{l}{\textbf{PC-sensitivity} (identified fraction of the 129 PC-outlier-QSMs)}\\
\addlinespace
\rotatebox{90}{Length} & \rotatebox{90}{LeafN} & \rotatebox{90}{Ext(A)} & \rotatebox{90}{Ext(a)} & \rotatebox{90}{Ext(M)} & \rotatebox{90}{Ext(m)} & \rotatebox{90}{Height} & \rotatebox{90}{InnerN} & \rotatebox{90}{mD} & \rotatebox{90}{VLD} & \rotatebox{90}{4thrt-rQi} & \rotatebox{90}{s-shape} & \rotatebox{90}{TIPL} & \rotatebox{90}{Volume} & \rotatebox{90}{Vol0-25} & \rotatebox{90}{B1} & \rotatebox{90}{sqrt-CLe} & \rotatebox{90}{Vol25-50} & \rotatebox{90}{CLln} & \rotatebox{90}{mI'} & \rotatebox{90}{mIw} & \rotatebox{90}{Sackin} & \rotatebox{90}{ALD} & \rotatebox{90}{AVD} & \rotatebox{90}{DBH} & \rotatebox{90}{Int-w(m)} & \rotatebox{90}{TC} & \rotatebox{90}{mWomD} & \rotatebox{90}{Int-w(A)} & \rotatebox{90}{Int-w(a)} & \rotatebox{90}{Int-w(M)} & \rotatebox{90}{Int-l(a)} & \rotatebox{90}{TPL} & \rotatebox{90}{Vol100-200} & \rotatebox{90}{Int-l(M)} & \rotatebox{90}{Int-l(m)} & \rotatebox{90}{mW} & \rotatebox{90}{Vol50-75} & \rotatebox{90}{B2} & \rotatebox{90}{Int-l(A)} & \rotatebox{90}{mDW} & \rotatebox{90}{TopRes} & \rotatebox{90}{Cherry} & \rotatebox{90}{ZeroCyl} & \rotatebox{90}{Vol75-100} & \rotatebox{90}{Vol200+} \\
 \rotatebox{90}{0.21} & \rotatebox{90}{0.18} & \rotatebox{90}{0.16} & \rotatebox{90}{0.16} & \rotatebox{90}{0.16} & \rotatebox{90}{0.16} & \rotatebox{90}{0.13} & \rotatebox{90}{0.12} & \rotatebox{90}{0.12} & \rotatebox{90}{0.1} & \rotatebox{90}{0.1} & \rotatebox{90}{0.09} & \rotatebox{90}{0.09} & \rotatebox{90}{0.09} & \rotatebox{90}{0.09} & \rotatebox{90}{0.09} & \rotatebox{90}{0.09} & \rotatebox{90}{0.08} & \rotatebox{90}{0.08} & \rotatebox{90}{0.07} & \rotatebox{90}{0.07} & \rotatebox{90}{0.07} & \rotatebox{90}{0.07} & \rotatebox{90}{0.07} & \rotatebox{90}{0.06} & \rotatebox{90}{0.06} & \rotatebox{90}{0.06} & \rotatebox{90}{0.06} & \rotatebox{90}{0.05} & \rotatebox{90}{0.05} & \rotatebox{90}{0.05} & \rotatebox{90}{0.05} & \rotatebox{90}{0.05} & \rotatebox{90}{0.05} & \rotatebox{90}{0.05} & \rotatebox{90}{0.05} & \rotatebox{90}{0.05} & \rotatebox{90}{0.04} & \rotatebox{90}{0.04} & \rotatebox{90}{0.03} & \rotatebox{90}{0.03} & \rotatebox{90}{0.03} & \rotatebox{90}{0.02} & \rotatebox{90}{0.01} & \rotatebox{90}{0} & \rotatebox{90}{0} \\
\noalign{\vskip 4pt}
\multicolumn{46}{l}{\textbf{DBHc-sensitivity} (identified fraction of the 441 DBHc-outlier-QSMs)}\\
\addlinespace
\rotatebox{90}{DBH} & \rotatebox{90}{Height} & \rotatebox{90}{VLD} & \rotatebox{90}{Vol25-50} & \rotatebox{90}{Int-w(m)} & \rotatebox{90}{mD} & \rotatebox{90}{mW} & \rotatebox{90}{Int-w(M)} & \rotatebox{90}{Ext(A)} & \rotatebox{90}{Ext(a)} & \rotatebox{90}{Ext(M)} & \rotatebox{90}{Ext(m)} & \rotatebox{90}{Volume} & \rotatebox{90}{LeafN} & \rotatebox{90}{Sackin} & \rotatebox{90}{mWomD} & \rotatebox{90}{InnerN} & \rotatebox{90}{B2} & \rotatebox{90}{mDW} & \rotatebox{90}{TC} & \rotatebox{90}{Vol50-75} & \rotatebox{90}{Int-w(a)} & \rotatebox{90}{mI'} & \rotatebox{90}{TIPL} & \rotatebox{90}{sqrt-CLe} & \rotatebox{90}{Int-w(A)} & \rotatebox{90}{Length} & \rotatebox{90}{CLln} & \rotatebox{90}{mIw} & \rotatebox{90}{Vol100-200} & \rotatebox{90}{Int-l(A)} & \rotatebox{90}{Int-l(m)} & \rotatebox{90}{TPL} & \rotatebox{90}{AVD} & \rotatebox{90}{Int-l(a)} & \rotatebox{90}{Int-l(M)} & \rotatebox{90}{Cherry} & \rotatebox{90}{ALD} & \rotatebox{90}{TopRes} & \rotatebox{90}{4thrt-rQi} & \rotatebox{90}{B1} & \rotatebox{90}{s-shape} & \rotatebox{90}{Vol0-25} & \rotatebox{90}{Vol75-100} & \rotatebox{90}{ZeroCyl} & \rotatebox{90}{Vol200+} \\
 \rotatebox{90}{0.12} & \rotatebox{90}{0.08} & \rotatebox{90}{0.07} & \rotatebox{90}{0.06} & \rotatebox{90}{0.06} & \rotatebox{90}{0.06} & \rotatebox{90}{0.06} & \rotatebox{90}{0.05} & \rotatebox{90}{0.05} & \rotatebox{90}{0.05} & \rotatebox{90}{0.05} & \rotatebox{90}{0.05} & \rotatebox{90}{0.05} & \rotatebox{90}{0.05} & \rotatebox{90}{0.05} & \rotatebox{90}{0.05} & \rotatebox{90}{0.05} & \rotatebox{90}{0.05} & \rotatebox{90}{0.05} & \rotatebox{90}{0.05} & \rotatebox{90}{0.05} & \rotatebox{90}{0.05} & \rotatebox{90}{0.05} & \rotatebox{90}{0.05} & \rotatebox{90}{0.05} & \rotatebox{90}{0.04} & \rotatebox{90}{0.04} & \rotatebox{90}{0.04} & \rotatebox{90}{0.04} & \rotatebox{90}{0.04} & \rotatebox{90}{0.04} & \rotatebox{90}{0.04} & \rotatebox{90}{0.04} & \rotatebox{90}{0.04} & \rotatebox{90}{0.03} & \rotatebox{90}{0.03} & \rotatebox{90}{0.03} & \rotatebox{90}{0.03} & \rotatebox{90}{0.03} & \rotatebox{90}{0.03} & \rotatebox{90}{0.03} & \rotatebox{90}{0.03} & \rotatebox{90}{0.02} & \rotatebox{90}{0.02} & \rotatebox{90}{0.01} & \rotatebox{90}{0} \\
\hline
\end{tabular}
\end{table}

When comparing the optimized method variations opt4, opt11, and opt15 to method version based on statistic sets of the same sizes 4, 11, and 15, respectively, there are partially clear improvements and in some parts both gains and losses. 
Variation op4 compared to only3D-fil70, can be interpreted as an overall improvement in sensitivity scores with only some increased $\#\beta_{PC}$ counts.

Method opt11 corresponds with version onlyQSM-fil70 and we can observe different functionalities: While opt11 detects more negSel-QSMs than onlyQSM-fil70 (see Table~\ref{tab:negSelQSM_count}) and in doing so has a higher census DBH-sensitivity, 44.4\% vs 39.8\%, onlyQSM-fil70's PC-sensitivity of 50\% outperforms opt11 with 44.2\% (see Table~\ref{tab:method_comparison}). It should be noted here once again that there were only 129 PC- compared to 441 DBHc-outlier-QSMs, which makes opt11's DBHc-sensitivity weigh more. This prompts us to be cautious when optimizing a set of statistics because apparently choosing optimal statistics based on the impact score alone can lead to underdetecting such known sources of flaws in QSMs. However, this also suggests that with high certainty there are a variety of other unknown flaws, apart from the ones noticeable by using the point cloud and census DBH data, which are better detected by opt11 than by onlyQSM-fil70. The QSM statistics (onlyQSM-fil70) seem to be particularly good in detecting PC-outlier-QSMs, Length and Height have among the highest PC-impact scores. However, consulting Table~\ref{tab:QSM_rejectingStats_count}, we can also conclude that this set contains several similar/redundant statistics that all detect the same QSMs as outliers, whereas for opt11 the number of rejecting statistics per QSM is kept small by the selection process of the 11 optimal statistics. The comparison of opt15 with its counterpart all-NOTnon3D-fil70 shows the same results as with opt11.

Although there are a few statistics that can be removed without big losses of sensitivity in the negative selection process, the general implication of the experiments with different sets of statistics is that the more indices are used, the better the results.

\paragraph{Relations and correlations between statistics}

\begin{figure}[ht]
\centering
\includegraphics[width=\textwidth]{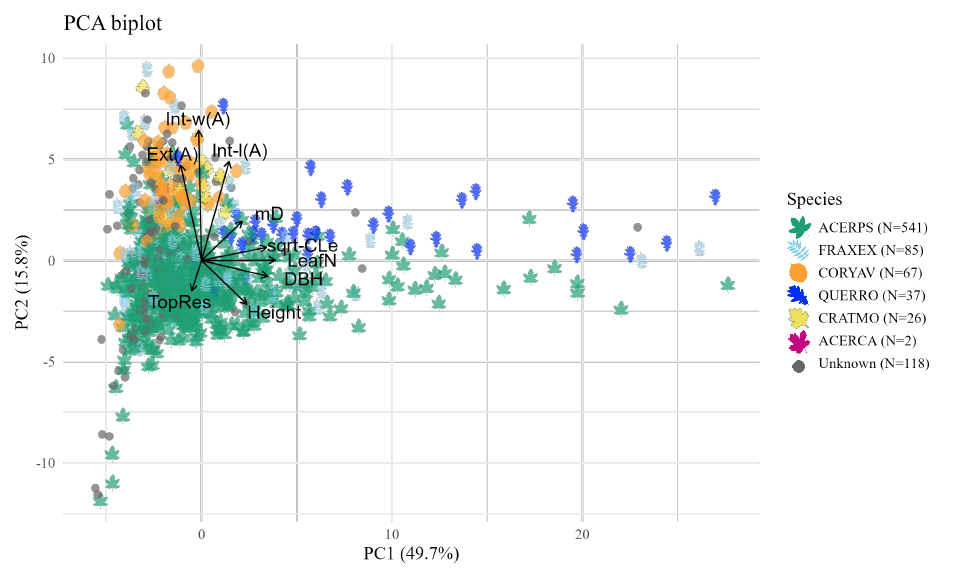} 
\vspace{-0.8cm}
\caption{PCA biplot of all 876 \enquote{best} QSMs based on all 49 numeric statistics $P_{all}$. Colors and shapes indicate tree species; the legend reports the total number of trees per species (N). For a selection of statistics, arrows indicate variable loadings (representing correlations between the original statistic and the first two principal components). One extreme observation (tree 446, one of the tallest trees) strongly influenced the scale of the PCA plot and was omitted from the visualization for clarity.}
\label{fig:PCA}
\end{figure}

The principal component analysis of the scaled statistics of the \enquote{best} QSMs revealed that the first two components explained 65.5\% of the total variance (PC1: 49.7\%, PC2: 15.8\%). PC1 is strongly associated with the number of vertices in the topology (LeafN, InnerN), several non-3D statistics which correlate with the size of the tree, Length, volume statistics, and CrownArea, suggesting that it captures variation in overall \enquote{size/complexity} of the 3D tree model, whereas PC2 loaded primarily on the Int-w, Int-l, and Ext statistics and may therefore reflect a \enquote{3D imbalance} axis (see Figure~\ref{fig:PCA}). Subsequent components each explained less than 10\% of the variance and were not interpreted further. Details on the loadings can be found in Table~\ref{tab:PCA_loadings}. The arrows indicating variable loadings in Figure~\ref{fig:PCA} are only illustrative. The respective arrows for the Int-w, Int-, and Ext statistics, for example, overlap almost perfectly and many (mostly shorter) arrows for the other statistics pointed roughly along the PC1-axis positioned at an angle between Height and mD. In the RF and GB model performance we could already see that some species classes could not be easily separated, for example CORYAV and Other (consisting of QUERRO, CRATMO, and ACERCA), which can also be sensed in this two-dimensional embedding since the CORYAV and CRATMO point clouds heavily overlap. QUERRO could have been better separated (it can already be separated well from CORYAV/CRATMO by using only PC1) and predicted, but its total tree count of 37 was too low to justify as a single class. In general, the fact that species classes can be partially separated by PC1 as a complexity/size axis, aligns with the initial observation that the tree height distributions differ between species (see Figure~\ref{fig:speciesDistrib}).

The two-dimensional non-linear UMAP embeddings of the \enquote{best} QSM data visualized the local neighborhood structure (see Figure~\ref{fig:umap_proj}). Similarly to the PCA-plot, most species at least partially overlap with the other species' value spaces and others mostly occupy the same region. ACERPS occupies the largest region of the feature space, overlapping with nearly any other species' region. By taking into account all instead of only the original statistics, Panel (b) compared to (a), the overlap between ACERPS and CORYAV/CRATMO visually decreases, which is consistent with the slight performance gains in the species classification. Using the full set of statistics, rather than only the original subset, also increased the compactness of the point clouds for each species. Specifically, points belonging to the same species form more cohesive clusters instead of splitting into multiple smaller clusters.

Statistics that covary according to the PCA (arrows pointing in the same direction) are also depicted as highly correlating in Figure~\ref{fig:correlations_all}. Some relations are especially interesting. Height, for example, does correlate moderately but not strongly with any variable, which might be counter-intuitive since one might think that, e.g., a thicker stem (more wider tree parts and thus more volume) would be a good indicator of a larger tree. It correlates negatively with 3D imbalance, external imbalance in particular. Examples like 180a and 180b (depicted in Figure~\ref{fig:workflow}), where the main stem is taller and more upright than the minor sister stem, might prompt that trees consisting of several stems could be a driving factor here. On average, a minor sister stem (ID ending in b or c) is around 2m smaller than the major sister stem (a), but there are no notable differences in internal and external 3D imbalance. Furthermore, a comparison of multi-stem trees with single-stem trees that have a volume comparable to the combined volume of all sister stems showed no systematic differences between the two groups.

Then, there is a large  subset of \enquote{twig-sensitive} statistics that highly correlate with the volume of thinner tree parts which have a diameter of 0-5cm. For, e.g., LeafN (the number of end points of the twigs) and the CrownArea, which is stretched across the twigs, this relation is not surprising. In the case of Length, this shows that the smaller branches make up a large part of the total tree length in many cases. Volume is also part of the subset but correlates most strongly with the Volume of thicker tree parts with diameter $>50$cm.

The high correlation of the Ext, Int-w, and Int-l 3D imbalance statistics suggests that one representative of each subset is sufficient, whereby the choice of representative hardly matters.

The small subset of VLD, mD, ALD, AVD -- statistics that depend on the number of branching nodes from the root to the tips of the branches -- do not correlate strongly with any of the \enquote{twig-sensitive} statistics. In other words, more leaves or twigs in the tree only moderately increase the distance of the leaves to the root, i.e., there are only more leaves or twigs of the same distance. From a phylogenetics perspective, the depth levels are rather densely filled out instead of having leaves in great depths -- reminiscent of the concept of maximally balanced $T^{mb}$ where each depth level is as full as possible creating a rooted tree with minimal mD and AVD \cite{fischer_tree_2023}.

All in all, the original statistics with the exception of the coordinates Locx and Locy, are closely correlated. Filtering with correlation threshold $<0.9$ left only Height, DBH, and Vol100-200, as well as Locx and Locy, while all other statistics were 3D and non-3D statistics (see Figure~\ref{fig:correlations_allfiltered}).

\paragraph{Variable importance for species classification}

Variable importance was investigated for the RF and GB models ii) and iii) (see Tables~\ref{tab:var_imp_GB} and \ref{tab:var_imp_RF}). As expected from the PCA which showed that complexity and 3D imbalance are the two axes that explain the most variance, the models based on ii) and iii) contained representatives of both as the most important variables. For GB iii), Int-w(m) and DBH and filled these rolls. The model GB ii) based on the original statistics regarded Height (the original variable that correlated most with 3D imbalance) as well as CrownArea as a typical complexity statistic as most crucial. This also holds for the RF variable importance.

As mentioned in Section~\ref{sec:results_classification}, the models based on the original statistics did not perform a lot worse than the models using all statistics. Height is the only original variable that correlates at least moderately with the 3D imbalance measures -- according to the PCA an important axis for understanding the dataset -- and seems to mitigate their absence to a degree.

\begin{table}[htbp!]
    \centering
    \caption{Variable importance of predictors for the gradient boosting models based on (a) $P_{ori}$ and (b) $P_{fil}$ averaged across the 10 outer cross-validation folds and shown in descending order. The standard deviation (SD) as well as the mean rank across folds (with corresponding SD of rank) quantify the stability of variable importance.}
    \label{tab:var_imp_GB}
\begin{subtable}[t]{0.495\textwidth}
\centering
\caption{GB model ii)  based on $P_{ori}$}
\begin{tabular}{@{\hskip 4pt}l@{\hskip 6pt}c@{\hskip 6pt}c@{\hskip 6pt}c@{\hskip 6pt}c@{\hskip 4pt}}
Variable & $\overline{\text{Imp}}$ & Imp$_\text{SD}$ & $\overline{\text{Rank}}$  & Rank$_\text{SD}$ \\
    \noalign{\vskip 2pt}\hline\noalign{\vskip 2pt}
Height & 1036.0 & 45.5 & 1.1 & 0.3 \\ 
CrownArea & 905.0 & 75.2 & 2.1 & 0.6 \\ 
Locx & 863.0 & 52.8 & 3.1 & 0.6 \\ 
DBH & 818.1 & 75.5 & 4.0 & 0.9 \\ 
Locy & 755.1 & 46.6 & 4.7 & 0.5 \\ 
Vol25-50 & 607.3 & 29.6 & 6.0 & 0 \\ 
Vol50-75 & 363.2 & 23.3 & 7.8 & 0.6 \\ 
Vol100-200 & 346.0 & 17.9 & 8.1 & 1.0 \\ 
Vol0-25 & 341.1 & 45.3 & 8.6 & 1.2 \\ 
Vol75-100 & 310.2 & 36.8 & 9.5 & 1.0 \\ 
Volume & 231.2 & 22.2 & 11.0 & 0 \\ 
Vol200+ & 165.3 & 32.2 & 12.0 & 0 \\ 
    \end{tabular}
\end{subtable}
    \begin{subtable}[t]{0.495\textwidth}
\centering
\caption{GB model iii)  based on $P_{fil}$}
\begin{tabular}{@{\hskip 4pt}l@{\hskip 6pt}c@{\hskip 6pt}c@{\hskip 6pt}c@{\hskip 6pt}c@{\hskip 4pt}}
Variable & $\overline{\text{Imp}}$ & Imp$_\text{SD}$ & $\overline{\text{Rank}}$  & Rank$_\text{SD}$ \\
    \noalign{\vskip 2pt}\hline\noalign{\vskip 2pt}
Int-w(m) & 1567.2 & 72.2 & 1.0 & 0 \\ 
DBH & 656.7 & 47.7 & 2.0 & 0 \\ 
Locx & 561.6 & 41.0 & 3.2 & 0.4 \\ 
Height & 532.7 & 47.0 & 3.8 & 0.4 \\ 
TopRes & 443.0 & 33.6 & 5.8 & 1.0 \\ 
Ext(a) & 439.8 & 47.5 & 6.0 & 1.1 \\ 
Int-l(m) & 431.8 & 34.8 & 6.4 & 0.7 \\ 
Locy & 394.5 & 44.8 & 7.8 & 0.4 \\ 
mWomD & 268.9 & 39.5 & 9.5 & 0.5 \\ 
mIw & 262.8 & 37.6 & 9.6 & 0.7 \\ 
Vol100-200 & 193.6 & 35.9 & 12.1 & 1.7 \\ 
VLD & 192.1 & 21.0 & 12.0 & 0.7 \\ 
sqrt-CLe & 177.0 & 19.2 & 12.8 & 1.5 \\ 
TC & 164.9 & 18.6 & 13.4 & 1.1 \\ 
StemCount & 147.7 & 12.3 & 15.1 & 0.9 \\ 
B2 & 129.9 & 18.4 & 15.7 & 0.8 \\ 
AVD & 108.3 & 18.7 & 16.8 & 0.4 \\
    \end{tabular}
\end{subtable}
\end{table}

\section{Discussion} \label{sec:discussion}

This study addressed three interrelated research questions regarding 1. the quality and reliability of 3D tree models in the form of quantitative structure models (QSMs), 2. the usefulness of QSM-derived statistics for species classification as an applied use case, and 3. the relative importance, redundancy, and relationships among different classes of statistics, including the relatively newly introduced 3D imbalance indices as well as topological/non-3D imbalance indices stemming from the field of phylogenetics. Overall, the results demonstrate that the majority of QSMs provide robust representations of tree structure, that informed quality control improves downstream analyses without introducing substantial bias, and that a broader range of statistics can meaningfully enhance performance and interpretability.

\paragraph{QSM quality}
A central finding of this work is that QSMs generated under the applied reconstruction framework are, in most cases, reliable representations of the underlying 3D tree architecture. Although quality differences between individual QSM realizations of the same tree are common, several non-deviating candidates are typically available per tree, and the presented method consistently identified QSMs that were supported by nearly all quality criteria. The results support the key assumptions that consensus among multiple QSMs proves effective in distinguishing genuine tree structure from reconstruction artifacts, both for absent and erroneously added tree parts.

Importantly, the detected flaws -— ranging from missing branches and internal imbalance to extreme cases of reconstruction failure -- rarely compromise the primary application of QSMs for estimating above-ground volume and biomass. The slight increase in total volume observed for the selected “best” QSMs compared to the mean over all QSMs suggests that volume estimates based on averaging may be mildly conservative, supporting previous indications that allometric models underestimate above-ground biomass in temperate forests in the UK \cite{calders_laser_2022}.

While the introduction of an advanced investigation right at the end of the QSM reconstruction might be advisable, including a final comparison with the point cloud that checks especially for the absence of tree parts, it would hardly be computationally feasible. The method presented here serves a computationally viable alternative, both because all mentioned statistics can be computed in linear time and also because the method of outlier-detection is easy to understand, reproduce, and adapt. Even a small number of statistics such as height and volume can already identify many of the most severe reconstruction failures. However, relying exclusively on these measures implies not detecting other deviation patterns. Adding 3D and non-3D statistics allows to scan for outliers regarding a more widespread range of 3D model properties, while also allowing a more relaxed negative selection in which few rejections of a QSM are accepted.

\paragraph{Species classification}
The species classification results demonstrate that the statistics contain sufficient discriminatory information to allow species class predictions with more than 80\% estimated accuracy, even in the presence of substantial class imbalance. Large parts of the unlabeled parts of the dataset could be assigned as species with high or moderate confidence. Although the performance differences between models based on different subsets of statistics were small, the consistent improvements observed across the two different modeling approaches and validation folds indicate that they are meaningful rather than incidental. We can draw multiple conclusions: Firstly, it is advisable to at least experiment with filtering redundant variables before using them for model training with the possibility of performance gains. 

Secondly, 3D and non-3D statistics add new information that improved model decision making, which makes the addition of these and the development of new information sources of high interest for future studies. With an accuracy of around 80\% there is still room for improvement. Future endeavors could uncover if the 3D architecture of a tree species has such unique patterns, which we are just not yet able to detect and quantify, that allow species identification just as we can determine the tree species from the leaves or the bark.

\paragraph{Importance of statistics}

Overall size and architectural complexity as well as 3D imbalance were the two main axes that explain the most variance. This separation aligns with both intuitive expectations and the observed behavior of species classification models, where representatives of both dimensions consistently ranked among the most important predictors.

The relatively weak correlations between height and other size-related measures challenge common assumptions. The distinction between “twig-sensitive” statistics and depth-related indices suggests that trees often increase structural complexity by growing more dense overall instead of growing along single branches.

The commonly used original tree metrics like DBH, tree height, volume, and crown area already encompass meaningful knowledge on the trees and might perform well enough for certain applications just by themselves.

Some subsets of statistics, representatives of the 3D imbalance measures in particular, are highly replaceable both for the outlier-QSM detection as well as the species classification. Others capture unique information that is not recoverable from traditional QSM descriptors alone.

\paragraph{Outlook}

Taken together, the results suggest that systematic quality control based on diverse structural statistics can substantially improve the reliability of QSM-based analyses without introducing significant bias at the dataset level. The newly employed 3D and non-3D statistics are computationally efficient and provide meaningful additional information, particularly for detecting unique reconstruction error patterns and enhancing species classification tasks. 

Moreover, the application of these statistics is not limited to species classification and understanding branching differences, but could also work for numerous other tasks. Examples could be the monitoring of growth and branching development over time within individual trees or the comparison of the 3D architecture of individuals of the same species located in different environments, which could point to adaptations in response to heat, drought, or nutrient availability. Another interesting factor would be an assessment of tree health or their status alive/dead, potentially even including statuses like infected/dying, which could not be investigated within the present study because of too few dead tree observations. With rising digitization and advancing technologies, \enquote{virtual forest} data will become more frequent, allowing both the improvement of such predictive models and the investigation of more complicated research questions but also the application of these models as a tool to reduce field work and check the datasets for potential erroneous entries.

The overall verdict regarding the range of statistics to use is \enquote{the more the merrier}, with the addition that filtering for redundancy might improve model performance for some use cases. The development of new meaningful statistics is highly encouraged.

\section{Acknowledgements}
MF and SK were supported by the project ArtIGROW, which is a part of the WIR!-Alliance “ArtIFARM – Artificial Intelligence in Farming”, and gratefully acknowledge the Bundesministerium für Forschung, Technologie und Raumfahrt (German Federal Ministry of Research, Technology and Space, FKZ: 03WIR4805) for financial support. We thank the Wytham Woods project team \cite{calders_laser_2022} for the dataset and Luise Kühn for bringing it to our attention. Furthermore, we thank Katharina Hoff, Mario Stanke, Michael Höhle, and Henriette Markwart for fruitful discussions about the statistical methods.

 
\bibliographystyle{plainnat}
\bibliography{literature}

\begin{thebibliography}{44}
\providecommand{\natexlab}[1]{#1}
\providecommand{\url}[1]{\texttt{#1}}
\expandafter\ifx\csname urlstyle\endcsname\relax
  \providecommand{\doi}[1]{doi: #1}\else
  \providecommand{\doi}{doi: \begingroup \urlstyle{rm}\Url}\fi

\bibitem[Agapow and Purvis(2002)]{agapow_power_2002}
P.-M. Agapow and A.~Purvis.
\newblock Power of eight tree shape statistics to detect nonrandom diversification: a comparison by simulation of two models of cladogenesis.
\newblock \emph{Systematic Biology}, 51\penalty0 (6):\penalty0 866--872, 2002.
\newblock \doi{10.1080/10635150290102564}.

\bibitem[Baltenberger et~al.(1987)Baltenberger, Ohm, and Foster]{baltenberger_reactions_1987}
D.~E. Baltenberger, H.~W. Ohm, and J.~E. Foster.
\newblock Reactions of oat, barley, and wheat to infection with barley yellow dwarf virus isolates.
\newblock \emph{Crop Science}, 27\penalty0 (2):\penalty0 195--198, 1987.

\bibitem[Blum and Fran{\c{c}}ois(2006)]{blum_which_2006}
M.~G.~B. Blum and O.~Fran{\c{c}}ois.
\newblock Which random processes describe the {Tree} of {Life}? {A} large-scale study of phylogenetic tree imbalance.
\newblock \emph{Systematic Biology}, 55\penalty0 (4):\penalty0 685--691, 2006.
\newblock ISSN 1076-836X, 1063-5157.

\bibitem[Butt et~al.(2009)Butt, Campbell, Malhi, Morecroft, Fenn, and Thomas]{butt_initial_2009}
N.~Butt, G.~Campbell, Y.~Malhi, M.~Morecroft, K.~Fenn, and M.~Thomas.
\newblock Initial results from establishment of a long-term broadleaf monitoring plot at wytham woods, oxford, uk.
\newblock \emph{University Oxford, Oxford, UK, Rep}, 2009.

\bibitem[Calders et~al.(2015)Calders, Newnham, Burt, Murphy, Raumonen, Herold, Culvenor, Avitabile, Disney, Armston, et~al.]{calders_nondestructive_2015}
K.~Calders, G.~Newnham, A.~Burt, S.~Murphy, P.~Raumonen, M.~Herold, D.~Culvenor, V.~Avitabile, M.~Disney, J.~Armston, et~al.
\newblock Nondestructive estimates of above-ground biomass using terrestrial laser scanning.
\newblock \emph{Methods in Ecology and Evolution}, 6\penalty0 (2):\penalty0 198--208, 2015.

\bibitem[Calders et~al.(2020)Calders, Adams, Armston, Bartholomeus, Bauwens, Bentley, Chave, Danson, Demol, Disney, et~al.]{calders_terrestrial_2020}
K.~Calders, J.~Adams, J.~Armston, H.~Bartholomeus, S.~Bauwens, L.~P. Bentley, J.~Chave, F.~Danson, M.~Demol, M.~Disney, et~al.
\newblock Terrestrial laser scanning in forest ecology: {E}xpanding the horizon.
\newblock \emph{Remote Sensing of Environment}, 251:\penalty0 112102, 2020.

\bibitem[Calders et~al.(2022)Calders, Verbeeck, Burt, Origo, Nightingale, Malhi, Wilkes, Raumonen, Bunce, and Disney]{calders_laser_2022}
K.~Calders, H.~Verbeeck, A.~Burt, N.~Origo, J.~Nightingale, Y.~Malhi, P.~Wilkes, P.~Raumonen, R.~G.~H. Bunce, and M.~Disney.
\newblock Laser scanning reveals potential underestimation of biomass carbon in temperate forest.
\newblock \emph{Ecological Solutions and Evidence}, 3\penalty0 (4):\penalty0 e12197, 2022.

\bibitem[Colijn and Gardy(2014)]{colijn_phylogenetic_2014}
C.~Colijn and J.~Gardy.
\newblock Phylogenetic tree shapes resolve disease transmission patterns.
\newblock \emph{Evolution, Medicine, and Public Health}, 2014\penalty0 (1):\penalty0 96--108, 2014.
\newblock ISSN 2050-6201.
\newblock \doi{10.1093/emph/eou018}.

\bibitem[Coronado et~al.(2019)Coronado, Mir, Rossell{\'{o}}, and Valiente]{coronado_balance_2019}
T.~M. Coronado, A.~Mir, F.~Rossell{\'{o}}, and G.~Valiente.
\newblock A balance index for phylogenetic trees based on rooted quartets.
\newblock \emph{Journal of Mathematical Biology}, 79\penalty0 (3):\penalty0 1105--1148, 2019.
\newblock \doi{10.1007/s00285-019-01377-w}.

\bibitem[Coronado et~al.(2020)Coronado, Mir, Rossell{\'{o}}, and Rotger]{coronado_sackins_2020}
T.~M. Coronado, A.~Mir, F.~Rossell{\'{o}}, and L.~Rotger.
\newblock On {S}ackin's original proposal: the variance of the leaves' depths as a phylogenetic balance index.
\newblock \emph{{BMC} Bioinformatics}, 21\penalty0 (1), 2020.
\newblock \doi{10.1186/s12859-020-3405-1}.

\bibitem[Dobrow and Fill(1999)]{dobrow_total_1999}
R.~P. Dobrow and J.~A. Fill.
\newblock Total path length for random recursive trees.
\newblock \emph{Combinatorics, Probability and Computing}, 8\penalty0 (4):\penalty0 317–333, 1999.
\newblock \doi{10.1017/S0963548399003855}.

\bibitem[Fischer et~al.(2023)Fischer, Herbst, Kersting, K{\"u}hn, and Wicke]{fischer_tree_2023}
M.~Fischer, L.~Herbst, S.~J. Kersting, L.~K{\"u}hn, and K.~Wicke.
\newblock \emph{Tree balance indices - a comprehensive survey}.
\newblock Springer, Berlin, 2023.
\newblock ISBN 978-3-031-39799-8.

\bibitem[Ford(2005)]{ford_probabilities_2005}
D.~J. Ford.
\newblock Probabilities on cladograms: introduction to the alpha model, 2005.

\bibitem[Fusco and Cronk(1995)]{fusco_new_1995}
G.~Fusco and Q.~C.~B. Cronk.
\newblock A new method for evaluating the shape of large phylogenies.
\newblock \emph{Journal of Theoretical Biology}, 175\penalty0 (2):\penalty0 235--243, 1995.
\newblock \doi{10.1006/jtbi.1995.0136}.

\bibitem[Hackenberg et~al.(2015)Hackenberg, Spiecker, Calders, Disney, and Raumonen]{hackenberg_simpletree_2015}
J.~Hackenberg, H.~Spiecker, K.~Calders, M.~Disney, and P.~Raumonen.
\newblock {SimpleTree} - an efficient open source tool to build tree models from {TLS} clouds.
\newblock \emph{Forests}, 6\penalty0 (11):\penalty0 4245--4294, 2015.

\bibitem[Hackenberg et~al.(2021)Hackenberg, Calders, Demol, Raumonen, Piboule, and Disney]{hackenberg_simpleforest_2021}
J.~Hackenberg, K.~Calders, M.~Demol, P.~Raumonen, A.~Piboule, and M.~Disney.
\newblock {SimpleForest} - a comprehensive tool for {3D} reconstruction of trees from forest plot point clouds.
\newblock \emph{bioRxiv}, 2021.

\bibitem[Hastie et~al.(2009)Hastie, Tibshirani, and Friedman]{hastie_elements_2009}
T.~Hastie, R.~Tibshirani, and J.~Friedman.
\newblock \emph{The elements of statistical learning: data mining, inference, and prediction}, volume~2.
\newblock Springer, 2009.

\bibitem[Hayati et~al.(2019)Hayati, Shadgar, and Chindelevitch]{hayati_new_2019}
M.~Hayati, B.~Shadgar, and L.~Chindelevitch.
\newblock A new resolution function to evaluate tree shape statistics.
\newblock \emph{{PLOS} {ONE}}, 14\penalty0 (11):\penalty0 e0224197, 2019.
\newblock \doi{10.1371/journal.pone.0224197}.

\bibitem[Hern\'{a}ndez-Garc\'{i}a et~al.(2010)Hern\'{a}ndez-Garc\'{i}a, Tu\u{g}rul, Alejandro~Herrada, Egu\'{i}luz, and Klemm]{hernandez_simple_2010}
E.~Hern\'{a}ndez-Garc\'{i}a, M.~Tu\u{g}rul, E.~Alejandro~Herrada, V.~M. Egu\'{i}luz, and K.~Klemm.
\newblock Simple models for scaling phylogenetic trees.
\newblock \emph{International Journal of Bifurcation and Chaos}, 20\penalty0 (03):\penalty0 805–811, March 2010.
\newblock \doi{10.1142/s0218127410026095}.

\bibitem[Herrada et~al.(2011)Herrada, Egu{\'{\i}}luz, Hern{\'{a}}ndez-Garc{\'{\i}}a, and Duarte]{herrada_scaling_2011}
E.~A. Herrada, V.~M. Egu{\'{\i}}luz, E.~Hern{\'{a}}ndez-Garc{\'{\i}}a, and C.~M. Duarte.
\newblock Scaling properties of protein family phylogenies.
\newblock \emph{{BMC} Evolutionary Biology}, 11\penalty0 (1), 2011.
\newblock \doi{10.1186/1471-2148-11-155}.

\bibitem[Jackson et~al.(2019)Jackson, Shenkin, Wellpott, Calders, Origo, Disney, Burt, Raumonen, Gardiner, Herold, et~al.]{jackson_finite_2019}
T.~Jackson, A.~Shenkin, A.~Wellpott, K.~Calders, N.~Origo, M.~Disney, A.~Burt, P.~Raumonen, B.~Gardiner, M.~Herold, et~al.
\newblock Finite element analysis of trees in the wind based on terrestrial laser scanning data.
\newblock \emph{Agricultural and Forest Meteorology}, 265:\penalty0 137--144, 2019.

\bibitem[James et~al.(2013)James, Witten, Hastie, Tibshirani, et~al.]{james_introduction_2013}
G.~James, D.~Witten, T.~Hastie, R.~Tibshirani, et~al.
\newblock \emph{An introduction to statistical learning: with applications in {R}}, volume 103.
\newblock Springer, 2013.

\bibitem[Kersting et~al.(2024)Kersting, K{\"u}hn, and Fischer]{kersting_measuring3D_2024}
S.~J. Kersting, A.~L. K{\"u}hn, and M.~Fischer.
\newblock Measuring {3D} tree imbalance of plant models using graph-theoretical approaches.
\newblock \emph{Ecological Informatics}, 80:\penalty0 102438, 2024.

\bibitem[Kirkpatrick and Slatkin(1993)]{kirkpatrick_searching_1993}
M.~Kirkpatrick and M.~Slatkin.
\newblock Searching for evolutionary patterns in the shape of a phylogenetic tree.
\newblock \emph{Evolution}, 47\penalty0 (4):\penalty0 1171--1181, 1993.
\newblock \doi{10.1111/j.1558-5646.1993.tb02144.x}.

\bibitem[Knuth(1998)]{knuth_volume3_1998}
D.~E. Knuth.
\newblock \emph{The art of computer programming volume 3: Sorting and searching}.
\newblock Addison-Wesley Professional, 2nd edition, 1998.
\newblock ISBN 0201896850.

\bibitem[Konopka(2023)]{konopka_umap_2023}
T.~Konopka.
\newblock \emph{umap: {U}niform {M}anifold {A}pproximation and {P}rojection}, 2023.
\newblock URL \url{https://CRAN.R-project.org/package=umap}.
\newblock R package version 0.2.10.0.

\bibitem[Kuhn(2008)]{kuhn_building_2008}
M.~Kuhn.
\newblock {B}uilding {P}redictive {M}odels in {R} {U}sing the caret {P}ackage.
\newblock \emph{Journal of Statistical Software}, 28\penalty0 (5):\penalty0 1–26, 2008.
\newblock \doi{10.18637/jss.v028.i05}.
\newblock URL \url{https://www.jstatsoft.org/index.php/jss/article/view/v028i05}.

\bibitem[Kunz et~al.(2019)Kunz, Fichtner, H{\"a}rdtle, Raumonen, Bruelheide, and Oheimb]{kunz_neighbour_2019}
M.~Kunz, A.~Fichtner, W.~H{\"a}rdtle, P.~Raumonen, H.~Bruelheide, and G.~Oheimb.
\newblock Neighbour species richness and local structural variability modulate aboveground allocation patterns and crown morphology of individual trees.
\newblock \emph{Ecology Letters}, 22\penalty0 (12):\penalty0 2130--2140, 2019.
\newblock ISSN 1461-023X, 1461-0248.
\newblock \doi{10.1111/ele.13400}.

\bibitem[Lau et~al.(2018)Lau, Bentley, Martius, Shenkin, Bartholomeus, Raumonen, Malhi, Jackson, and Herold]{lau_quantifying_2018}
A.~Lau, L.~P. Bentley, C.~Martius, A.~Shenkin, H.~Bartholomeus, P.~Raumonen, Y.~Malhi, T.~Jackson, and M.~Herold.
\newblock Quantifying branch architecture of tropical trees using terrestrial {LiDAR} and {3D} modelling.
\newblock \emph{Trees}, 32\penalty0 (5):\penalty0 1219--1231, 2018.

\bibitem[Liaw and Wiener(2002)]{liaw_classification_2002}
A.~Liaw and M.~Wiener.
\newblock Classification and regression by random{F}orest.
\newblock \emph{R News}, 2\penalty0 (3):\penalty0 18--22, 2002.
\newblock URL \url{https://CRAN.R-project.org/doc/Rnews/}.

\bibitem[McInnes et~al.(2018)McInnes, Healy, and Melville]{mcinnes_umap_2018}
L.~McInnes, J.~Healy, and J.~Melville.
\newblock Umap: Uniform manifold approximation and projection for dimension reduction.
\newblock \emph{arXiv e-prints}, art. arXiv:1802.03426, 2018.

\bibitem[McKenzie and Steel(2000)]{mckenzie_distributions_2000}
A.~McKenzie and M.~Steel.
\newblock Distributions of cherries for two models of trees.
\newblock \emph{Mathematical Biosciences}, 164\penalty0 (1):\penalty0 81--92, 2000.
\newblock \doi{10.1016/s0025-5564(99)00060-7}.

\bibitem[Mir et~al.(2013)Mir, Rossell{\'{o}}, and Rotger]{mir_new_2013}
A.~Mir, F.~Rossell{\'{o}}, and L.~Rotger.
\newblock A new balance index for phylogenetic trees.
\newblock \emph{Mathematical Biosciences}, 241\penalty0 (1):\penalty0 125--136, 2013.
\newblock \doi{10.1016/j.mbs.2012.10.005}.

\bibitem[Mir et~al.(2018)Mir, Rotger, and Rossell{\'{o}}]{mir_sound_2018}
A.~Mir, L.~Rotger, and F.~Rossell{\'{o}}.
\newblock Sound {C}olless-like balance indices for multifurcating trees.
\newblock \emph{{PLOS} {ONE}}, 13\penalty0 (9):\penalty0 e0203401, 2018.
\newblock \doi{10.1371/journal.pone.0203401}.

\bibitem[Purvis et~al.(2002)Purvis, Katzourakis, and Agapow]{purvis_evaluating_2002}
A.~Purvis, A.~Katzourakis, and P.-M. Agapow.
\newblock Evaluating phylogenetic tree shape: two modifications to {F}usco \& {C}ronk's method.
\newblock \emph{Journal of Theoretical Biology}, 214\penalty0 (1):\penalty0 99--103, 2002.
\newblock \doi{10.1006/jtbi.2001.2443}.

\bibitem[{R Core Team}(2025)]{RCoreTeam2025}
{R Core Team}.
\newblock \emph{{R}: a language and environment for statistical computing}.
\newblock R Foundation for Statistical Computing, Vienna, Austria, 2025.
\newblock URL \url{https://www.R-project.org/}.

\bibitem[Raumonen et~al.(2013)Raumonen, Kaasalainen, {\AA}kerblom, Kaasalainen, Kaartinen, Vastaranta, Holopainen, Disney, and Lewis]{raumonen_fast_2013}
P.~Raumonen, M.~Kaasalainen, M.~{\AA}kerblom, S.~Kaasalainen, H.~Kaartinen, M.~Vastaranta, M.~Holopainen, M.~Disney, and P.~Lewis.
\newblock Fast automatic precision tree models from terrestrial laser scanner data.
\newblock \emph{Remote Sensing}, 5\penalty0 (2):\penalty0 491--520, 2013.
\newblock ISSN 2072-4292.
\newblock \doi{10.3390/rs5020491}.

\bibitem[Rid et~al.(2016)Rid, Mesca, Ayasse, and Gross]{rid_apple_2016}
M.~Rid, C.~Mesca, M.~Ayasse, and J.~Gross.
\newblock Apple proliferation phytoplasma influences the pattern of plant volatiles emitted depending on pathogen virulence.
\newblock \emph{Frontiers in Ecology and Evolution}, 3:\penalty0 152, 2016.

\bibitem[Ridgeway and {GBM Developers}(2026)]{ridgeway_gbm_2026}
G.~Ridgeway and {GBM Developers}.
\newblock \emph{gbm: {G}eneralized {B}oosted {R}egression {M}odels}, 2026.
\newblock URL \url{https://CRAN.R-project.org/package=gbm}.
\newblock R package version 2.2.3.

\bibitem[Sackin(1972)]{sackin_good_1972}
M.~J. Sackin.
\newblock ``{G}ood'' and ``bad'' phenograms.
\newblock \emph{Systematic Biology}, 21\penalty0 (2):\penalty0 225--226, 1972.
\newblock \doi{10.1093/sysbio/21.2.225}.

\bibitem[Schelhaas et~al.(2007)Schelhaas, Kramer, Peltola, Van~der Werf, and Wijdeven]{schelhaas_introducing_2007}
M.~J. Schelhaas, K.~Kramer, H.~Peltola, D.~C. Van~der Werf, and S.~M.~J. Wijdeven.
\newblock Introducing tree interactions in wind damage simulation.
\newblock \emph{Ecological Modelling}, 207\penalty0 (2-4):\penalty0 197--209, 2007.

\bibitem[Shao and Sokal(1990)]{shao_tree_1990}
K.-T. Shao and R.~R. Sokal.
\newblock Tree balance.
\newblock \emph{Systematic Zoology}, 39\penalty0 (3):\penalty0 266, 1990.
\newblock \doi{10.2307/2992186}.

\bibitem[Takacs(1992)]{takacs_total_1992}
L.~Takacs.
\newblock On the total heights of random rooted trees.
\newblock \emph{Journal of Applied Probability}, 29\penalty0 (3):\penalty0 543–556, 1992.
\newblock \doi{10.2307/3214892}.

\bibitem[Takacs(1994)]{takacs_total_1994}
L.~Takacs.
\newblock On the total heights of random rooted binary trees.
\newblock \emph{Journal of Combinatorial Theory, Series B}, 61\penalty0 (2):\penalty0 155--166, 1994.
\newblock ISSN 0095-8956.
\newblock \doi{10.1006/jctb.1994.1041}.

\end{thebibliography}

\newpage
{
\appendix
\renewcommand{\thesection}{\Alph{section}}
\section{Software: \textsf{R} package \textsf{treeDbalance} (Version 1.2.0)} \label{sec:software}

Our software package \textsf{treeDbalance -- Computation of 3D Tree Imbalance} written in the free and openly available programming language \textsf{R} \citep{RCoreTeam2025} and publicly available on CRAN (see \url{https://CRAN.R-project.org/package=treeDbalance}) has been updated and extended: Version 1.2.0$^+$ now provides the means to transform QSMs into rooted 3D trees and to extract their non-3D topology. The following commands in the gray box show an example of the execution of these two procedures (the package contains an \texttt{exampleQSM.mat} file) while also giving some information on how to deal with the resulting objects.

\begin{tcolorbox}[colback=verylightgray, bottom=0.05pt, top=0.05pt, colframe=verylightgray, frame empty]
\begin{lstlisting}
library("treeDbalance")
# The rooted 3D tree -------------------------------------------------------------
r3Dtree <- qsm2phylo3D(file = "PATH-TO-FILE/exampleQSM.mat", 
      version = "2.4.x", setConnection2zero = TRUE
    )
# Can be visualized with
plot(r3Dtree); plotPhylo3D(r3Dtree)
# Example computation of a 3D imbalance index:
A_Index(r3Dtree)
# Its extracted non-3D topology  ------------------------------------------------- 
treetop <- extractTopology(r3Dtree)
# Can be visualized with
ape::plot.phylo(treetop,
      type = "c", direction = "upwards", use.edge.length = FALSE
    )
# Example computation of a topological index:
treebalance::B2I(treetop)
\end{lstlisting}
\end{tcolorbox}
}

\begin{figure}[htbp]
	\centering 
    \includegraphics[width=0.8\textwidth]{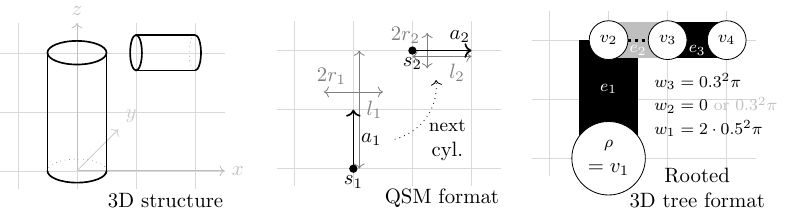} 
	\caption{Sketch of the main components of the QSM and rooted 3D format to capture the exemplary 3D structure.}
	\label{fig:ex_transformation}
\end{figure}

Since pairs of consecutive cylinders in QSMs typically do not share a mutual end/start position (see, for example, Figure~\ref{fig:ex_transformation} on the left), connection edges are introduced in the transformation to a rooted 3D tree. For this, the parameter \texttt{setConnection2zero} allows the user to decide if the connections of such pairs of cylinders should have a \underline{width $>0$ or not} (see the gray edge in the right part of Figure~\ref{fig:ex_transformation}). Specifying this is important since consecutive cylinders in QSMs often do not form smooth branch lines -- this issue of cylinder offsets is discussed in Section~\ref{sec:results_quality} (see also Figure~\ref{fig:CylinderOffsets}) --, which would increase internal 3D imbalance if these connection edges had widths $>0$, creating noise which might distort the information of the internal 3D imbalance of the cylinder-supported parts of the rooted 3D tree model. The following paragraphs also discuss this issue using the example shown in Figure~\ref{fig:ex_transformation}.

\paragraph{Transforming a QSM into a rooted 3D model in \texttt{phylo3D} format}

This paragraph explains the basic idea of the transformation from QSM to rooted 3D tree (function \texttt{qsm2phylo3D()}). A QSM (usually given as a Matlab-file) contains among others the following information: For each cylinder its radius, its length, its start coordinates, its axis (direction vector of length 1), its parent cylinder, and its child/extension cylinder.

A rooted 3D tree holds similar information: For every edge the start and end vertex, its length, and its radius, and for every vertex its coordinates. The basic idea of the two formats and the transformation can be best described using an example (see Figure~\ref{fig:ex_transformation}): We consider a simple 3D model that consists of two cylinders, $c_1$ from (0,0,0) to (0,0,2) with radius 0.5 and $c_2$ from (1,0,2) to (2,0,2) with radius 0.3. As with most QSMs these cylinders do not share an start/end point, which allows us to also address the above mentioned decision which width to give the connection edges.

In the QSM format, this information would be given as follows: radii $r = (0.5,0.3)$, lengths $l = (2,1)$, start coordinates $s = ((0,0,0),(1,0,2))$, axes $a = ((0,0,1),(1,0,0))$, parents $p = (0,1)$ (where 0 is no parent), and extensions/children $c = (2,0)$ (where 0 is no extension). From here, the end point of cylinders $c_i$ can be calculated with $s_i+l_i\cdot a_i$.

The rooted 3D tree format on the other hand contains three edges $e = ((1,2),(2,3),(3,4))$, where the first and the last correspond to $c_1$ and $c_2$ and the middle is a connection edge that connects the end of $c_1$ with the start of $c_2$, with their radii $r^*=(1,\ \underline{0 \text{ or } 0.5},\ 0.5)$ (as specified by the user in \texttt{setConnection2zero}) and lengths $l^*=(2,1,1)$, as well as the vertex coordinates $((0,0,0),(0,0,2),(1,0,2),(2,0,2))$. 

Please note that this notation with the ordered lists is chosen here for a better comparison of the two formats and since it is closer to the implemented version of this format, \texttt{phylo3D}, in the \textsf{R}-package. The established graph-theoretical notation \cite{kersting_measuring3D_2024} of this rooted 3D tree is as follows: $\mathsf{T}=(T,w)$ with the topology $T=(V,E)$, where $V=\{\rho=v_1=(0,0,0),v_2=(0,0,2),v_3=(1,0,2),v_4=(2,0,2)\}$ and $E=\{e_1=(v_1,v_2),e_2=(v_2,v_3),e_3=(v_3,v_4)\}$ and the weight function $w$ with $w_i\coloneqq w(e_i)=l^*_i \cdot \pi \cdot r^{*2}_i$ for $i=1,2,3$ gives the volume of the edge by default. Following the strict definition, the weights $w(e_i)$ would have to be positive. However, with regards to the structure of the QSMs, we decided for connecting edges having \underline{width $0$} (and with that weight $0$) in order to not influence the tree volume estimations and to not have the cylinder offsets (see Figure~\ref{fig:CylinderOffsets}) distort the internal 3D imbalance.

This should convey the basic idea of the transformation, some technical details like creating the correct connection edges for the starting cylinders of consecutive cylinder lines in the QSM format are a bit more involved (for such details we refer the reader to the annotated source code).

\paragraph{Extracting the non-3D toplogy of a rooted 3D model}

This paragraph provides explanations how the non-3D topology $T^*$ is extracted (function \texttt{extractTopology()}). Let $\mathsf{T}=(T,w)$ be a rooted 3D tree with a 3D topology $T=(V,E)$, root $\rho\in V$, and the weight function $w$. Since the edge lengths can be calculated from the coordinates of the start and end vertex of each edge (and they are typically available in the \texttt{phylo3D} format), we consider them to be given as well. Although the non-3D indices considered in this study do not take edge lengths into account, \texttt{extractTopology()} also extracts edge lengths to enable future investigations of this aspect. 

\begin{figure}[htbp]
	\centering 
    \includegraphics[width=0.8\textwidth]{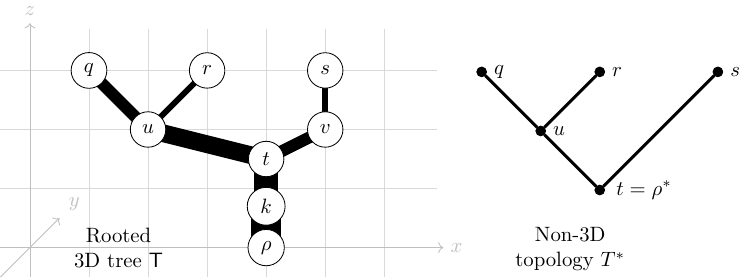} 
	\caption{Example of a rooted 3D tree and its extracted non-3D topology (as used in this study without edge length information). The old root $\rho$ and $k$ along with their edges up to the new root $\rho^*=t$ are deleted and $v$ is suppressed.}
	\label{fig:ex_extraction}
\end{figure}

\begin{figure}[ht]
\centering
\begin{subfigure}[t]{0.4\textwidth}
\centering
\includegraphics[width=1\textwidth]{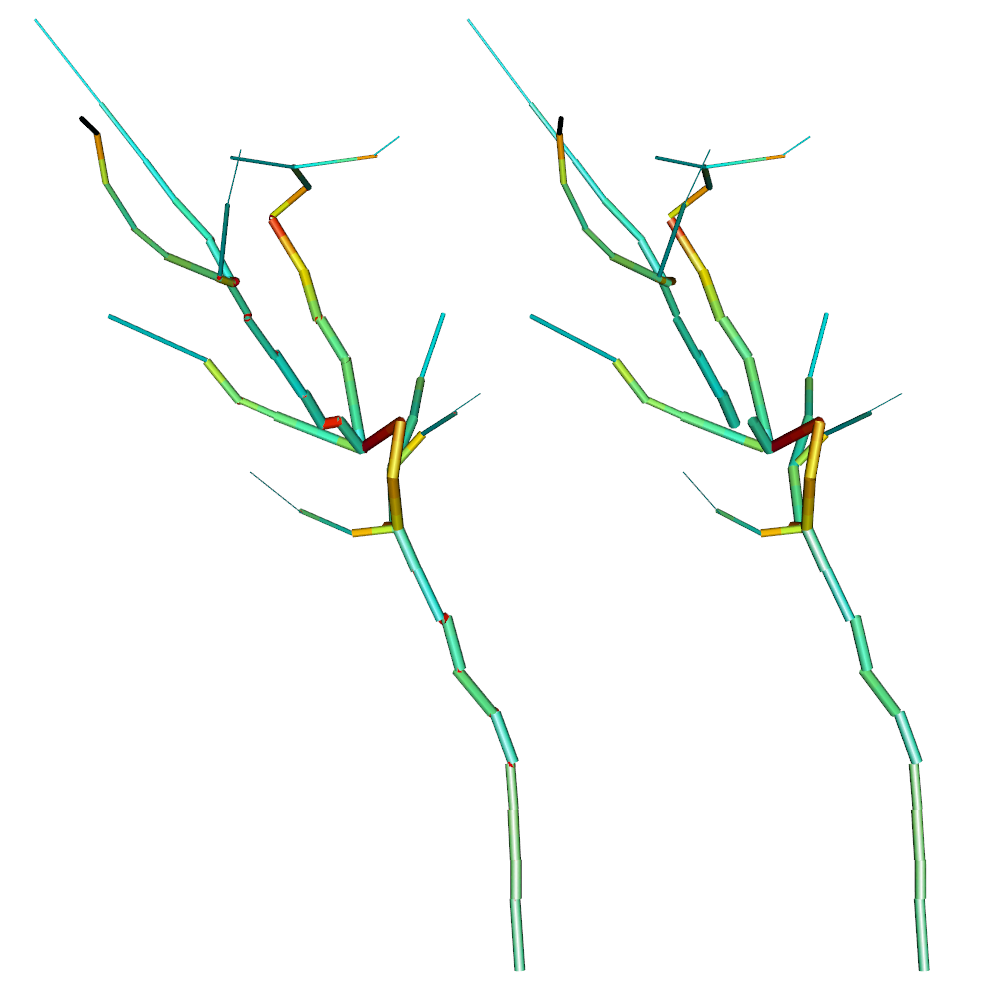}
\caption{}
\end{subfigure}
\begin{subfigure}[t]{0.55\textwidth}
\centering
\includegraphics[width=1\textwidth]{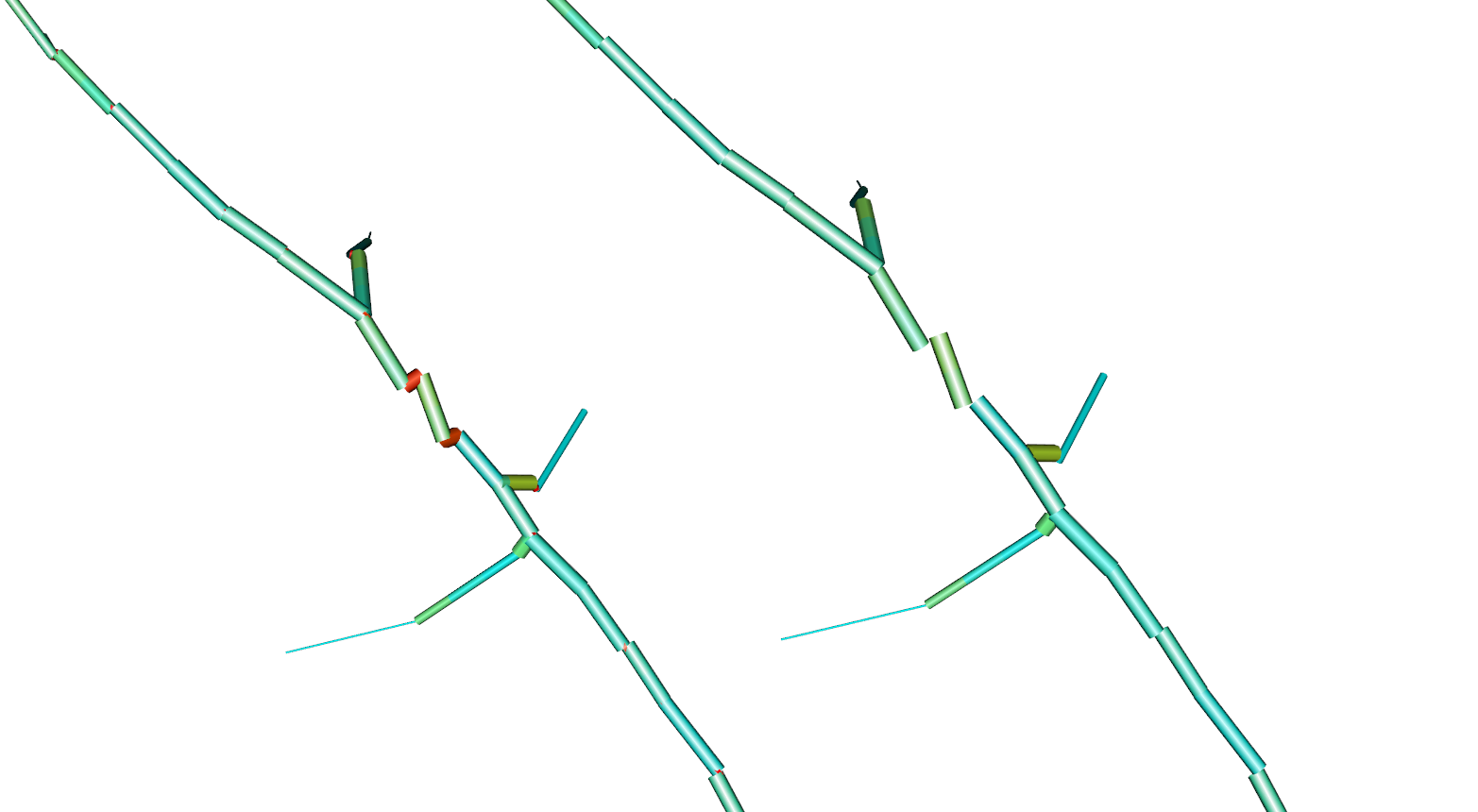}
\caption{}
\end{subfigure} \vspace{-0.1cm}
\caption{Cylinder offsets, i.e., the non-matching endpoints of cylinders that form a branch in a QSM, exemplified by QSM 180b$^7$ (a) and a section of 8145c$^0$ (b). On the left sides each with width $>0$ for the connecting cylinders (parameter \texttt{setConnection2zero} set to \texttt{FALSE}) and on the right sides the respective tree parts with zero-width connecting cylinders (\texttt{setConnection2zero = TRUE}). The QSMs are colored according to their internal $\mathcal{A}$ imbalance (see Section~\ref{sec:statistics3DT}) which highlights how this parameter choice can affect the total 3D imbalance of a tree as all these small highly-imbalanced red cylinders increase the average imbalance over all cylinders when edge weights/volume is used as the weighting method.\\
Furthermore, 180b$^7$ (a) displays the faulty branch reconstruction in higher detail, where the cylinders form a triangular peak instead of building the branch relatively straight as the other QSMs of 180b do (see Figure~\ref{fig:workflow}). }
\label{fig:CylinderOffsets}
\end{figure}

Now, regarding the extraction process (see Figure~\ref{fig:ex_extraction}):

The aim is to drop all 3D information (the coordinates of the vertices and the weights (and radii) of the edges) and to keep only multifurcating nodes, i.e., nodes that have multiple child nodes. In the first step the new root $\rho^*$ is identified as the vertex closest to the old root $\rho$ that is multifurcating. All vertices from the old root up to the new root and the edges between them are deleted. With that any \enquote{stem} is removed. Next, on each path from $\rho^*$ to a leaf all vertices with one ancestor and one child are suppressed, i.e., such a vertex and its two incident edges are deleted, and instead a new edge is created from its parent to its child node. The new edge gets a length equal to the sum of the two deleted edges. The remaining vertices and edges form the non-3D topology $T^*$ (see Figure~\ref{fig:ex_extraction}).

\newpage
\section{Further figures and tables} \label{sec:tablesAndfigures}

\small
\begin{center}
\begin{longtable}[t]{lp{4cm}c@{\hskip 1pt}c@{\hskip 1pt}c@{\hskip 1pt}c@{\hskip 1pt}c}
\caption{\normalsize List of all statistics used in this study. The first column contains the respective abbreviated names used throughout this manuscript. If a statistic was removed or replaced, the replacement is indicated with a $\rightarrow$ and the row is shown with a gray background. The second the corresponding column name in the published datasets (\texttt{tls\_summary.csv} and \texttt{trees\_summary.csv} of \cite{calders_laser_2022}, and \texttt{df\_ww\_all.csv}, \texttt{df\_ww\_prep.csv}, etc.). The third column clarifies if it is an existing feature in the original datasets of \cite{calders_laser_2022} and the fourth if its values are specific for each QSM (otherwise all QSMs have the same value). Columns five, six, and seven clarify whether the statistic was used for the corresponding research question/methods. Note that all statistics that are not grayed out or marked with - would have been suitable numeric statistics for 2a). Notation: Y = Yes, empty field = No, - = not applicable. Descriptions of the statistics can be found in Section~\ref{sec:statistics} in the main part of the manuscript.}
\label{tab:statisticsColumns}\\

\hline
\addlinespace
Abbreviated name & \parbox[t]{4cm}{Column name in \\ the datasets}  & Original & 
\parbox[t]{1.5cm}{\centering QSM\\specific} & 
\parbox[t]{1.5cm}{\centering1)\\QSM\\Quality} & 
\parbox[t]{1.8cm}{\centering2a) Species\\Class.\\($<0.9$ corr.)} & 
\parbox[t]{1.8cm}{\centering2b) Species\\Class.\\\enquote{original}}\\
\addlinespace
\hline
\endfirsthead

\hline
\addlinespace
Abbreviated name & \parbox[t]{4cm}{Column name in \\ the datasets} & Original & 
\parbox[t]{1.5cm}{\centering QSM\\specific} & 
\parbox[t]{1.5cm}{\centering1)\\QSM\\Quality} & 
\parbox[t]{1.8cm}{\centering2a) Species\\Class.\\($<0.9$ corr.)} & 
\parbox[t]{1.8cm}{\centering2b) Species\\Class.\\\enquote{original}}\\
\addlinespace
\hline
\endhead

\hline\noalign{\vskip 2pt}
\multicolumn{7}{r}{\small Continued on next page.} \\
\hline\noalign{\vskip 1pt}
\endfoot

\hline
\endlastfoot

\addlinespace
Tree ID & \texttt{Tree\_ID}, \texttt{TLS\_ID} & Y &  & - & - & - \\
QSM iteration & \texttt{QSM\_iteration} & Y & Y & - & - & - \\
Species & \texttt{species} & Y &  & - & - & - \\
 & \texttt{latin\_name} & Y &  & - & - & - \\
 & \texttt{common\_name} & Y &  & - & - & - \\
StemCount & \texttt{stem\_count} &  &  & - & Y &  \\
\rowcolor{gray!30}  ID type & \texttt{ID\_type} &  &  & - & - & - \\
Dead & \texttt{Dead} & Y &  & - & - & - \\
Locx & \texttt{stemlocx\_m} & Y &  &  & Y & Y \\
Locy & \texttt{stemlocy\_m} & Y &  &  & Y & Y \\
\grayhline
DBH & \texttt{DBH\_QSM\_m} &  & Y & Y & Y & Y \\
\rowcolor{gray!30}  $\rightarrow$ DBH & \texttt{DBH\_QSM\_avg\_m} & Y &  &  &  &  \\
\rowcolor{gray!30}  $\rightarrow$ DBH & \texttt{DBH\_pts\_m} & Y &  &   &   &   \\
\rowcolor{gray!30}  $\rightarrow$ DBH & \texttt{DBH\_TLS\_m} & Y &  &   &   &   \\
\rowcolor{gray!30}  DBHc  $\rightarrow$ DBH  & \texttt{DBH\_census\_m} & Y &  &   &   &   \\
DBHsd & \texttt{DBH\_QSM\_sd\_m} & Y &  &  & - & - \\
CrownArea & \texttt{VerticalCrownProjected} \phantom{----}\texttt{Area\_pts\_m2} & Y &  &  &  & Y \\
Height & \texttt{height\_m} &  & Y & Y & Y & Y \\
\rowcolor{gray!30}  $\rightarrow$ Height & \texttt{Hgt\_pts\_m} & Y &  &   &   &   \\
Length & \texttt{tot\_length\_m} &  & Y & Y &  &   \\
ZeroCyl & \texttt{zeroWeightCyl\_count} &  & Y & Y & - & - \\
\grayhline
Volume & \texttt{tot\_volume\_m3} &  & Y & Y &  & Y \\
\rowcolor{gray!30}  $\rightarrow$ Volume & \texttt{Vol\_QSM\_avg\_m3} & Y &  &   &   &   \\
Vol0-25 & \texttt{Vol\_QSM\_D0\_25mm\_m3} &  & Y & Y &  & Y \\
Vol25-50 & \texttt{Vol\_QSM\_D25\_50mm\_m3} &  & Y & Y &  & Y \\
Vol50-75 & \texttt{Vol\_QSM\_D50\_75mm\_m3} &  & Y & Y &  & Y \\
Vol75-100 & \texttt{Vol\_QSM\_D75\_100mm\_m3} &  & Y & Y &  & Y \\
Vol100-200 & \texttt{Vol\_QSM\_D100\_200mm\_m3} &  & Y & Y & Y & Y \\
Vol200+ & \texttt{Vol\_QSM\_D200mm\_m3} &  & Y & Y &  & Y \\
\rowcolor{gray!30}  $\rightarrow$ Vol0-25 & \texttt{Vol\_QSM\_D0\_25mm\_avg\_m3} & Y &  &  &   &   \\
\rowcolor{gray!30}  $\rightarrow$ Vol25-50 & \texttt{Vol\_QSM\_D25\_50mm\_avg\_m3} & Y &  &  &   &   \\
\rowcolor{gray!30}  $\rightarrow$ Vol50-75 & \texttt{Vol\_QSM\_D50\_75mm\_avg\_m3} & Y &  &  &   &   \\
\rowcolor{gray!30}  $\rightarrow$ Vol75-100 & \texttt{Vol\_QSM\_D75\_100mm\_avg\_m3} & Y &  &  &   &   \\
\rowcolor{gray!30}  $\rightarrow$ Vol100-200 & \texttt{Vol\_QSM\_D100\_200mm\_avg\_m3} & Y &  &  &  &  \\
\rowcolor{gray!30}  $\rightarrow$ Vol200+ & \texttt{Vol\_QSM\_D200\_500mm\_avg\_m3} & Y &  &   &   &   \\
\rowcolor{gray!30}  $\rightarrow$ Vol200+ & \texttt{Vol\_QSM\_D500\_1000mm\_avg\_m3} & Y &  &   &   &   \\
\rowcolor{gray!30}  $\rightarrow$  Vol200+ & \texttt{Vol\_QSM\_D1000mm\_avg\_m3} & Y &  &   &   &   \\
Volsd & \texttt{Vol\_QSM\_sd\_m3} & Y &  &  & - & - \\
Vol0-25sd & \texttt{Vol\_QSM\_D0\_25mm\_sd\_m3} & Y &  &  & - & - \\
Vol25-50sd & \texttt{Vol\_QSM\_D25\_50mm\_sd\_m3} & Y &  &  & - & - \\
Vol50-75sd & \texttt{Vol\_QSM\_D50\_75mm\_sd\_m3} & Y &  &  & - & - \\
Vol75-100sd & \texttt{Vol\_QSM\_D75\_100mm\_sd\_m3} & Y &  &  & - & - \\
Vol100-200sd & \texttt{Vol\_QSM\_D100\_200mm\_sd\_m3} & Y &  &  & - & - \\
Vol200-500sd & \texttt{Vol\_QSM\_D200\_500mm\_sd\_m3} & Y &  &  & - & - \\
\rowcolor{gray!30}   & \texttt{Vol\_QSM\_D500\_1000mm\_sd\_m3} & Y &  &   &   &   \\
\rowcolor{gray!30}   & \texttt{Vol\_QSM\_D1000mm\_sd\_m3} & Y &  &   &   &   \\
\grayhline
Int-w(A) & \texttt{A\_w} &  & Y & Y &  &  \\
Int-w(a) & \texttt{alpha\_w} &  & Y & Y & Y &   \\
Int-w(M) & \texttt{M\_w} &  & Y & Y &  &   \\
Int-w(m) & \texttt{mu\_w} &  & Y & Y &  &   \\
Int-l(A) & \texttt{A\_l} &  & Y & Y &  &   \\
Int-l(a) & \texttt{alpha\_l} &  & Y & Y &  &   \\
Int-l(M) & \texttt{M\_l} &  & Y & Y &  &   \\
Int-l(m) & \texttt{mu\_l} &  & Y & Y & Y &  \\
\grayhline
Ext(A) & \texttt{root\_A} &  & Y & Y &  &  \\
Ext(a) & \texttt{root\_alpha} &  & Y & Y & Y &   \\
Ext(M) & \texttt{root\_M} &  & Y & Y &  &   \\
Ext(m) & \texttt{root\_mu} &  & Y & Y &  &   \\
\grayhline
LeafN & \texttt{n\_leaves} &  & Y & Y &  &   \\
InnerN & \texttt{n\_innerN} &  & Y & Y &  &   \\
TopRes & \texttt{resolution\_n} &  & Y & Y & Y &  \\
\rowcolor{gray!30} & \texttt{is\_binary} &  & Y & - & - & -  \\
\grayhline
B1 & \texttt{B1I} &  & Y & Y &  &   \\
B2 & \texttt{B2I} &  & Y & Y & Y &  \\
Cherry & \texttt{CherryI} &  & Y & Y &  &   \\
\rowcolor{gray!30}  $\rightarrow$ sqrt-CLe & \texttt{Coll\_likeI\_e} &  & Y &   &   &   \\
sqrt-CLe & \texttt{Coll\_likeI\_e\_sqrt} &  & Y & Y & Y &  \\
CLln & \texttt{Coll\_likeI\_ln} &  & Y & Y &  &   \\
mD & \texttt{maxDepth} &  & Y & Y &  &   \\
mW & \texttt{maxWidth} &  & Y & Y &  &   \\
mDW & \texttt{modMaxDiffW} &  & Y & Y &  &  \\
mI' & \texttt{IbasedI\_meanP} &  & Y & Y &  &  \\
mIw & \texttt{IbasedI\_meanW} &  & Y & Y & Y &   \\
\rowcolor{gray!30} $\rightarrow$ 4thrt-rQi & \texttt{rQuartetI} &  & Y &   &  &   \\
4thrt-rQi & \texttt{rQuartetI\_4rt} &  & Y & Y &  &   \\
Sackin & \texttt{SackinI} &  & Y & Y &  &   \\
s-shape & \texttt{sshapeI} &  & Y & Y &  &   \\
TPL & \texttt{totPathL} &  & Y & Y &  &   \\
TC & \texttt{totalCophI} &  & Y & Y & Y &  \\
VLD & \texttt{VarLDI} &  & Y & Y & Y &  \\
ALD & \texttt{AvgLDI} &  & Y & Y &  &   \\
AVD & \texttt{AvgVertD} &  & Y & Y & Y &  \\
TIPL & \texttt{totIntPathL} &  & Y & Y &  &   \\
mWomD & \texttt{maxWoMaxD} &  & Y & Y & Y &   \\
\end{longtable}
\end{center}

\begin{figure}[ht]
\centering
\includegraphics[width=0.98\textwidth]{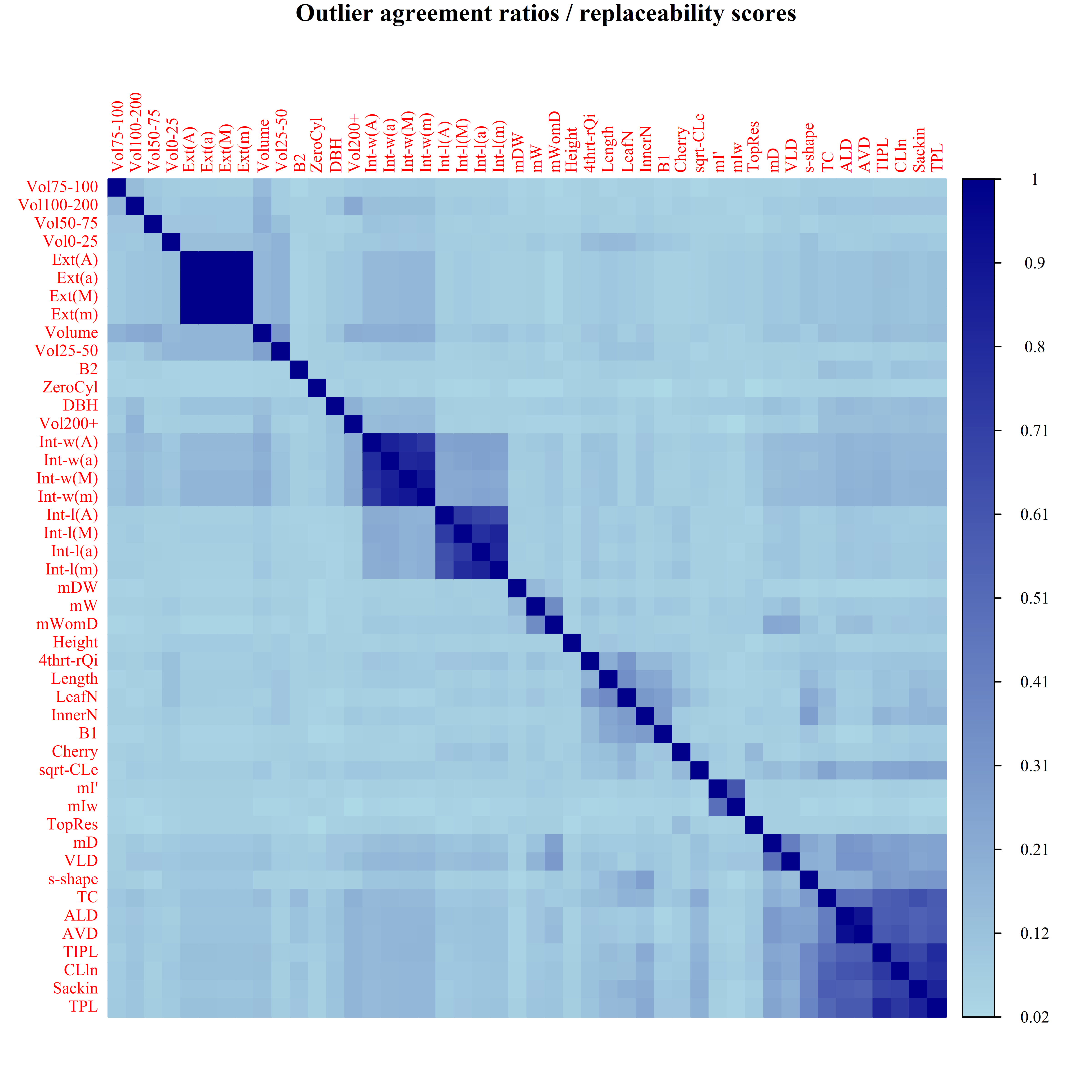}
\caption{The pairwise outlier agreement ratios (or replaceability scores) of all QSM-specific statistics. Such an agreement ratio $a(i,j)$ of statistics $i$ and $j$ is the number of times $j$ agrees with $i$ that a QSM is an outlier divided by the number of times $i$ flags a QSM as outlier. The closer $a(i,j)$ is to 1, the better $i$ can be replaced by $j$ as $j$ detects $i$'s outliers just as well. The row $i$ shows how well $i$ can be replaced by the other statistics and, thus, this matrix is not symmetric. The minimal replaceability score of 0.018 has mIw with Vol200+ -- in other words, Vol200+ catches only around 2$\%$ of mIw's detected outliers. The statistics were reordered using hierarchical clustering with complete linkage, grouping features with similar score patterns together.}
\label{fig:agreementRatios}
\end{figure}

\begin{figure}[ht]
\centering
\includegraphics[width=0.98\textwidth]{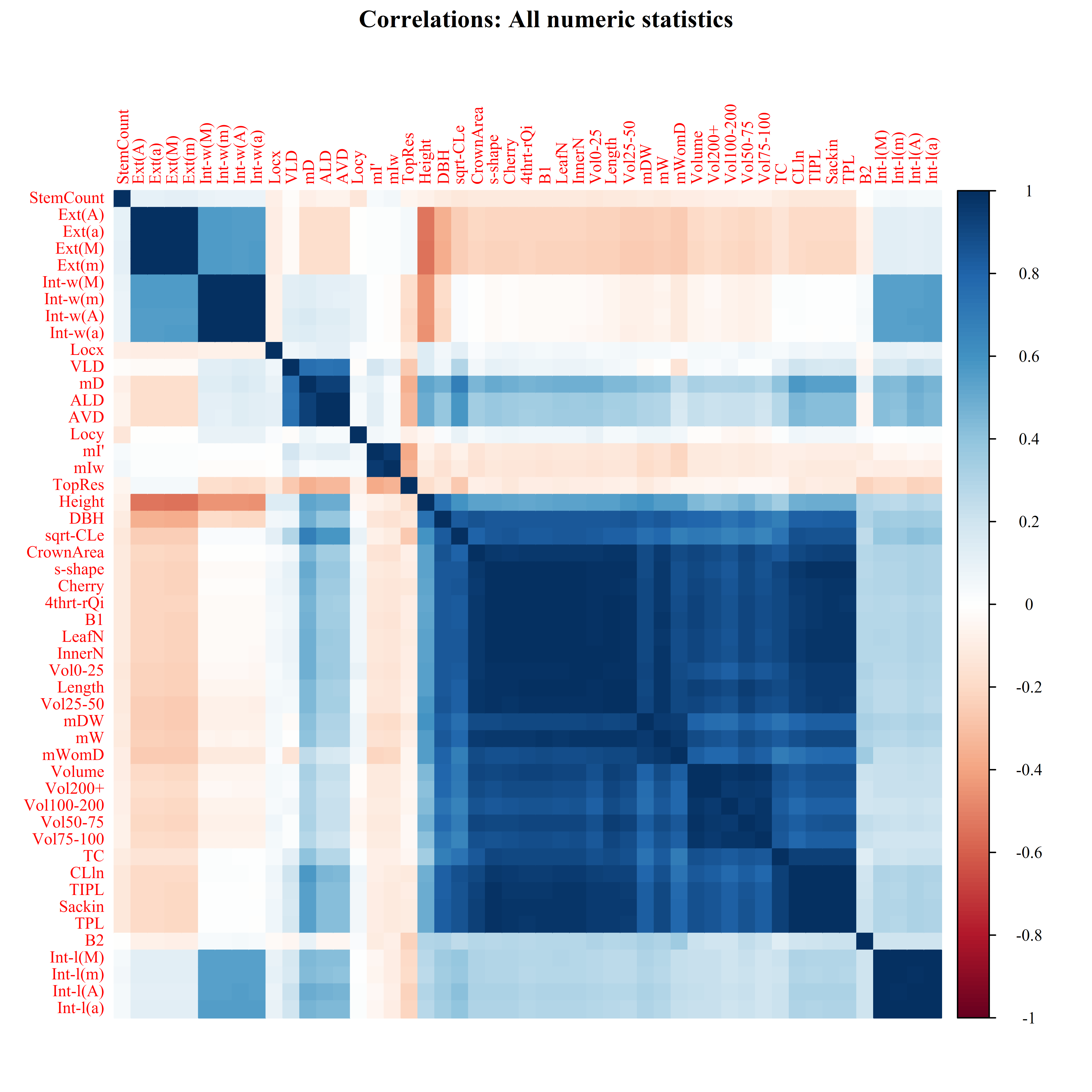}
\caption{Correlations of all statistics based on the Pearson correlation coefficient. The statistics in the correlation plot were reordered using hierarchical clustering with complete linkage, grouping features with similar correlation patterns together.}
\label{fig:correlations_all}
\end{figure}

\begin{figure}[ht]
\centering
\includegraphics[width=0.78\textwidth]{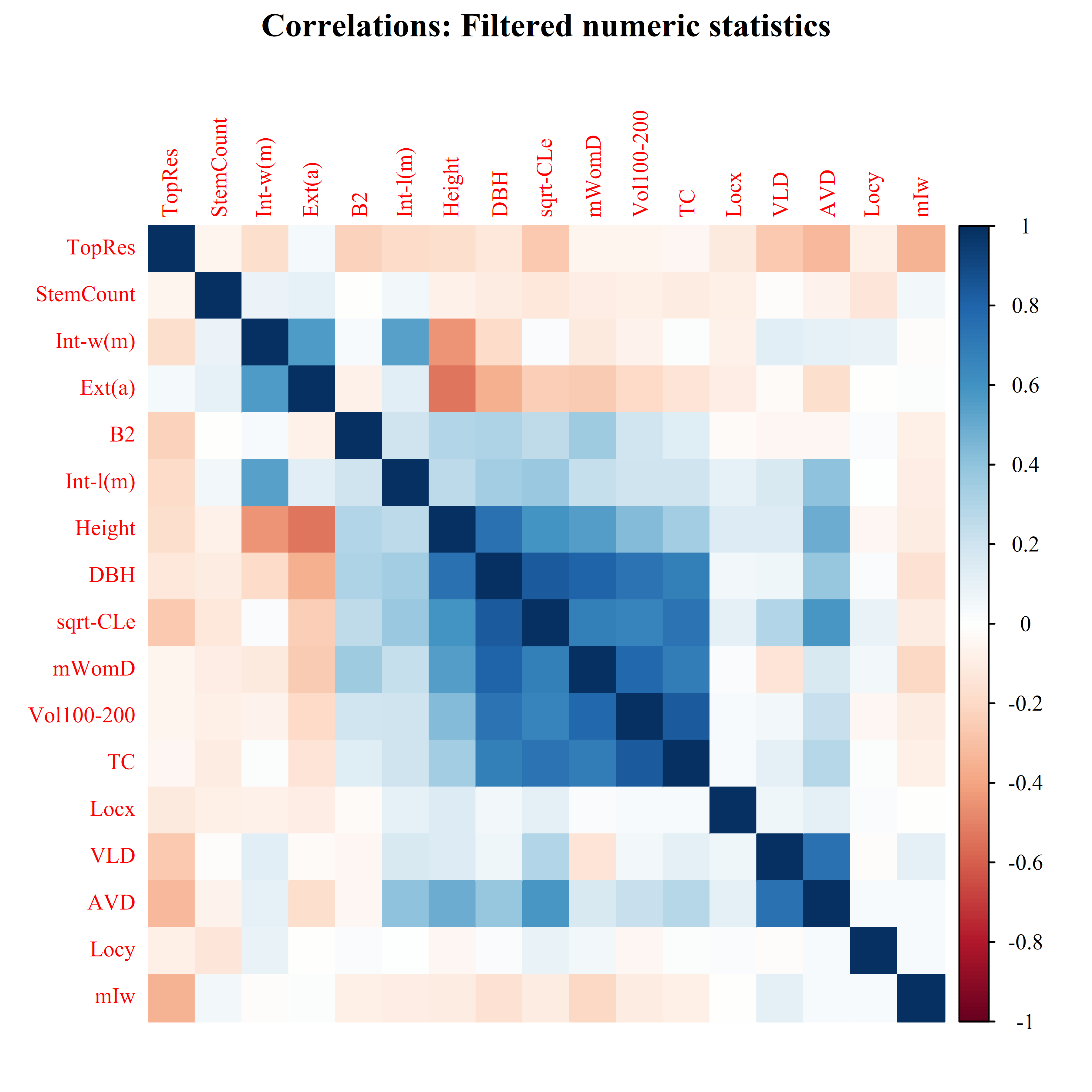}
\caption{Correlations based on the Pearson correlation coefficient of the filtered predictor set used in the species classification. The statistics in the correlation plot were reordered using hierarchical clustering with complete linkage, grouping features with similar correlation patterns together. All pairwise correlations are below the threshold of $<0.9$ with TC and Vol100-200 having the maximal absolute correlation of $\approx 0.83$.}
\label{fig:correlations_allfiltered}
\end{figure}

\begin{table}[htbp!]
    \centering
    \caption{Confusion matrices for the random forest model i) based on all 49 available numeric statistics $P_{all}$ showing (a) absolute classification counts and (b) row-normalized percentages (per true species). Overall performance: Accuracy: 80.5\%, log-loss: 0.52, Cohen’s $\kappa$: 0.5, macro-averaged sensitivity: 55.5\%.}
    \label{tab:confusionMat_all}
    \begin{subtable}[t]{0.495\textwidth}
\centering
\caption{}
\begin{tabular}{@{\hskip 4pt}c@{\hskip 6pt}l@{\hskip 6pt}c@{\hskip 6pt}c@{\hskip 6pt}c@{\hskip 6pt}c@{\hskip 4pt}}
 &  & \multicolumn{4}{c}{Prediction} \\
 &  & ACERPS & CORYAV & FRAXEX & Other \\
    \noalign{\vskip 2pt}\hline\noalign{\vskip 2pt}
\multirow{4}{*}{\rotatebox{90}{Reference}}
& ACERPS & \textbf{527} & 5 & 4 & 5 \\ 
& CORYAV & 19 & \textbf{42} & 0 & 6 \\ 
& FRAXEX & 73 & 7 & \textbf{3} & 2 \\ 
& Other & 12 & 14 & 1 & \textbf{38} \\ 
    \end{tabular}
\end{subtable}
    \begin{subtable}[t]{0.495\textwidth}
\centering
\caption{}
\begin{tabular}{@{\hskip 4pt}c@{\hskip 6pt}l@{\hskip 6pt}c@{\hskip 6pt}c@{\hskip 6pt}c@{\hskip 6pt}c@{\hskip 4pt}}
 &  & \multicolumn{4}{c}{Prediction} \\
 &  & ACERPS & CORYAV & FRAXEX & Other \\
    \noalign{\vskip 2pt}\hline\noalign{\vskip 2pt}
\multirow{4}{*}{\rotatebox{90}{Reference}}
& ACERPS & \textbf{97.4} & 0.9 & 0.7 & 0.9 \\ 
& CORYAV & 28.4 & \textbf{62.7} & 0 & 9 \\ 
& FRAXEX & 85.9 & 8.2 & \textbf{3.5} & 2.4 \\ 
& Other & 18.5 & 21.5 & 1.5 & \textbf{58.5} \\
    \end{tabular}
\end{subtable}
\end{table}

\begin{table}[htbp!]
    \centering
    \caption{Confusion matrices for the random forest model ii) based on the 12 original numeric statistics $P_{ori}$ showing (a) absolute classification counts and (b) row-normalized percentages (per true species). Overall performance: Accuracy: 79\%, log-loss: 0.53, Cohen’s $\kappa$: 0.46, macro-averaged sensitivity: 53.9\%.}
    \label{tab:confusionMat_ori}
    \begin{subtable}[t]{0.495\textwidth}
\centering
\caption{}
\begin{tabular}{@{\hskip 4pt}c@{\hskip 6pt}l@{\hskip 6pt}c@{\hskip 6pt}c@{\hskip 6pt}c@{\hskip 6pt}c@{\hskip 4pt}}
 &  & \multicolumn{4}{c}{Prediction} \\
 &  & ACERPS & CORYAV & FRAXEX & Other \\
    \noalign{\vskip 2pt}\hline\noalign{\vskip 2pt}
\multirow{4}{*}{\rotatebox{90}{Reference}}
& ACERPS & \textbf{517} & 8 & 11 & 5 \\ 
& CORYAV & 20 & \textbf{42} & 2 & 3 \\ 
& FRAXEX & 68 & 4 & \textbf{12} & 1 \\ 
& Other & 27 & 10 & 0 & \textbf{28} \\
    \end{tabular}
\end{subtable}
    \begin{subtable}[t]{0.495\textwidth}
\centering
\caption{}
\begin{tabular}{@{\hskip 4pt}c@{\hskip 6pt}l@{\hskip 6pt}c@{\hskip 6pt}c@{\hskip 6pt}c@{\hskip 6pt}c@{\hskip 4pt}}
 &  & \multicolumn{4}{c}{Prediction} \\
 &  & ACERPS & CORYAV & FRAXEX & Other \\
    \noalign{\vskip 2pt}\hline\noalign{\vskip 2pt}
\multirow{4}{*}{\rotatebox{90}{Reference}}
& ACERPS & \textbf{95.6} & 1.5 & 2 & 0.9 \\ 
& CORYAV & 29.9 & \textbf{62.7} & 3 & 4.5 \\ 
& FRAXEX & 80 & 4.7 & \textbf{14.1} & 1.2 \\ 
& Other & 41.5 & 15.4 & 0 & \textbf{43.1} \\ 
    \end{tabular}
\end{subtable}
\end{table}

\begin{table}[htbp!]
    \centering
    \caption{Confusion matrices for the gradient boosting model i) based on all 49 available numeric statistics $P_{all}$ showing (a) absolute classification counts and (b) row-normalized percentages (per true species). Overall performance: Accuracy: 81.3\%, log-loss: 0.48, Cohen’s $\kappa$: 0.53, macro-averaged sensitivity: 57.9\%.}
    \label{tab:confusionMat_boost_all}
    \begin{subtable}[t]{0.495\textwidth}
\centering
\caption{}
\begin{tabular}{@{\hskip 4pt}c@{\hskip 6pt}l@{\hskip 6pt}c@{\hskip 6pt}c@{\hskip 6pt}c@{\hskip 6pt}c@{\hskip 4pt}}
 &  & \multicolumn{4}{c}{Prediction} \\
 &  & ACERPS & CORYAV & FRAXEX & Other \\
    \noalign{\vskip 2pt}\hline\noalign{\vskip 2pt}
\multirow{4}{*}{\rotatebox{90}{Reference}}
& ACERPS & \textbf{523} & 6 & 8 & 4 \\ 
& CORYAV & 17 & \textbf{41} & 3 & 6 \\ 
& FRAXEX & 62 & 4 & \textbf{17} & 2 \\ 
& Other & 17 & 11 & 2 & \textbf{35} \\  
    \end{tabular}
\end{subtable}
    \begin{subtable}[t]{0.495\textwidth}
\centering
\caption{}
\begin{tabular}{@{\hskip 4pt}c@{\hskip 6pt}l@{\hskip 6pt}c@{\hskip 6pt}c@{\hskip 6pt}c@{\hskip 6pt}c@{\hskip 4pt}}
 &  & \multicolumn{4}{c}{Prediction} \\
 &  & ACERPS & CORYAV & FRAXEX & Other \\
    \noalign{\vskip 2pt}\hline\noalign{\vskip 2pt}
\multirow{4}{*}{\rotatebox{90}{Reference}}
& ACERPS & \textbf{96.7} & 1.1 & 1.5 & 0.7 \\ 
& CORYAV & 25.4 & \textbf{61.2} & 4.5 & 9 \\ 
& FRAXEX & 72.9 & 4.7 & \textbf{20} & 2.4 \\ 
& Other & 26.2 & 16.9 & 3.1 & \textbf{53.8} \\
    \end{tabular}
\end{subtable}
\end{table}

\begin{table}[htbp!]
    \centering
    \caption{Confusion matrices for the gradient boosting model ii) based on the 12 original numeric statistics $P_{ori}$ showing (a) absolute classification counts and (b) row-normalized percentages (per true species). Overall performance: Accuracy: 80.7\%, log-loss: 0.52, Cohen’s $\kappa$: 0.51, macro-averaged sensitivity: 57.8\%.}
    \label{tab:confusionMat_boost_ori}
    \begin{subtable}[t]{0.495\textwidth}
\centering
\caption{}
\begin{tabular}{@{\hskip 4pt}c@{\hskip 6pt}l@{\hskip 6pt}c@{\hskip 6pt}c@{\hskip 6pt}c@{\hskip 6pt}c@{\hskip 4pt}}
 &  & \multicolumn{4}{c}{Prediction} \\
 &  & ACERPS & CORYAV & FRAXEX & Other \\
    \noalign{\vskip 2pt}\hline\noalign{\vskip 2pt}
\multirow{4}{*}{\rotatebox{90}{Reference}}
& ACERPS & \textbf{520} & 5 & 11 & 5 \\ 
& CORYAV & 17 & \textbf{47} & 1 & 2 \\ 
& FRAXEX & 69 & 2 & \textbf{12} & 2 \\ 
& Other & 23 & 9 & 0 & \textbf{33} \\
    \end{tabular}
\end{subtable}
    \begin{subtable}[t]{0.495\textwidth}
\centering
\caption{}
\begin{tabular}{@{\hskip 4pt}c@{\hskip 6pt}l@{\hskip 6pt}c@{\hskip 6pt}c@{\hskip 6pt}c@{\hskip 6pt}c@{\hskip 4pt}}
 &  & \multicolumn{4}{c}{Prediction} \\
 &  & ACERPS & CORYAV & FRAXEX & Other \\
    \noalign{\vskip 2pt}\hline\noalign{\vskip 2pt}
\multirow{4}{*}{\rotatebox{90}{Reference}}
& ACERPS & \textbf{96.1} & 0.9 & 2 & 0.9 \\ 
& CORYAV & 25.4 & \textbf{70.1} & 1.5 & 3 \\ 
& FRAXEX & 81.2 & 2.4 & \textbf{14.1} & 2.4 \\ 
& Other & 35.4 & 13.8 & 0 & \textbf{50.8} \\ 
    \end{tabular}
\end{subtable}
\end{table}

\begin{table}[htbp!]
    \centering
    \caption{Variable loadings rounded to two decimal places of the first five principal components (explaining 49.7, 15.8, 9, 5.2, and 3.7\% of the total variance, respectively). The five variables with the largest absolute loadings for PC1 and PC2 and additional variables highlighted in Figure~\ref{fig:PCA} are listed along with all variables with absolute loadings $\geq 0.2$ for PC1-5 for completeness. For each variable, the largest absolute loading across components is shown in bold.}
    \label{tab:PCA_loadings}
    \begin{tabular}{@{\hskip 4pt}l@{\hskip 6pt}c@{\hskip 6pt}c@{\hskip 6pt}c@{\hskip 6pt}c@{\hskip 6pt}c@{\hskip 4pt}}
Variable & PC1 & PC2 & PC3 & PC4 & PC5 \\
    \noalign{\vskip 2pt}\hline\noalign{\vskip 2pt}
Height &  0.12 & -0.11 & \textbf{-0.23} &  0.13 &  0 \\
DBH & \textbf{ 0.18} & -0.04 & -0.06 &  0.08 &  0.02 \\
Int-w(A) & -0.01 & \textbf{ 0.33} &  0.04 &  0.01 &  0.05 \\
Int-w(a) & -0.01 & \textbf{ 0.33} &  0.05 &  0.01 &  0.05 \\
Int-w(M) & -0.01 & \textbf{ 0.33} &  0.05 &  0.01 &  0.06 \\
Int-w(m) & -0.01 & \textbf{ 0.33} &  0.05 &  0.02 &  0.06 \\
Int-l(A) &  0.08 & \textbf{ 0.25} & -0.18 &  0.24 &  0.10 \\
Int-l(a) &  0.07 & \textbf{ 0.26} & -0.17 &  0.25 &  0.11 \\
Int-l(M) &  0.07 &  0.25 & -0.17 & \textbf{ 0.26} &  0.11 \\
Int-l(m) &  0.07 &  0.25 & -0.16 & \textbf{ 0.26} &  0.12 \\
Ext(A) & -0.06 & \textbf{ 0.24} &  0.24 & -0.16 & -0.12 \\
Ext(a) & -0.06 & \textbf{ 0.24} &  0.24 & -0.16 & -0.12 \\
Ext(M) & -0.06 & \textbf{ 0.24} &  0.24 & -0.16 & -0.12 \\
Ext(m) & -0.06 & \textbf{ 0.24} &  0.24 & -0.16 & -0.12 \\
LeafN & \textbf{ 0.20} &  0 &  0.05 & -0.03 &  0 \\
InnerN & \textbf{ 0.20} &  0 &  0.05 & -0.02 &  0 \\
B1 & \textbf{ 0.20} &  0 &  0.05 & -0.02 &  0 \\
B2 & 0.06 & 0.02 & 0.01 & 0.17 & \textbf{0.21} \\
Cherry & \textbf{ 0.20} &  0 &  0.05 & -0.03 &  0 \\
mD &  0.11 &  0.10 & \textbf{-0.31} & -0.17 & -0.20 \\
mI' & -0.03 &  0.01 & -0.12 & -0.41 & \textbf{ 0.47} \\
mIw & -0.03 &  0 & -0.09 & -0.40 & \textbf{ 0.50} \\
s-shape & \textbf{ 0.20} &  0 &  0.04 & -0.04 & -0.01 \\
VLD &  0.03 &  0.09 & -0.28 & \textbf{-0.29} & -0.27 \\
ALD &  0.09 &  0.09 & \textbf{-0.34} & -0.16 & -0.23 \\
AVD &  0.09 &  0.09 & \textbf{-0.34} & -0.17 & -0.23 \\
sqrt-CLe & \textbf{ 0.18} &  0.03 & -0.10 & -0.05 & -0.07 \\
TopRes & -0.03 & -0.08 &  0.18 &  0.19 & \textbf{-0.30} \\
    \end{tabular}
\end{table}

\begin{figure}[ht]
\centering
\begin{subfigure}[t]{0.95\textwidth}
\centering
\includegraphics[width=0.97\textwidth]{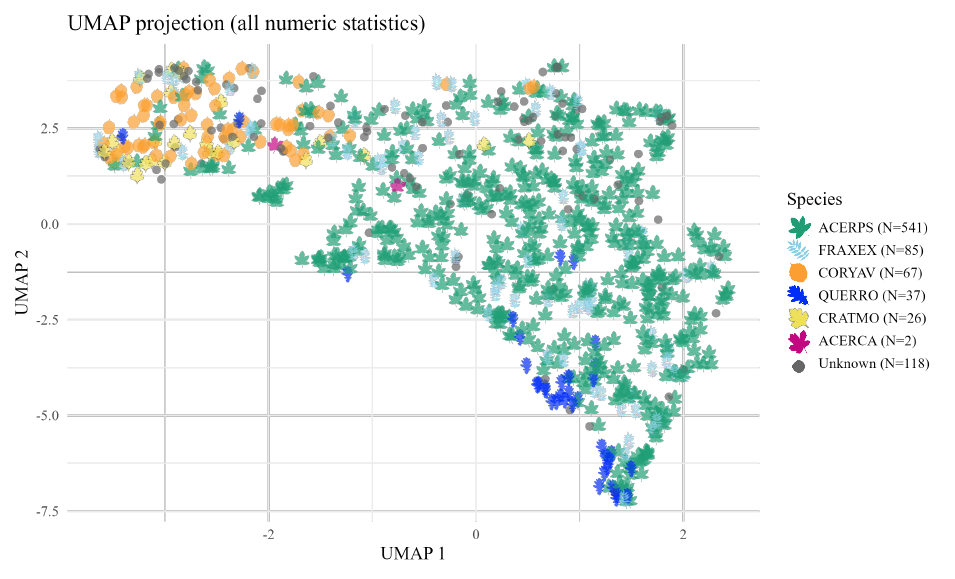} \vspace{-0.2cm}
\caption{}
\end{subfigure}
\begin{subfigure}[t]{0.495\textwidth}
\centering
\includegraphics[width=0.98\textwidth]{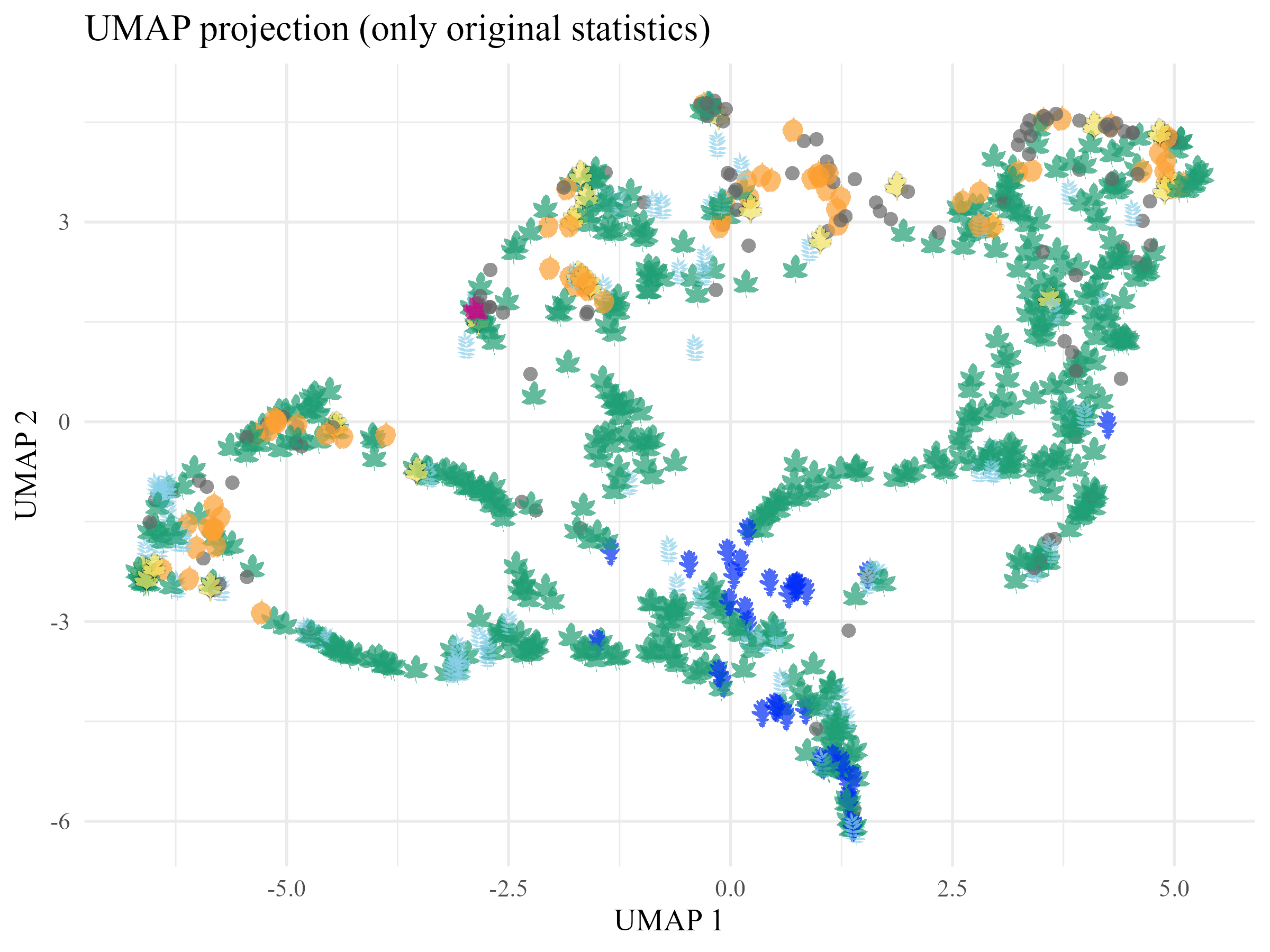}
\caption{}
\end{subfigure}
\begin{subfigure}[t]{0.495\textwidth}
\centering
\includegraphics[width=0.98\textwidth]{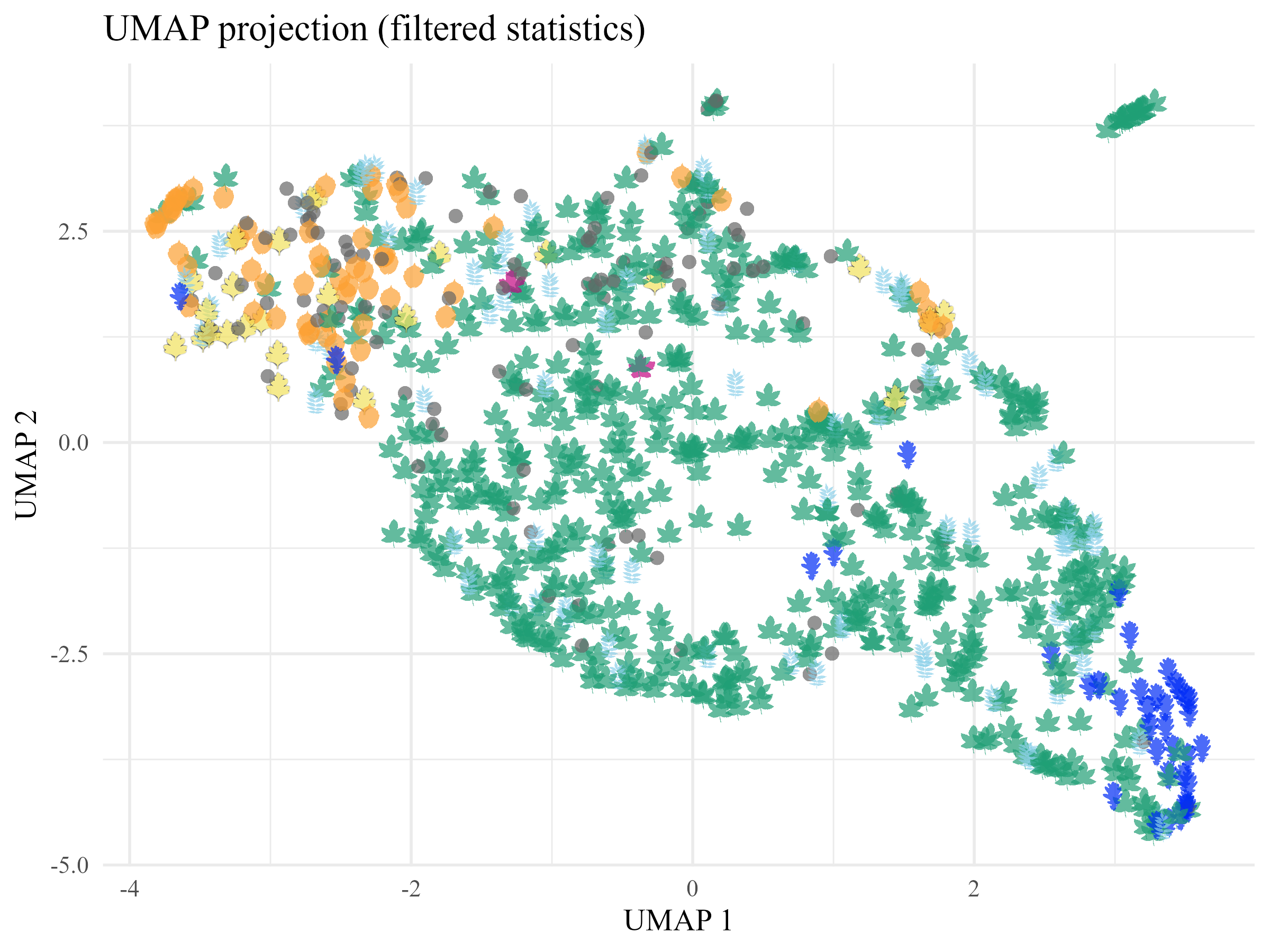}
\caption{}
\end{subfigure} \vspace{-0.1cm}
\caption{Two-dimensional projection of the selected \enquote{best} QSMs based on (a) all 49 numeric statistics, (b) only the 12 original QSM/TLS statistics, and (c) only the 17 filtered statistics. Colors and shapes indicate tree species for all subfigures; the legend reports the total number of trees per species (N).}
\label{fig:umap_proj}
\end{figure}

\begin{table}[htbp!]
    \centering
    \caption{Variable importance of predictors for the random forest models based on (a) $P_{ori}$ and (b) $P_{fil}$ averaged across the 10 outer cross-validation folds and shown in descending order. The standard deviation (SD) as well as the mean rank across folds (with corresponding SD of rank) quantify the stability of variable importance.}
    \label{tab:var_imp_RF}
    \begin{subtable}[t]{0.495\textwidth}
\centering
\caption{RF model ii) based on $P_{ori}$}
\begin{tabular}{@{\hskip 4pt}l@{\hskip 6pt}c@{\hskip 6pt}c@{\hskip 6pt}c@{\hskip 6pt}c@{\hskip 4pt}}
Variable & $\overline{\text{Imp}}$ & Imp$_\text{SD}$ & $\overline{\text{Rank}}$  & Rank$_\text{SD}$ \\
    \noalign{\vskip 2pt}\hline\noalign{\vskip 2pt}
Height & 79.9 & 8.3 & 1 & 0 \\ 
CrownArea & 52.8 & 2.7 & 2.8 & 0.8 \\ 
DBH & 52.1 & 2.6 & 2.7 & 0.7 \\ 
Vol25-50 & 49.8 & 2.6 & 3.5 & 0.8 \\ 
Vol0-25 & 41 & 2.4 & 5.6 & 0.7 \\ 
Vol75-100 & 39.1 & 2.3 & 6.4 & 1.1 \\ 
Vol100-200 & 37.6 & 2.2 & 6.9 & 1.2 \\ 
Locx & 37.5 & 3.1 & 7.3 & 1.3 \\ 
Vol50-75 & 33.4 & 1.5 & 9 & 0.7 \\ 
Vol200+ & 30.4 & 1.7 & 10.2 & 0.6 \\ 
Locy & 28.2 & 2.8 & 10.9 & 0.9 \\ 
Volume & 25.5 & 2.7 & 11.7 & 0.7 \\
    \end{tabular}
\end{subtable}
    \begin{subtable}[t]{0.495\textwidth}
\centering
\caption{RF model iii)  based on $P_{fil}$}
\begin{tabular}{@{\hskip 4pt}l@{\hskip 6pt}c@{\hskip 6pt}c@{\hskip 6pt}c@{\hskip 6pt}c@{\hskip 4pt}}
Variable & $\overline{\text{Imp}}$ & Imp$_\text{SD}$ & $\overline{\text{Rank}}$  & Rank$_\text{SD}$ \\
    \noalign{\vskip 2pt}\hline\noalign{\vskip 2pt}
Int-w(m) & 67.9 & 3.9 & 1 & 0 \\ 
Int-l(m) & 39.7 & 1.4 & 2.1 & 0.3 \\ 
DBH & 37 & 2.1 & 3.1 & 0.6 \\ 
Ext(a) & 35.3 & 2.5 & 4.1 & 0.6 \\ 
Height & 32.9 & 1.2 & 4.7 & 0.7 \\ 
Vol100-200 & 29.9 & 1.1 & 6 & 0 \\ 
Locx & 25.3 & 1.7 & 8.1 & 1.4 \\ 
TC & 25.3 & 1.1 & 8.1 & 0.6 \\ 
sqrt-CLe & 25.2 & 1.6 & 8.4 & 1.2 \\ 
TopRes & 23 & 1.5 & 9.7 & 1.2 \\ 
mWomD & 20.3 & 1.7 & 11.1 & 0.9 \\ 
mIw & 18.6 & 1.7 & 12 & 0.9 \\ 
AVD & 16.9 & 1.2 & 12.7 & 0.7 \\ 
VLD & 14.2 & 1.3 & 14.9 & 1.1 \\ 
StemCount & 14.1 & 1.9 & 15.1 & 0.9 \\ 
Locy & 13.8 & 1.7 & 14.9 & 0.7 \\ 
B2 & 6.2 & 1.5 & 17 & 0 \\ 
    \end{tabular}
\end{subtable}
\end{table}

\end{document}